\documentclass[fleqn,usenatbib]{mnras}

\usepackage{amsmath}	
\usepackage{amssymb}	
\usepackage{newtxtext,newtxmath}

\usepackage[T1]{fontenc}
\usepackage{ae,aecompl}

\usepackage{graphicx}	

\usepackage{times}

\title[SPH \& MCRT]{Modelling of ionising feedback with Smoothed Particle Hydrodynamics and Monte Carlo Radiative Transfer on a Voronoi grid}

\author[M. A. Petkova et al.]{Maya A. Petkova,$^{1,2}$\thanks{E-mail: maya.petkova@uni-heidelberg.de (MAP)}
Bert Vandenbroucke,$^{2,3}$
Ian A. Bonnell,$^{2}$
\newauthor
J. M. Diederik Kruijssen$^{1}$
\\
$^{1}$Astronomisches Rechen-Institut, Zentrum f{\"u}r Astronomie der Universit{\"a}t Heidelberg, M{\"o}nchhofstra{\ss}e 12-14, 69120 Heidelberg, Germany\\
$^{2}$School of Physics and Astronomy, University of St Andrews, North Haugh, KY16 8SS, St Andrews, UK\\
$^{3}$Sterrenkundig Observatorium, Universiteit Gent, Krijgslaan 281, B-9000 Gent, Belgium
}
\date{Accepted XXX. Received YYY; in original form ZZZ}

\pubyear{2021}

\begin{document}
\label{firstpage}
\pagerange{\pageref{firstpage}--\pageref{lastpage}}
\maketitle

\begin{abstract}
The ionising feedback of young massive stars is well known to influence the dynamics of the birth environment and hence plays an important role in regulating the star formation process in molecular clouds. For this reason, modern hydrodynamics codes adopt a variety of techniques accounting for these radiative effects. A key problem hampering these efforts is that the hydrodynamics are often solved using smoothed particle hydrodynamics (SPH), whereas radiative transfer is typically solved on a grid. Here we present a radiation-hydrodynamics (RHD) scheme combining the SPH code \textsc{Phantom} and the Monte Carlo Radiative Transfer (MCRT) code \textsc{CMacIonize}, using the particle distribution to construct a Voronoi grid on which the MCRT is performed. 
We demonstrate that the scheme successfully reproduces the well-studied problem of D-type H~II region expansion in a uniform density medium. Furthermore, we use this simulation setup to study the robustness of the RHD code with varying choice of grid structure, density mapping method, and mass and temporal resolution. To test the scheme under more realistic conditions, we apply it to a simulated star-forming cloud reminiscing those in the Central Molecular Zone of our galaxy, in order to estimate the amount of ionised material that a single source could create. We find that a stellar population of several $10^3~\rm{M_{\odot}}$ is needed to noticeably ionise the cloud. Based on our results, we formulate a set of recommendations to guide the numerical setup of future and more complex simulations of star forming clouds.
\end{abstract}

\begin{keywords}
radiation:dynamics -- methods:numerical -- HII regions -- Galaxy:centre
\end{keywords}



\section{Introduction}
The radiative feedback of high-mass stars can have profound impact on the formation of stellar and planetary systems, as well as on the interstellar environment. Ionising radiation has been shown to be able to externally photoevaporate protoplanetary discs \citep{O'dell1994,Scally2001}, to restrict the mass accretion onto the source star \citep{Kuiper2018,Sartorio2019} and deplete some of the available gas for further star formation \citep{Dale2012,Walch2012,Ali2019,Decataldo2019}. The presence of high-mass stars has also been argued to affect the star formation rate (SFR) in their surroundings. Star formation may be triggered via the collect-and-collapse mechanism, which occurs when the shock swept up by an expanding H~II region collects sufficient amounts of mass to become gravitationally unstable and collapse. This mechanism was first introduced by \cite{Elmegreen1977}, and was expanded by later studies with numerical models, which showed that the masses of the newly produced stars depend on the thickness of the shocked shell \citep{Wunsch2010,Dale2011b}.
Another way of triggering star formation can be achieved through radiation-driven implosion, during which pre-existing cloud cores become unstable under gravity as they get compressed by an expanding H~II region \citep{Kessel-Deynet2003,Mellema2006,Bisbas2011}. However, ionising feedback is mainly expected to have a negative effect on the SFR by altogether dispersing the cloud containing the source stars \citep{Dale2015,Kruijssen2019b,Chevance2020}.

In order to study the above mechanisms, many different methods of incorporating ionising feedback into hydrodynamics simulations have been developed. Early work by \cite{Lucy1977} and \cite{Brookshaw1994} described the propagation of light as a diffusive process, which could be included directly into the hydrodynamics equations. The diffusion approximation only holds for an optically thick medium, and it was later equipped with a flux-limiter in order to handle optically thin environments, giving rise to the flux-limited diffusion method \citep[e.g.][]{Levermore1981}. Other authors have traced rays of light through the diffuse medium and computed the ionisation state of the material along them. This can be achieved either by following the ray from the source to a given point \citep[e.g.][]{Kessel-Deynet2000,Abel2002,Dale2005,Pawlik2008, Wise2011, Haid2019}, or by starting from a chosen point in space and following a ray back to its source \citep{Grond2019}.

In recent years, there have been efforts towards using Monte Carlo Radiative Transfer (MCRT) in combination with hydrodynamics codes, since MCRT treats radiative effects (such as casting of shadows) with high accuracy \citep{Harries2015, Vandenbroucke2018}. Such radiation-hydrodynamics (RHD) schemes have been successfully implemented for grid-based \citep{Harries2015,Smith2017} and moving mesh \citep{Vandenbroucke2018,Smith2020} hydro codes, but are still under development for Smoothed Particle Hydrodynamics (SPH) codes. This is due to the particle nature of SPH and the fact that, with a few exceptions \citep{Forgan2010,Lomax2016}, MCRT algorithms are typically performed on a grid. 

In this work we present for the first time a numerical scheme combining SPH and MCRT for modelling ionising radiation "on the fly". In order to reconcile the particle- and grid-based descriptions of the diffuse medium we have employed the use of Voronoi grids, which have been successfully used by other authors for post-processing SPH snapshots \citep[e.g.][]{Hubber2016,Koepferl2017}. We test the scheme by successfully applying it to the well-studied problem of spherical expansion of an H~II region. Furthermore,  we perform simulations with varying time step associated with the radiative feedback, spatial resolution, and properties of the interconnecting Voronoi grid, in order to fine tune the scheme's performance. 

Finally, we apply our RHD method to a non-uniform density medium, in order to study the effects that an ionising source would have on the galactic centre cloud G0.253+0.016 (also known as the Brick). 
The Brick is a high--density cloud in the Central Molecular Zone (CMZ) of our galaxy, and it is believed to be a likely progenitor of a future high--mass stellar cluster \citep{Longmore2012,Rathborne2015,Walker2015}. Despite its high density, the cloud shows limited signs of ongoing star formation \citep{Walker2021}, and does not yet contain any identified H~II regions. Therefore, the Brick is a unique system that may provide the initial conditions for H~II region formation and early evolution. Furthermore, similarly to the rest of the CMZ, the Brick has high density, pressure and temperature \citep{Ginsburg2016,Krieger2017,Barnes2020}, which are representative features of high-redshift star-forming environments, and hence we can probe the early evolution and impact of H~II regions under conditions characteristic of high-redshift galaxies \citep{Kruijssen2013}. In particular, the following questions of interest arise:
\begin{enumerate}
 \item What will be the ionised mass of a newly formed H~II region?
 \item How rapidly will the ionised mass increase?
\end{enumerate}

This paper is organised as follows. We describe the numerical setup and the simulation parameters in Section~\ref{sec:numerics}. We then present the simulations in uniform density medium in Section~\ref{sec:results}, followed by the simulations in clumpy medium in Section~\ref{sec:clumpy}. Finally, we summarise our findings and give recommendations for future RHD simulations in Section~\ref{sec:conclusion}.

\section{Numerical setup}
\label{sec:numerics}
\subsection{Code Overview}
\begin{figure}
	\centering
	\includegraphics[width=\columnwidth]{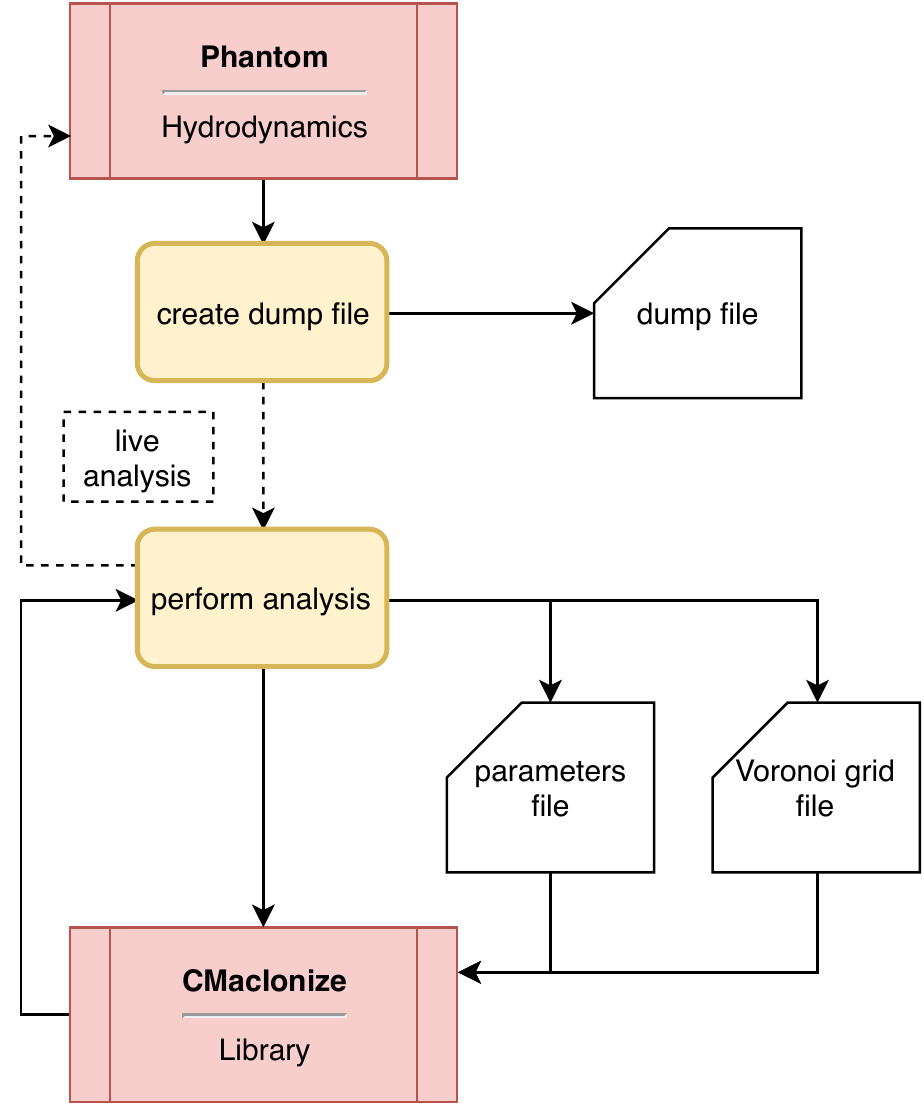}
	\caption{Code structure of the RHD scheme. During each simulation, \textsc{Phantom} performs hydrodynamics calculations and it ouputs its particle parameters in dump files at regular time intervals. We use the ``live analysis'' feature to externally modify some particle properties, each time a dump file is created. The modification is performed by calling \textsc{CMacIonize} in its library form and passing on particle positions and smoothing lengths. In addition, our analysis script outputs a parameters file and a Voronoi grid file, which are necessary for the function of CMI. Finally, CMI performs a radiative transfer calculation and it returns ionic fractions for all particles, which are used for modifying the particles' internal energies.}
	\label{fig:code-schematics}
\end{figure}

In this paper, we combine a Smoothed Particle Hydrodynamics (SPH) code and a Monte Carlo Radiative Transfer (MCRT) code in order to produce a radiation-hydrodynamics (RHD) scheme. This is achieved by linking the two codes in the following way. During the runtime of the SPH code its particle distribution is used to construct a Voronoi grid \citep{Voronoi1908}. The particle positions, smoothing lengths and masses are then used for calculating the density of each cell. The grid information is passed on to the MCRT code, which computes the ionic fraction of each grid cell by propagating a large number of photon packets through the Voronoi grid. The ionic fraction is mapped from the grid cells to the SPH particles, and this information is returned to the SPH code. Finally, the particles' internal energies are adjusted according to the ionic fraction. Each of the above steps is explained in greater detail later in this section.

We use the \textsc{Phantom} SPH code \citep{Price2018} for our RHD scheme. As a typical SPH code, it follows a set of Lagrangian particles and it integrates the corresponding fluid equations. \textsc{Phantom} was created with a focus on stellar, planetary and galactic astrophysics. The code is well suited for the study of star formation as it is three-dimensional and it includes the treatment of self-gravity, sink particles and gas-dust mixtures. Additionally, it can be run as both ideal and non-ideal MHD. Crucially for our work, \textsc{Phantom} can execute an analysis module, which is typically used to post-process the SPH outputs (dump files). There is a "live analysis" option, which ensures that the analysis module is executed during \textsc{Phantom}'s runtime.

The MCRT code that we have used is \textsc{CMacIonize} \citep{Vandenbroucke2018}. \textsc{CMacIonize} (CMI) models photoionisation on a variety of 3D grid structures, including Cartesian, AMR and Voronoi grid. The code uses two Voronoi grid construction algorithms: an incremental construction algorithm, based on the publicly available Voro++ library \citep{Rycroft2009}, and a Delaunay tessellation based algorithm, similar to \citet{Springel2010} and \citet{Vandenbroucke2016}. CMI contains a radiation hydrodynamics scheme in itself, allowing the user to choose between a fixed grid and a moving mesh grid hydrodynamics. The code models the photoionisation of hydrogen and helium self-consistently in the energy range between 13.6 and 54.4 eV. The ionisation states of a number of metal ions are modelled approximately, as they contribute to the cooling of the gas. Presently there is no treatment of lower energy photons, dust or radiation pressure (for more information see \citealp{Vandenbroucke2018}). Additionally, CMI exists in a library form, which allows it to be called from other C or Fortran programs.

The MCRT algorithm works as follows \citep[see Sec.~2.1.1 of][]{Vandenbroucke2018}. A large number of photon packets are emitted from one or multiple sources. The properties of each photon packet (i.e. origin, direction and wavelength) are sampled using the Monte Carlo technique. Each photon packet is then propagated through the density grid until a randomly sampled optical depth is reached or until the photon packet exits the grid. The algorithm accumulates the path lengths of the photon packets that pass through each cell to obtain an estimate of the local ionising field \citep{Lucy1999} from which the ionisation state can be calculated. Once a photon packet reaches its sampled optical depth, it is absorbed, and it may be re-emitted if a recombination event occurs locally. This newly emitted photon packet is propagated further if its frequency is high enough to cause ionisation (otherwise the photon packet escapes the simulation without contributing to the ionisation state of the diffuse medium). The photon packets created by recombination events are collectively known as the ``diffuse field''. In the simulations that we present here we omit the diffuse field for consistency with the StarBench test (see Section~\ref{sec:results}). However, it is trivial to take it into account in future work, as it is controlled by a standard \textsc{CMacIonize} input parameter.

The process described above, involving the random sampling and propagation of photon packets, is performed iteratively. At the beginning of the first iteration, \textsc{CMacIonize} assumes that all of the grid cells are fully ionised, and they are assigned an appropriately high temperature. After the propagation the of photon packets, the ionic fraction and the temperature of each grid cell is updated. After a number of iterations (typically $\sim 10$), the ionic fractions and temperatures converge. Note that if we run \textsc{CMacIonize} with a diffuse medium consisting only of hydrogen, the temperature of the ionised gas is set to a constant value (usually $10^4$~K), while if we assume a mix of elements, the temperature is computed iteratively. Despite the fact that \textsc{CMacIonize} has its own cell temperatures, we do not take these values into account in our simulations and we leave it up to the hydrodynamics to assign the gas temperatures based on a particle's ionic fraction. This is done for consistency with the StarBench setup.

The connection between the two codes is established by using \textsc{Phantom}'s option for performing live analysis and \textsc{CMacIonize}'s library functionality (see Figure~\ref{fig:code-schematics}). The CMI library is called as part of \textsc{Phantom}'s analysis module, which passes along information about the SPH particle positions, masses and smoothing lengths, and receives the neutral hydrogen fraction of each particle. The module then uses the neutral hydrogen fractions in order to modify the internal energies of the SPH particles in \textsc{Phantom}. Additionally, the analysis module generates a \textsc{CMacIonize} parameters file necessary for the execution of the MCRT, together with a file containing the positions of a set of Voronoi cell generating sites.

This particular way of linking the two codes uses a fixed time step for the radiative feedback, which coincides with the SPH output frequency. Care has been taken when choosing the appropriate time step, and we will address this further in Section~\ref{sec:results}.

\subsection{Voronoi grid}
\begin{figure*}
	\centering
	\includegraphics[width=\columnwidth]{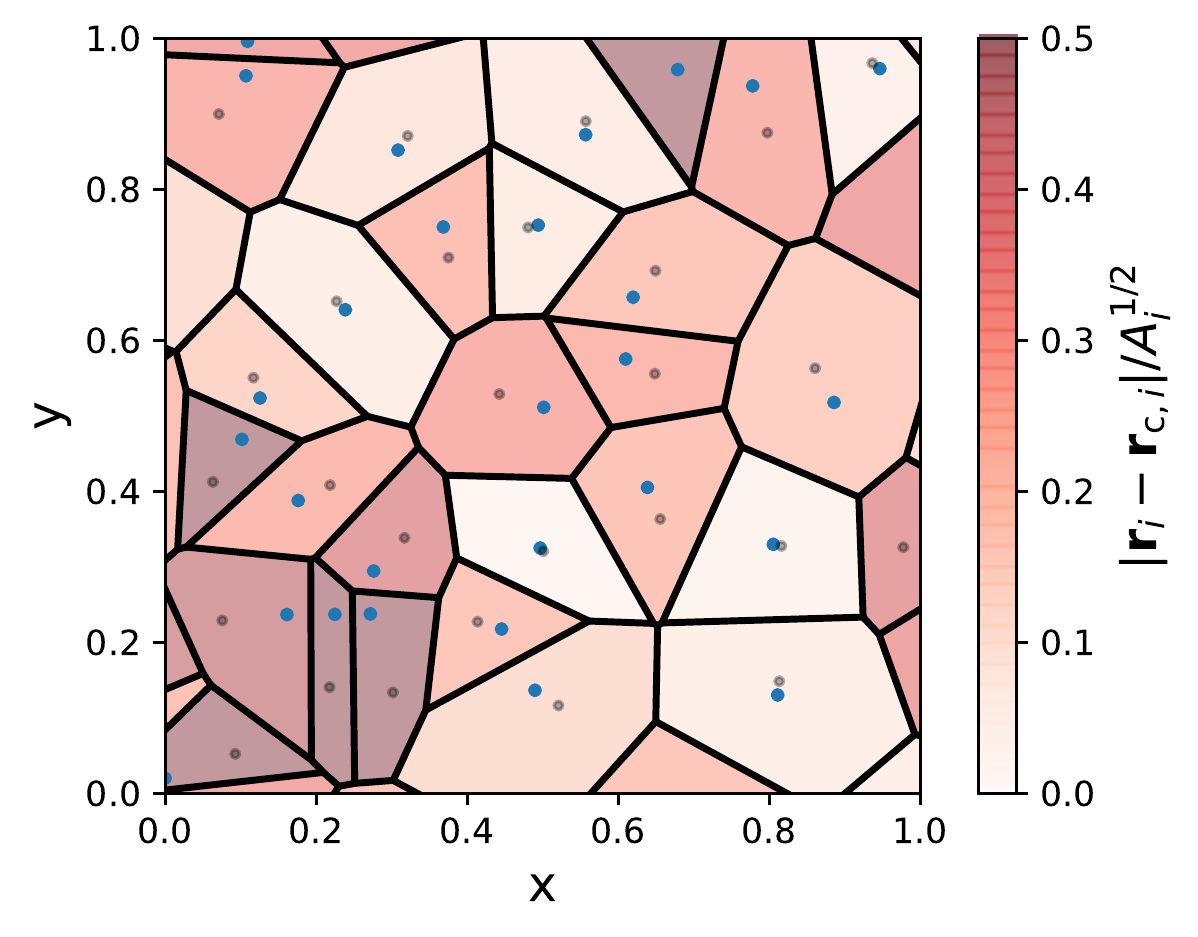}
	\includegraphics[width=\columnwidth]{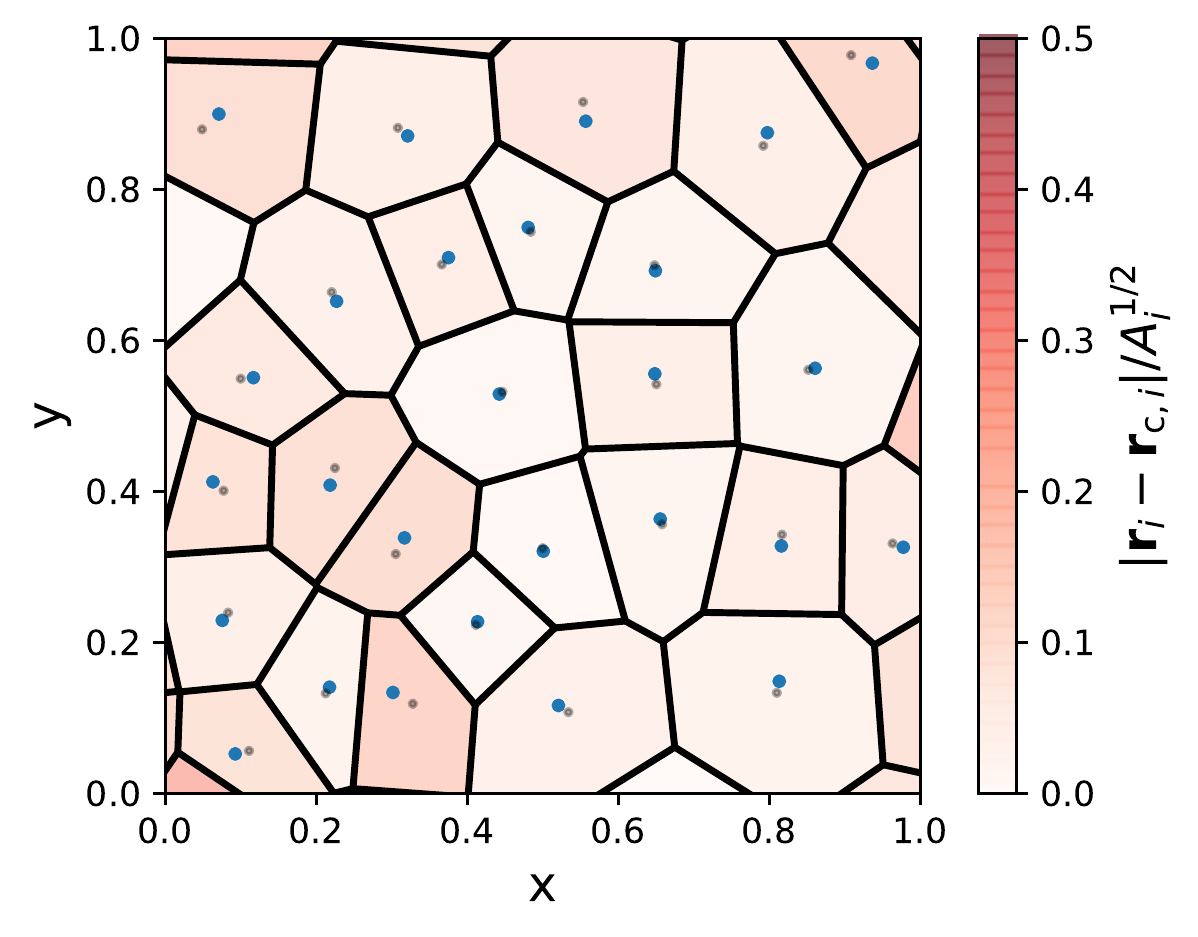}
	\caption{An example of a 2D Voronoi grid generated around randomly sampled points (\textit{left}), and its regularised counterpart (\textit{right}). In both panels, the grid generating sites are marked with blue circles, and the centroid of each grid cell is marked with a grey circle. The centroid points of the left panel serve as generating sites for the right panel (i.e. one Lloyd iteration has been performed going from left to right). The colour of each grid cell indicates the distance between the generating site and the centroid in units of the cell size.}
	\label{fig:voro-example}
\end{figure*}
A Voronoi grid is a structure consisting of polyhedral cells built around a set of generating sites, such that each cell wall bisects the distance between two neighbouring sites \citep{Voronoi1908}. This means that the choice of generating sites uniquely defines the grid and its structural properties. In this work we have used two different choices of generating sites and have compared how the simulation outcome depends on the grid.

The first type of grid that we have used is one where the generating sites coincide with the particle positions. This is the most intuitive and basic way of constructing a Voronoi grid from a set of particles, and it ensures that the concentration of grid cells follows that of the particles. This type of grid, however, has a known issue when large density gradients are present. The cells at the interface between low density and high density material become elongated and under-resolve the interface \citep{Koepferl2017}.

Since a large density gradient is to be expected in our simulations, we have also considered a regularisation of the grid in order to improve the resolution. The regularisation has been performed using Lloyd's algorithm \citep{Lloyd1982}, which is an iterative process. At each iteration the generating sites of the cells are moved to the cell centroids and the grid is subsequently reconstructed. For the simulations presented here we have used five Lloyd's iterations, which have been sufficient to alter the grid structure significantly (see Section~\ref{sec:clumpy}).

A 2D illustration of the Lloyd's algorithm is shown in Figure~\ref{fig:voro-example}. In the left panel of the figure, we have randomly sampled x- and y-positions for a set of generating sites (marked with blue circles), and we have constructed a Voronoi grid around them. Note that we have sampled 100 generating sites with coordinates between $-0.5$ and 1.5, but we only show the [0,1] range in x- and y-coordinates, in order to avoid boundary effects. The left panel also contains the positions of the cell centroids, shown in grey circles. We can see that the less circular cells typically have greater distances between their generating sites and their centroids. This is also emphasised by the cell colour, which corresponds to the distance between a cell's generating site (positioned at $\mathbf{r}_{i}$) and its centroid (positioned at $\mathbf{r}_{{\rm c},i}$) divided by the cell size ($A_i^{1/2}$, where $A_i$ is the cell area). In the right panel of Figure~\ref{fig:voro-example} we plot the same grid after one iteration of Lloyd's algorithm. This means that we take the generating sites from the left panel and we move them to the centroids of their corresponding cells. We see that the grid cells in the right panel are a lot more circular than those in the top one, and have centroids which are much closer to their generating sites.

For both choices of generating sites in our simulations, the grids have been constructed using the direct incremental construction algorithm that is part of \textsc{CMacIonize}, and that is based on the publicly available Voro++ library\footnote{\url{http://math.lbl.gov/voro\%2B\%2B/}} \citep{Rycroft2009}.

\subsection{Density mapping}
Once the Voronoi grid is constructed, we need to assign densities to the cells. We have employed three different types of density mapping, in order to test the robustness of the simulation outcome. The first density mapping takes advantage of the fact that there is one grid cell per particle, and the cell density is given as the particle mass divided by the cell volume, or
\begin{equation}
\rho_{{\rm m},i} = \frac{m_a}{V_i},
\label{eq:MonV}
\end{equation}
where `$a$' is the index of the particle and `$i$' is the index of its corresponding cell. This method (labelled as `M/V') is very quick and easy to compute, however it is prone to large uncertainties since there is no direct correspondence between a particle's smoothing length and the volume of its Voronoi cell. Also note that we can only apply it to the non-regularised Voronoi grid because after Lloyd's algorithm is applied not all cells necessarily contain exactly one particle, and we lose the correspondence between `$a$' and `$i$'.

The second density mapping method (labelled as `Centroid') uses the SPH density estimated at the cell's centroid. The cell density is given by:

\begin{equation}
\rho_{{\rm c},i} = \sum_{a=1}^{N} m_a W(|\mathbf{r}_a - \mathbf{r}_{{\rm c},i}|, h_a),
\label{eq:centroid}
\end{equation}
where `$i$' is the index of the cell with a centroid position at $\mathbf{r}_{{\rm c},i}$, $\mathbf{r}_a$ and $h_a$ are the position and smoothing length of particle `$a$', $N$ is the total number of particles\footnote{Note that when we use a kernel function which goes to zero for large radii (i.e. a kernel function with compact support), only some of the particles have non-zero contributions to the above sum. If $\mathbf{r}_{{\rm c},i}$ was the position of a particle, we would refer to the particles with non-zero contributions as that particle's neighbours. $N$ could then be replaced by $N_{\rm neigh}$ in the above sum. A particle in our simulations has on average 58 neighbours.}, and $W$ is the kernel function. For all of our simulations we have used a cubic spline form for $W$ \citep{Monaghan1985}, which goes to zero for distances to the particle greater than $2h_a$. This mapping method is more consistent with how SPH uses densities, and it only requires slightly longer to compute than the previous method. The downside is that it does not conserve mass when mapping the particle densities onto the cells. Note that the magnitude of the error in the total mass depends on the particle distribution, and the Voronoi cell arrangement. Moreover, the error in the total mass is typically different at each ionisation time step, as the particles have rearranged themselves, and it is not cumulative as the simulation progresses.

Finally, we have also used density mapping (labelled as `Exact') which computes the exact average SPH density within the grid cell, as derived in \cite{Petkova2018}. The method calculates the integral of the kernel function over the cell volume, and it relates it to the cell density by:
\begin{eqnarray}
\rho_{{\rm p},i} & =  & \frac{1}{V_i} \int_{V_i}  \rho(\mathbf{r'}) \mathrm{d}V' \\
 & = & \frac{1}{V_i} \int_{V_i}  \sum_{a=1}^N m_a W(|\mathbf{r'}-\mathbf{r}_a|,h_a) \mathrm{d}V'\\
 & = & \frac{1}{V_i} \sum_{a=1}^N m_a \int_{V_i}  W(|\mathbf{r'}-\mathbf{r}_a|,h_a) \mathrm{d}V'.
\label{eq:exact}
\end{eqnarray}
Unlike the previous density mapping method, this one conserves the total mass when computing cell densities, which ensures better accuracy when performing the radiative transfer. For computational efficiency the values of the integral of the cubic spline kernel function have been pre--computed and used for interpolation. Even so, however, this method remains many times slower then the first two (see Appendix~\ref{sec:computing-time}).
\subsection{Ionic fraction mapping}
After the CMI library computes the ionic fraction of each grid cell, these values have to be mapped onto the particles. Analogously to the previous section, we have used three different ways of performing the ionic fraction mapping depending on the choice of density mapping.

The first type of density mapping (`M/V'; given by eq.~\ref{eq:MonV}) assumes that a particle and its corresponding cell contain the same amount of mass, and hence we can assume that they also contain the same amount of ionised mass. This leads to their ionic fractions being identical:
\begin{equation}
f_{{\rm m},a} = f_{i}.
\label{eq:MonV-inv}
\end{equation}

In the second type of density mapping (`Centroid'; eq.~\ref{eq:centroid}) we can think of each particle as contributing to the density of each cell that it overlaps with. If we multiply these density contributions with the corresponding cell volumes and ionic fractions, they become ionised mass contributions. The inverse mapping for this method is then given by:

\begin{equation}
f_{{\rm c},a} = \frac{\sum_{i=1}^{N} f_{i} V_i m_a W(|\mathbf{r}_a - \mathbf{r}_{{\rm c},i}|, h_a)}{\sum_{i=1}^{N} V_i m_a W(|\mathbf{r}_a - \mathbf{r}_{{\rm c},i}|, h_a)}.
\label{eq:centroid-inv}
\end{equation}
Note that the denominator\footnote{It is possible, in rare cases, that the denominator could be zero. This happens if a particle's kernel function does not overlap with the centroid of any of the cells. In these cases, we conservatively assume that the affected particles are neutral. If we have one odd neutral particle within an H II region, the dynamics of the gas will not change significantly, but even a single ionised particle in the middle of a neutral region can drive some local expansion.} is included in order to ensure that the particle's ionic fraction could not exceed the value of 1. Additionally, since the corresponding method of density mapping does not conserve mass, this type of ionic fraction mapping does not conserve ionised mass either. 

Finally, in the exact density mapping (`Exact'; eq.~\ref{eq:exact}) each particle contributes a fraction of its mass to the grid cells it overlaps with. By multiplying this fractional mass contribution with the ionic fraction of the corresponding grid cell, we can construct a mapping which conserves the ionised mass:
\begin{equation}
f_{{\rm p},a} = \sum_{i=1}^{N} f_{i} \int_{V_i}  W(|\mathbf{r'}-\mathbf{r}_a|,h_a) \mathrm{d}V'.
\label{eq:exact-inv}
\end{equation}

\subsection{Simulation setup}
We present two sets of simulations -- one reproducing the D-type expansion of an H~II region in a uniform medium, and one ionising a clumpy, star-forming cloud in the galactic centre. For consistency and ease of comparison we have kept most of the physical parameters the same between the two sets (see Table \ref{table:d-type-setup}). The individual setups are presented in more detail in the beginning of Section~\ref{sec:results} and~\ref{sec:clumpy}.

\begin{table}
\caption{Physical parameters for the two sets of simulations.}
  \centering
  \begin{tabular}{llll}
    \hline
    Parameter & Units & StarBench test (`SB') & The Brick (`B')\\
    \hline
    \hline
    $\dot Q$ & s$^{-1}$ & $10^{49}$ & $5\times 10^{50}$ \\
    $\mu_{\rm o}$ & & 1 & 1 \\
    $\mu_{\rm i}$ & & 0.5 & 0.5 \\
    $\rho_{\rm o}$ & g cm$^{-3}$ & $5.21 \times 10^{-21}$ & - \\
    $T_{\rm i}$ & K & $10^{4}$ & $10^{4}$ \\
    $c_{\rm i}$ & km s$^{-1}$ & 12.85 & 12.85 \\
    $T_{\rm o}$ & K & $100$ & 65 \\
    $c_{\rm o}$ & km s$^{-1}$ & 0.91 & 0.73 \\
    \hline
  \end{tabular}
\label{table:d-type-setup}
\end{table}

In all simulations \textsc{CMacIonize} has been performed with 10 iterations of $10^6$ photon packets each, in order to compute accurate ionic fractions. In all of our simulations we disregard the diffuse field and invoke the B-case recombination coefficient, $\alpha_B = 2.7 \times 10^{-13}$ cm$^3$ s$^{-1}$, which corresponds to an ionising temperature of $T_{\rm i} = 10^4$~K \citep{Osterbrock1989}. This assumes that any ionising photon released by a recombination event is absorbed locally (i.e. the so-called ``on-the-spot'' approximation). The photoionisation cross-section is assumed to be $\sigma = 6.3 \times 10^{-18}$ cm$^2$.

For the hydrodynamics we have used a polytropic equation of state with $\gamma=1.00011$ in order to mimic two-temperature isothermal gas, even though the exact value of $\gamma$ does not affect the H~II region expansion significantly, according to the models included in \cite{Bisbas2015}. The gas is composed purely of atomic hydrogen ($\mu_{\rm o} = 1$). After each time \textsc{CMacIonize} is called, the particles with ionic fraction higher than 0.5 have their internal energies set to correspond to a temperature of $T_{\rm i} = 10^4$~K and a mean molecular weight of $\mu_{\rm i} = 0.5$. 
This creates a high temperature (and energy) contrast between some neighbouring particles, which is commonly known in SPH as ``contact discontinuity''. In order to minimise numerical issues that may arise from the presence of this contact discontinuity, we include artificial thermal conductivity (with $\alpha=1$), which has been introduced as a necessary numerical correction by \citet{Price2008}.

\begin{table*}
  \centering
  \caption{List of the RHD simulations included in this paper, and their setup. The particle mass is denoted as $m_{\rm{part}}$, the time step is $\delta t$, and the final simulation time is $t_{\rm{end}}$. The mass of the full simulation box is $M_{\rm{cl}}$, and it has a side of $R_{\rm{cl}}$. Note that for all `SB' simulations the mean initial density, $\rho_{\rm o}$, is kept the same, and hence the ratio $M_{\rm{cl}}/R_{\rm{cl}}^3$ remains constant.}
  \begin{tabular}{lcccccccr}
    \hline
    Simulation name & Initial conditions & $m_{\rm{part}}$ [M$_{\rm{\odot}}$] & $\delta t$ [Myr] & $M_{\rm{cl}}$ [M$_{\rm{\odot}}$] & $R_{\rm{cl}}$ [pc] & $t_{\rm{end}}$ [Myr] & Density mapping & Lloyd iterations\\
    \hline
    \hline
    SB I & glass $128^3$ & $10^{-3}$ & $8.5 \times 10^{-4}$ & 2097 & 3 & 0.14 & Exact & yes\\
    SB Ia & glass $512^3$ & $10^{-3}$ & $1.36 \times 10^{-3}$ & 134218 & 12 & 2.5 & Centroid & yes\\
    SB II & glass $64^3$ & $10^{-3}$ & $1.7 \times 10^{-4}$ & 262 & 1.5 & 0.04 & Exact & no\\
    SB III & glass $64^3$ & $10^{-3}$ & $3.4 \times 10^{-4}$ & 262 & 1.5 & 0.04 & Exact & no\\
    SB IV & glass $64^3$ & $10^{-3}$ & $8.5 \times 10^{-4}$ & 262 & 1.5 & 0.04 & Exact & no\\
    SB V & glass $64^3$ & $10^{-3}$ & $1.02 \times 10^{-3}$ & 262 & 1.5 & 0.04 & Exact & no\\
    SB VI & glass $64^3$ & $10^{-3}$ & $1.36 \times 10^{-3}$ & 262 & 1.5 & 0.04 & Exact & no\\
    SB VII & glass $64^3$ & $10^{-3}$ & $1.36 \times 10^{-3}$ & 262 & 1.5 & 0.04 & Exact & yes\\
    SB VIII & glass $64^3$ & $10^{-3}$ & $1.36 \times 10^{-3}$ & 262 & 1.5 & 0.04 & Centroid & no\\
    SB IX & glass $64^3$ & $10^{-3}$ & $1.36 \times 10^{-3}$ & 262 & 1.5 & 0.04 & Centroid & yes\\
    SB X & glass $64^3$ & $10^{-3}$ & $1.36 \times 10^{-3}$ & 262 & 1.5 & 0.04 & M/V & no\\
    SB XI & glass $64^3$ & $5.12 \times 10^{-1}$ & $2.72 \times 10^{-3}$ & 134218 & 12 & 0.04 & Exact & no\\
    SB XII & glass $64^3$ & $5.12 \times 10^{-1}$ & $2.72 \times 10^{-3}$ & 134218 & 12 & 0.04 & Exact & yes\\
    SB XIII & glass $64^3$ & $6.4 \times 10^{-2}$ & $2.16 \times 10^{-3}$ & 16777 & 6 & 0.04 & Exact & no\\
    SB XIV & glass $64^3$ & $6.4 \times 10^{-2}$ & $2.16 \times 10^{-3}$ & 16777 & 6 & 0.04 & Exact & yes\\
    SB XV & glass $64^3$ & $8 \times 10^{-3}$ & $1.71 \times 10^{-3}$ & 2097 & 3 & 0.04 & Exact & no\\
    SB XVI & glass $64^3$ & $8 \times 10^{-3}$ & $1.71 \times 10^{-3}$ & 2097 & 3 & 0.04 & Exact & yes\\
    SB XVII & glass $128^3$ & $1.25 \times 10^{-4}$ & $1.08 \times 10^{-3}$ & 262 & 1.5 & 0.04 & Exact & no\\
    SB XVIII & glass $128^3$ & $1.25 \times 10^{-4}$ & $1.08 \times 10^{-3}$ & 262 & 1.5 & 0.04 & Exact & yes\\
    B I & Brick sim. & 0.4463 & $5 \times 10^{-4}$ & 45133 & 6 & 0.04 & Exact & no\\
    B II & Brick sim. & 0.4463 & $5 \times 10^{-4}$ & 45133 & 6 & 0.04 & Exact & yes\\
    B III & Brick sim. & 0.4463 & $5 \times 10^{-4}$ & 45133 & 6 & 0.04 & Centroid & no\\
    B IV & Brick sim. & 0.4463 & $5 \times 10^{-4}$ & 45133 & 6 & 0.04 & Centroid & yes\\
    B V & Brick sim. & 0.4463 & $5 \times 10^{-4}$ & 45133 & 6 & 0.04 & M/V & no\\
    \hline
  \end{tabular}
\label{table:sim-params}
\end{table*}

\section{Simulations in uniform density medium}
\label{sec:results}
In this section, we present a set of simulations of a single ionising source in a uniform density medium (labelled as `SB'; see Table \ref{table:sim-params}). 
We adopt the setup of \citep[][StarBench]{Bisbas2015}, who established a benchmark test by performing simulations with a number of different 1D and 3D hydrodynamics codes coupled with ionising feedback. To reproduce their setup we have constructed a cube of uniform density ($\rho_{\rm o}=5.21 \times 10^{-21}$ g cm$^{-3}$) with a source of constant ionising luminosity ($\dot Q = 10^{49}$ s$^{-1}$) in the centre. The SPH particles have been initially arranged in a glass distribution \citep[available as part of the public simulation code SWIFT;][]{Borrow2018} to reduce numerical noise. All particles are initially at rest and have a temperature of $T_{\rm o} = 100$~K. The simulation volume has a side $R_{\rm{cl}}$, and total mass $M_{\rm{cl}}$, where $M_{\rm{cl}} = \rho_{\rm o} R_{\rm{cl}}^3$ and the total number of particles, N, is chosen so that the particle mass is $m_{\rm{part}} \approx 10^{-3}$ M$_{\rm{\odot}}$. We have used periodic boundary conditions to ensure that the cubic volume of gas does not diffuse out due to a pressure gradient. In order to avoid numerical inaccuracies, we have stopped each simulation before the point where the shock reaches the edge of the cube.

The ionising emission of the source creates a spherical H~II region, which undergoes rapid expansion. Determining the time evolution of the H~II region is a very well known problem, which has been studied extensively, both theoretically \citep{Spitzer1978,Hosokawa2006,Raga2012b,Raga2012a,Williams2018} and numerically \citep{Bisbas2015}. We begin this section by reviewing the theoretical work, which lays the foundation for interpreting our simulation results. We then demonstrate that our main simulation (`SB I') is in good agreement with the theory, as well as previous numerical work \citep{Bisbas2015}. For completeness, we also perform a much longer simulation (`SB Ia') which matches the `late test' from \citet{Bisbas2015}. In the remaining subsections we present variations\footnote{In all of the simulations derived from `SB I', we only focus on the first 0.04~Myr of the early StarBench test, since this is the time of most rapid expansion and we expect to see greatest variation with the choice of time step, grid/density mapping and resolution.} of `SB I', in which we alter the ionisation time step (Section~\ref{sec:timestepping}), grid type and density mapping method (Section~\ref{sec:grid&dens}), and resolution (Section~\ref{sec:resolution}). These additional simulations allow us to formulate recommendations for future applications of the radiation-hydrodynamics scheme, which we list in Section~\ref{sec:conclusion}.

\subsection{D-type expansion of an H~II region}
\label{sec:D-type}
\begin{figure*}
	\centering
	\includegraphics[width=\textwidth]{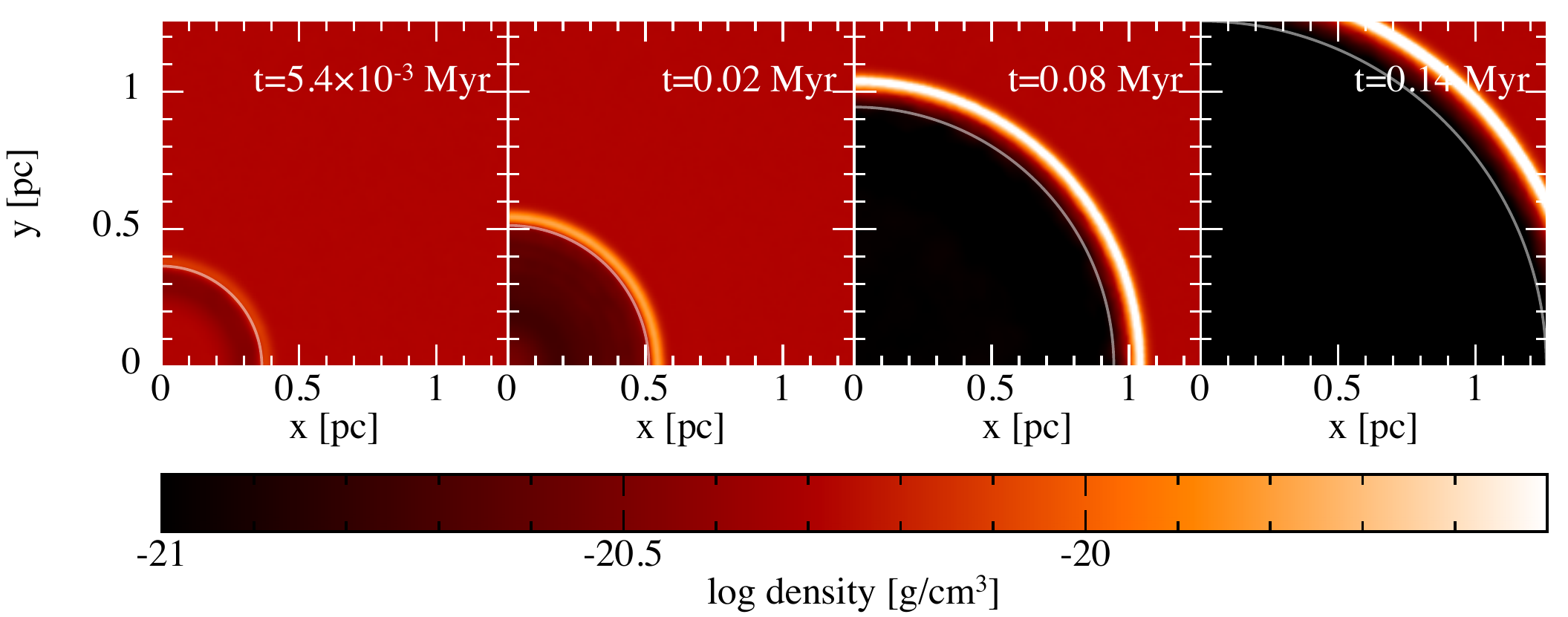}
	\caption{Density slices at $z=0$ of a simulation of the early StarBench test, computed with our RHD scheme (`SB I'; see Table \ref{table:sim-params}). Each panel corresponds to a different simulation snapshot, as labelled in the figure. The grey line marks the position of the ionisation front. Note that only one quadrant of the density slices is displayed.}
	\label{fig:early-expansion-density}
\end{figure*}
Let us consider an idealised scenario of a pure atomic hydrogen cloud of constant density with an embedded high mass star. The cloud is initially completely at rest, and we will ignore the effects of gravity. As the star begins to emit ionising radiation, an increasingly large spherical region around it will become ionised, until the amount of ionised gas is large enough that the hydrogen recombination rate will balance out the ionisation. This is achieved at the Str\"omgren radius \citep{Stromgren1939}:

\begin{equation}
R_{\rm{St}}= \left (\frac{3\dot Q m_p^2}{4\pi \alpha_B \rho_{\rm o}^2}\right )^{1/3},
\label{Stromgren}
\end{equation}
where $\dot Q$ is the number of ionising photons coming from the star per unit time, $m_p$ is the hydrogen mass, $\rho_{\rm o}$ is the density of the medium and $\alpha_B$ is the recombination coefficient.

The D-type expansion of an H~II region begins after the Str\"omgren radius is reached, and it is caused by the pressure imbalance between the hot, ionised material, and the cold, neutral gas, forcing the ionised gas to expand. The expanding material moves supersonicly into the neutral gas and it sweeps up material to form a thin shock that precedes the ionisation front \citep{Kahn1954}.

The rate of expansion of the H~II region described above has been the subject of many studies. By considering the pressure of the ionised gas and the ram pressure experienced by the neutral gas as the shock sweeps through it, we can arrive at the Spitzer solution for the radius of the ionisation front \citep{Spitzer1978}:

\begin{equation}
R_{\rm{Sp}}(t)= R_{\rm{St}} \left (1 + \frac{7}{4}\frac{c_{\rm i}t}{R_{\rm{St}}}\right )^{4/7},
\label{Spitzer}
\end{equation}
where $c_{\rm i}$ is the sound speed in the ionised medium.

The Spitzer solution was later improved by \citet{Hosokawa2006}, who noticed that the inertia of the shock promotes additional expansion. By solving the equation of motion of the shock, they obtained the Hosokawa-Inutsuka solution \citep{Hosokawa2006}:

\begin{equation}
R_{\rm{HI}}(t)= R_{\rm{St}} \left (1 + \frac{7}{4}\sqrt{\frac{4}{3}}\frac{c_{\rm i}t}{R_{\rm{St}}}\right )^{4/7}.
\label{Hosokawa-Inutsuka}
\end{equation}

The above equations provide good description of the early time expansion of an H~II region. If they are followed to later times, however, their functional form predicts that the H~II region will continue to expand infinitely. This is not what we expect, since the ionised material should eventually reach a pressure equilibrium with the neutral cloud. \cite{Raga2012a} have addressed this issue by including the thermal pressure of the neutral gas into the equation, governing the expansion of the ionisation front, $R_{\rm{IF}}$:

\begin{equation}
\frac{1}{c_{\rm i}}\frac{\mathrm{d}R_{\rm{IF}}(t)}{\mathrm{d}t} = \left (\frac{R_{\rm{St}}}{R_{\rm{IF}}(t)} \right )^{3/4} - \frac{\mu_{\rm i}T_{\rm o}}{\mu_{\rm o}T_{\rm i}} \left (\frac{R_{\rm{St}}}{R_{\rm{IF}}(t)} \right )^{-3/4}.
\label{RagaI}
\end{equation}
In the above, $\mu_{\rm o}$ and $\mu_{\rm i}$ are the mean molecular weights of the neutral and the ionised gas, and $T_{\rm o}$ and $T_{\rm i}$ are the neutral and ionised temperatures respectively. We can argue that typically $\mu_{\rm i}T_{\rm o}/(\mu_{\rm o}T_{\rm i}) \ll 1$ since $T_{\rm i} \gg T_{\rm o}$. Therefore, at early times $\mu_{\rm i}T_{\rm o}/(\mu_{\rm o}T_{\rm i}) (R_{\rm{St}}/R_{\rm{IF}}(t))^{-3/4}$ can be neglected and then equation~\ref{RagaI} leads to the Spitzer solution.

\cite{Raga2012b} improved on the work of \cite{Raga2012a} by including the inertia of the expanding shock, as it propagates through the neutral medium. Their consideration leads to an expansion equation:

\begin{equation}
\frac{1}{c_{\rm i}}\frac{\mathrm{d}R_{\rm{IF}}(t)}{\mathrm{d}t} = \sqrt{\frac{4}{3} \left (\frac{R_{\rm{St}}}{R_{\rm{IF}}(t)} \right )^{3/2} - \frac{\mu_{\rm i}T_{\rm o}}{2\mu_{\rm o}T_{\rm i}}}.
\label{RagaII}
\end{equation}
Similarly, at early times the term $\mu_{\rm i}T_{\rm o}/(2\mu_{\rm o}T_{\rm i})$ can be neglected, which is equivalent to neglecting the thermal pressure of the neutral gas. Equation~\ref{RagaII} then leads to the Hosokawa--Inutsuka solution.

The late time behaviour of an H~II region was further studied by \cite{Williams2018}, who developed a thick--shell solution capturing the decoupling of the ionisation front and the shock front. Even though this solution was intended for improving the late time evolutionary model of an H~II region, it is also relevant for our early time numerical studies with particle--based hydrodynamics, where we see thick shocks forming early on. \citet{Williams2018} point out that the Hosokawa--Inutsuka solution does not distinguish between the position of the shock front and the position of the ionisation front, since the shock is initially very thin. As we will see further in the paper, the thick shocks produced by the particle--based hydrodynamics will result in slightly smaller H~II region sizes than expected. Finally, \citet{Williams2018} use an acoustic approximation to demonstrate that at first the stagnation radius is exceeded by the expanding H~II region, which subsequently shrinks until it reaches its final size through a strongly damped oscillation.

\subsection{Comparison to StarBench}
\begin{figure}
	\centering
	\includegraphics[width=\columnwidth]{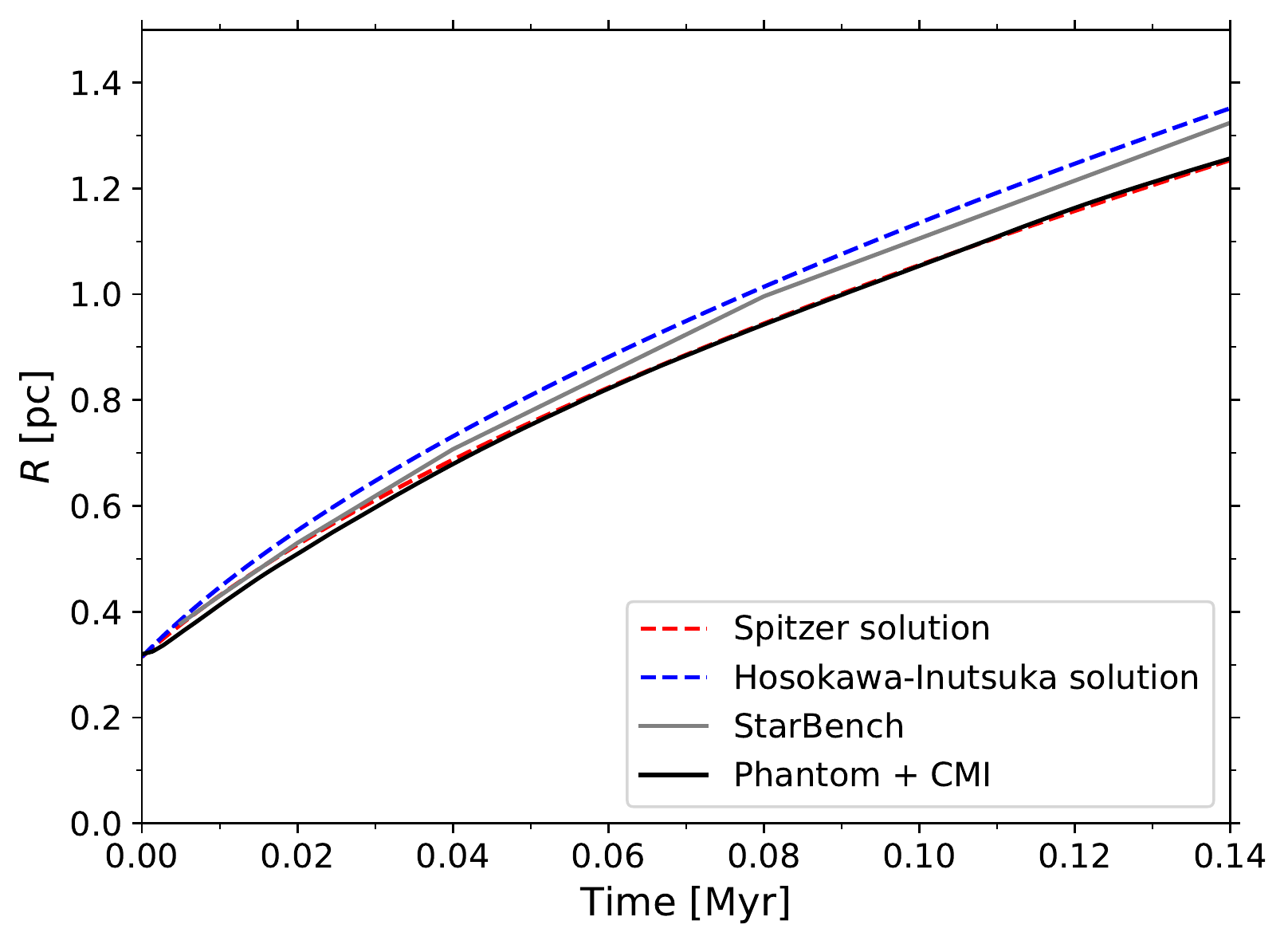}	
	\includegraphics[width=\columnwidth]{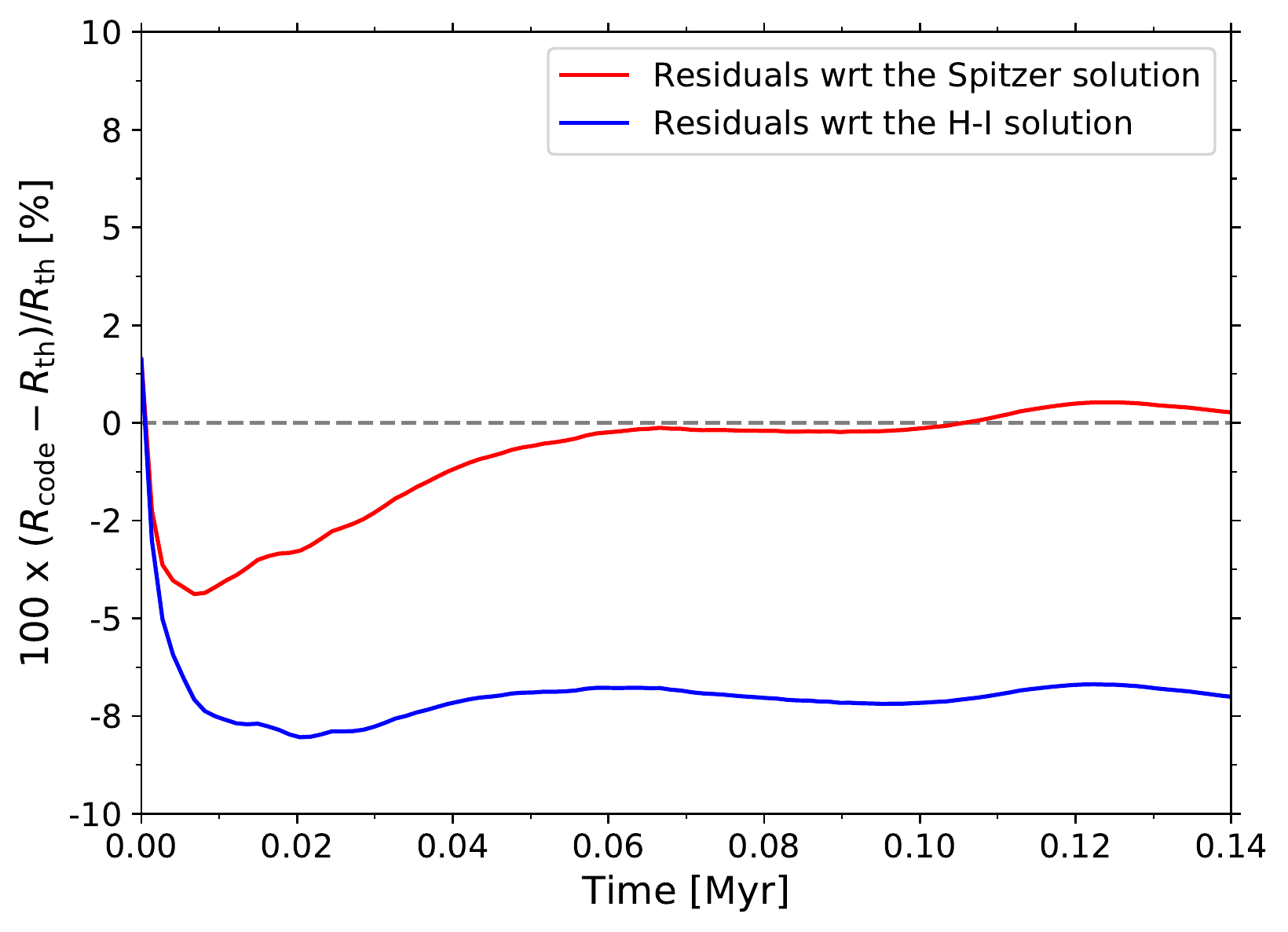}	
	\caption{The expansion of an H~II region, following the StarBench setup (`SB I'; see Table \ref{table:sim-params}). \textit{Top:} ionisation front radius as a function of time. The Spitzer and Hosokawa-Inutsuka solutions are shown in dashed lines, the averaged 3D StarBench solution is in solid grey \citep{Bisbas2015}, and the \textsc{Phantom} + \textsc{CMacIonize} live radiation hydrodynamics scheme is shown in solid black line. \textit{Bottom:} the relative error of the \textsc{Phantom} + \textsc{CMacIonize} solution compared to the Spitzer (red) and Hosokawa-Inutsuka (blue) solutions. In the label, $R_{\rm code}$ refers to the \textsc{Phantom} + \textsc{CMacIonize} solution, while $R_{\rm th}$ refers to one of the theoretical solutions.}
	\label{fig:early-expansion}
\end{figure}

We now demonstrate that the new radiation-hydrodynamics scheme presented in this paper reproduces the well-established StarBench `early test' \citep{Bisbas2015}, using a simulation setup with $N = 128^3$, $m_{\rm{part}} = 10^{-3}$~M$_{\rm{\odot}}$ and ionisation time step $\delta t = 8.5 \times 10^{-4}$~Myr (`SB I'; see Table \ref{table:sim-params}). Figure~\ref{fig:early-expansion-density} shows density slices at $z=0$ of the simulation at different times. We can see that qualitatively the D-type expansion is reproduced correctly, as an initially spherical region expands rapidly and sweeps up a shell of shocked gas. It can also be noticed that the shock is relatively thick even at the early stages of its formation. This is due to the use of SPH, and results in a slight suppression of the ionisation front, as discussed by \citet{Bisbas2015}.

In order to quantify the rate of expansion, we have determined the position of the ionisation front by selecting particles with ionic fractions between 0.2 and 0.8, and averaging their radial distance from the source. Note that there is a sharp transition between fully-ionised and neutral particles as a function of radius, and therefore the exact range of partial ionisation used to determine the position of the ionisation front, does not alter the result significantly. This procedure has been repeated for each SPH snapshot to produce the time evolution of the ionisation front presented in Figure~\ref{fig:early-expansion} (the position of the ionisation front is also shown in the four snapshots of Figure~\ref{fig:early-expansion-density} as a grey line). Additionally, the top panel of Figure~\ref{fig:early-expansion} contains the analytic curves of the Spitzer and Hosokawa-Initsuka solutions (see equations \ref{Spitzer} and \ref{Hosokawa-Inutsuka}), and the average, empirically derived expansion curve from the StarBench paper (\cite{Bisbas2015}). In the bottom panel we present the relative error of the ionisation front radius compared to the two theoretical solutions. Note that in the label $R_{\rm code}$ refers to the \textsc{Phantom} + \textsc{CMacIonize} solution, while $R_{\rm th}$ refers to one of the theoretical solutions.

Initially, the expansion of the ionisation front in our simulation happens more slowly than the one predicted by the theoretical models, however, the size of the ionised region soon reaches a value between the Spitzer and Hosokawa-Initsuka solutions. Similar behaviour can be seen in the average StarBench solution, even though their curve remains slightly above ours at all times (see Figure~\ref{fig:early-expansion}). This slight discrepancy is attributed mainly to the thicker shocks produced by SPH, as already mentioned above, but it could also be affected by the choice of ionisation scheme. In Figure~\ref{fig:early-expansion-density} we see that the peak of the shock and the position of the ionisation front are separated by a medium-density shock tail. A different ionisation scheme might place the ionisation front closer to the shock, making the results more consistent with the StarBench paper.

In addition to the `early test', we also perform the `late test' from the StarBench study (`SB Ia'; see Table \ref{table:sim-params}). The simulation setup uses $N=512^3$ particles with masses $m_{\rm part}=10^{-3}$~M$_{\rm{\odot}}$. The time step is $\delta t = 1.36 \times 10^{-3}$~Myr, and we evolve the simulation for 2.5~Myr, which allows us to observe the equilibrium state of the H~II region. Note that the initial gas temperature in this test ($T_{\rm o} = 1000$~K) is higher than in `SB I' in order for the equilibrium state to be reached sooner.

\begin{figure*}
	\centering
	\includegraphics[width=\textwidth]{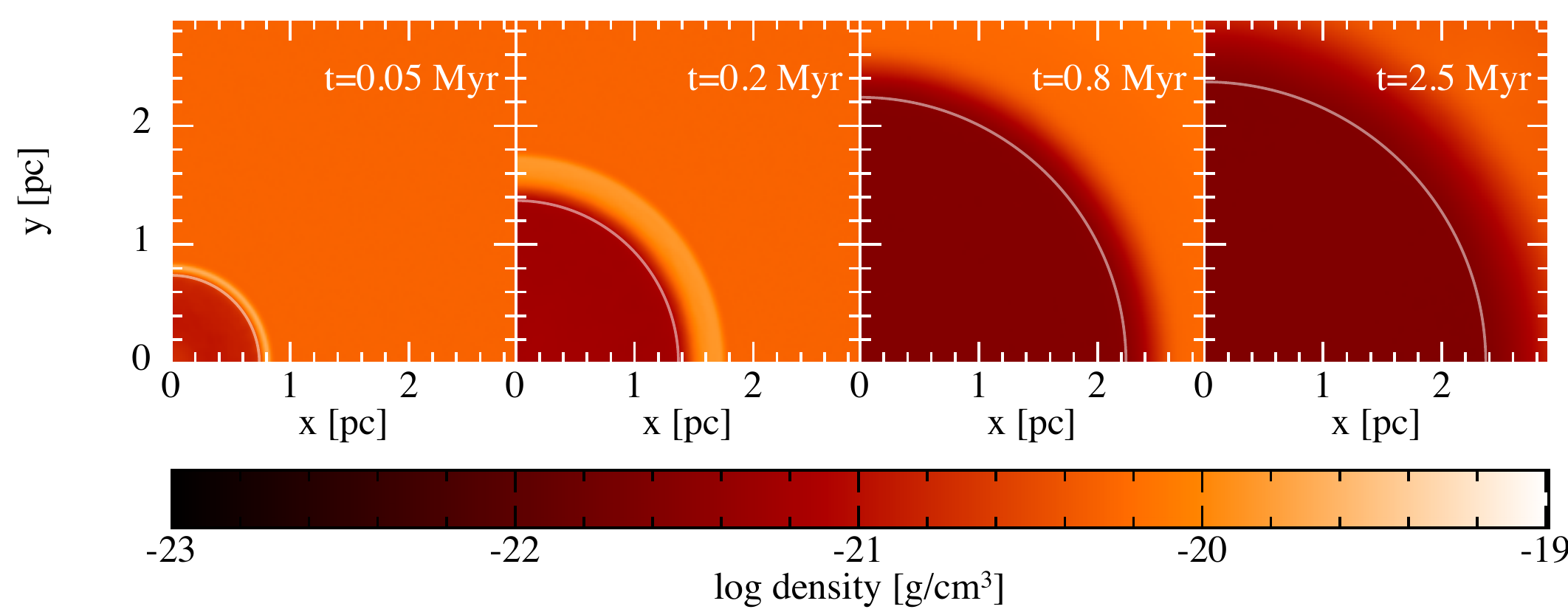}
	\caption{Density slices at $z=0$ of a simulation of the late StarBench test, computed with our RHD scheme (`SB Ia'; see Table \ref{table:sim-params}). Each panel corresponds to a different simulation snapshot, as labelled in the figure. The grey line marks the position of the ionisation front. Note that only one quadrant of the density slices is displayed.}
	\label{fig:late-expansion-density}
\end{figure*}

\begin{figure}
	\centering
	\includegraphics[width=\columnwidth]{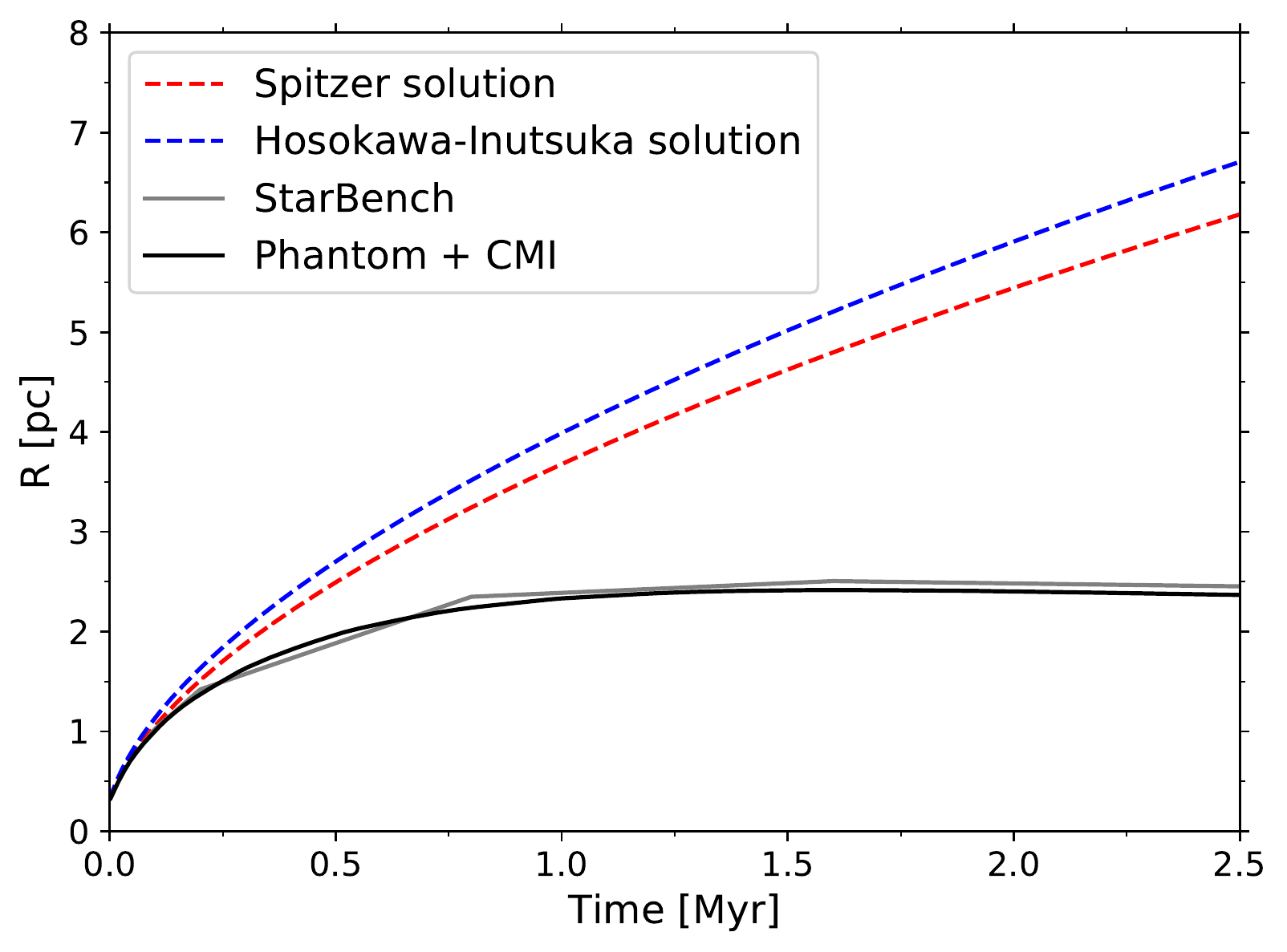}
	\caption{The late-time expansion of an H~II region, following the StarBench setup (`SB Ia'; see Table \ref{table:sim-params}). The Spitzer and Hosokawa-Inutsuka solutions are shown in dashed lines, the averaged 3D StarBench solution is in solid grey \citep{Bisbas2015}, and the \textsc{Phantom} + \textsc{CMacIonize} live radiation hydrodynamics scheme is shown in solid black line.}
	\label{fig:late-expansion}
\end{figure}

The late test captures the decoupling of the shock and the ionisation front. While the former continues to propagate through the neutral medium, the latter settles into an equilibrium state. This behaviour can be seen in Figure~\ref{fig:late-expansion-density}, which shows density slices at different times, similarly to Figure~\ref{fig:early-expansion-density} for the early test. The decoupling mentioned above occurs between the second and the third panel of Figure~\ref{fig:late-expansion-density}. In addition, we show the time evolution of the radius of the H~II region in Figure~\ref{fig:late-expansion}. The figure demonstrates an excellent agreement between our results (in black) and the StarBench data (in grey). Furthermore, we see that the expansion curve from our simulation follows the Spitzer and the Hosokawa-Inutsuka solutions at early times. As the shock thickens (at $\sim 0.2$~Myr), the expansion curve deviates from these solutions, as expected.

\subsection{Timestepping}
\label{sec:timestepping}
\begin{figure}
	\centering
	\includegraphics[width=\columnwidth]{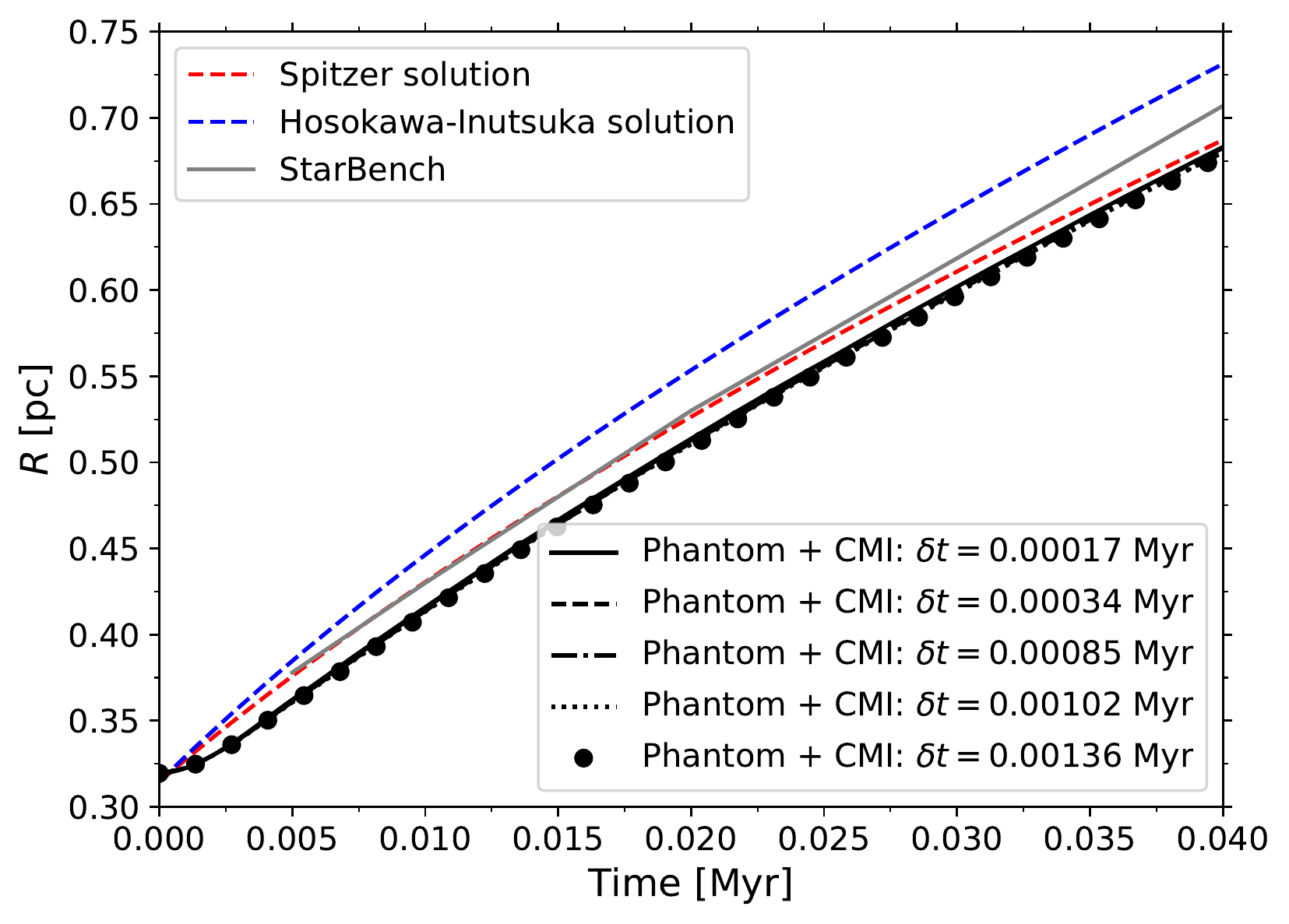}	
	\includegraphics[width=\columnwidth]{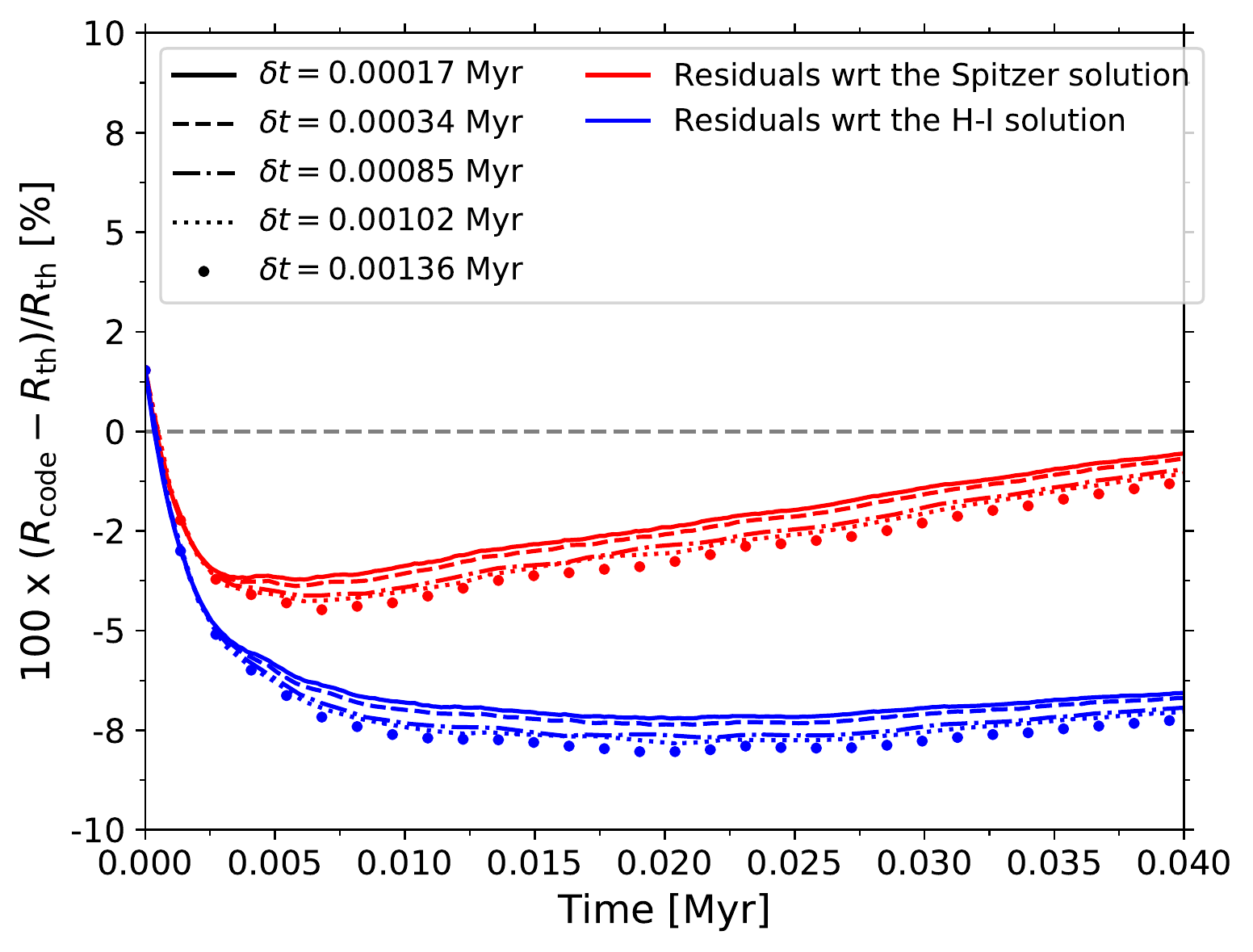}	
	\caption{The early-time expansion of an H~II region, calculated using different ionisation timesteps (`SB II--VI'; see Table \ref{table:sim-params}). The figure axes and the lines have the same meaning as in Figure~\ref{fig:early-expansion}. The lines demonstrate good convergence between the different time steps that have been used.}
	\label{fig:timesteps}
\end{figure}

Next we will examine the sensitivity of the simulation outcome to the choice of ionisation time step, $\delta t$. We have performed a series of small-scale simulations ($N = 64^3$, $m_{\rm{part}} = 10^{-3}$ M$_{\rm{\odot}}$), with $\delta t$ ranging between $1.7 \times 10^{-4}$ Myr and $1.36 \times 10^{-3}$ Myr (`SB II' to `SB VI'; see Table \ref{table:sim-params}). Note that all time steps have been chosen to be smaller than $1.7 \times 10^{-3}$ Myr, since using time steps larger than this critical value causes poor energy conservation and prompts errors in \textsc{Phantom}. This is a byproduct of the initial energy discontinuity between the ionised and the neutral medium (see \citealt{Price2018} for more details).

\begin{figure}
	\centering
	\includegraphics[width=\columnwidth]{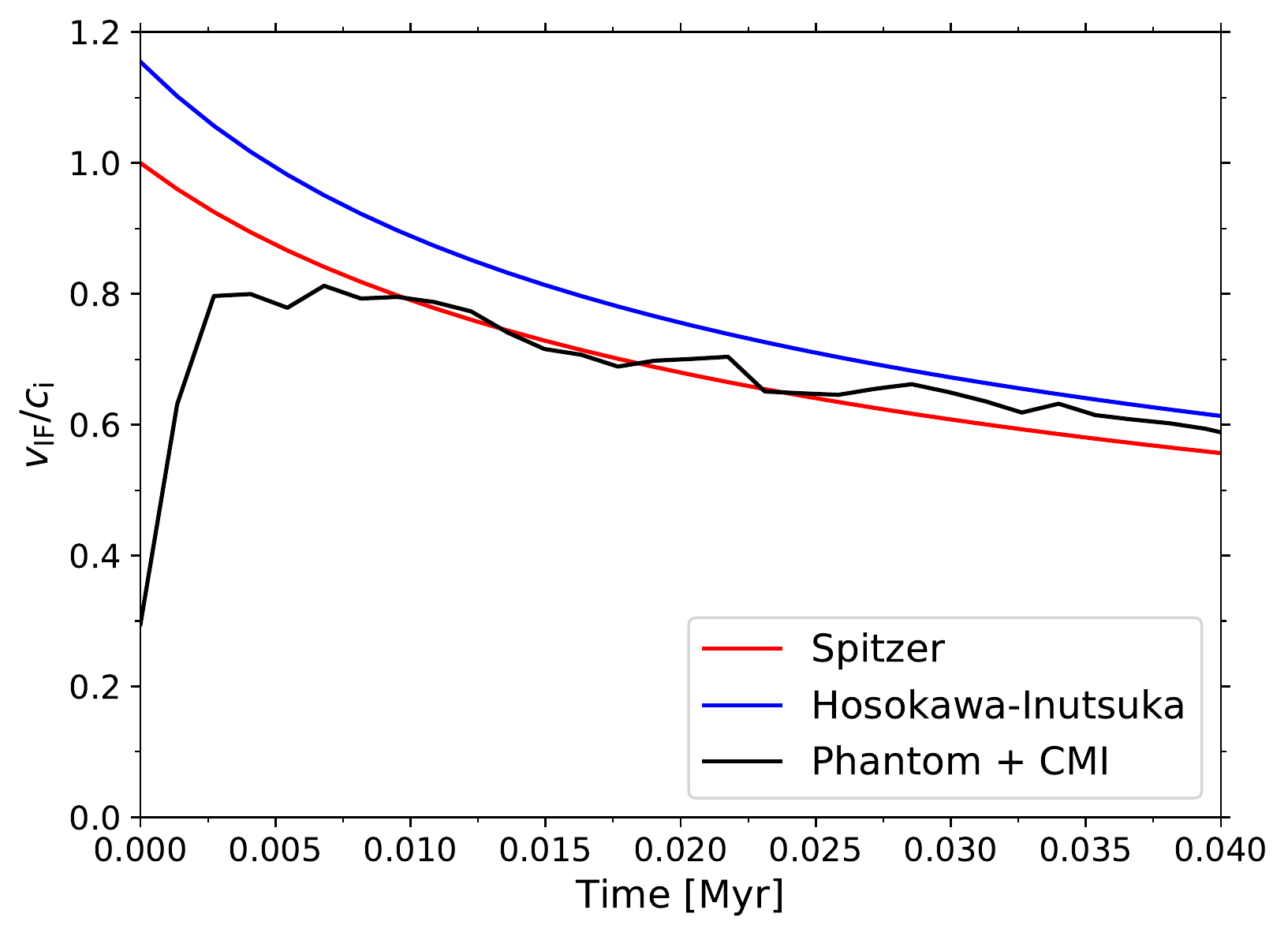}
	\includegraphics[width=\columnwidth]{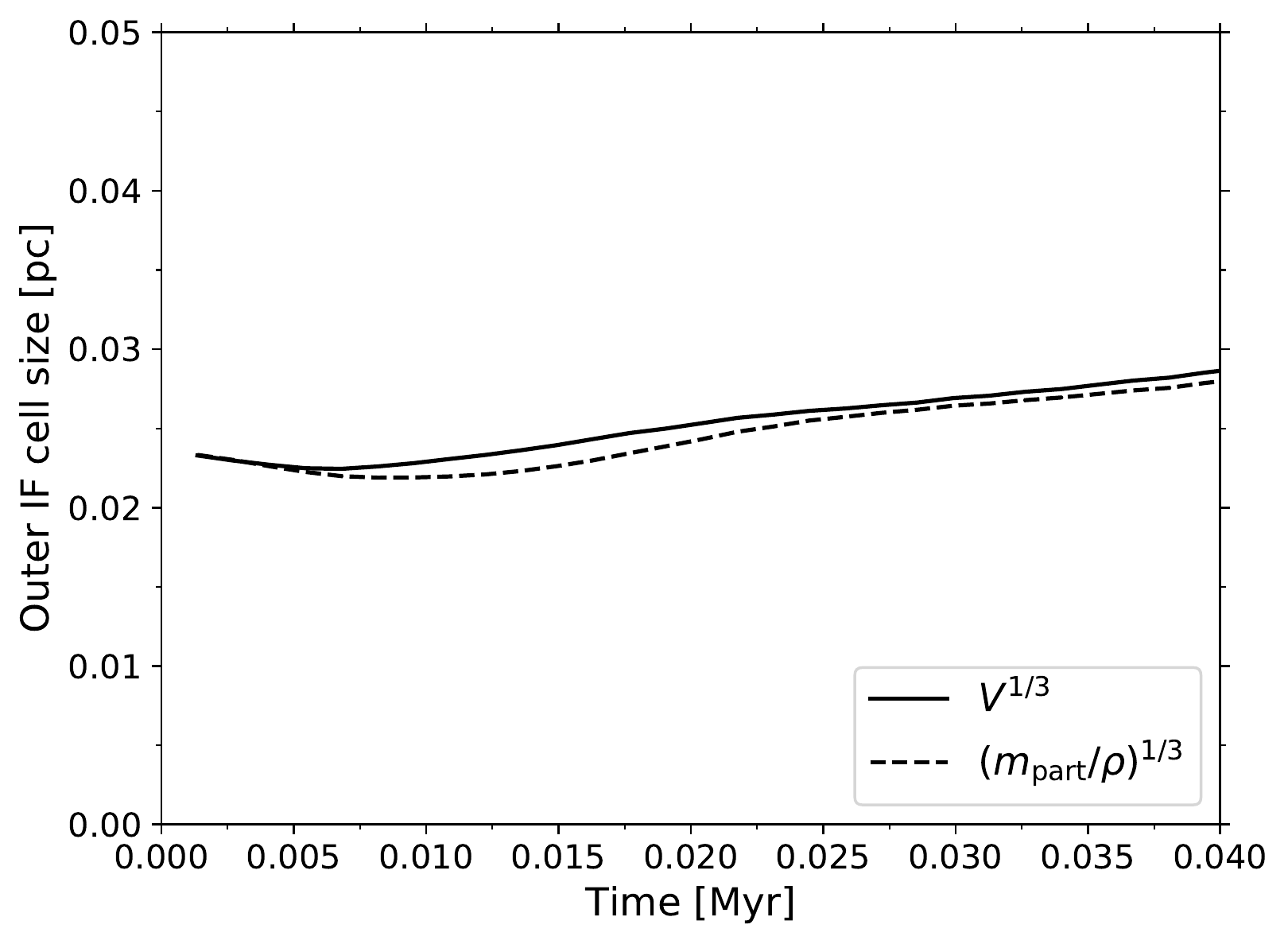}
	\includegraphics[width=\columnwidth]{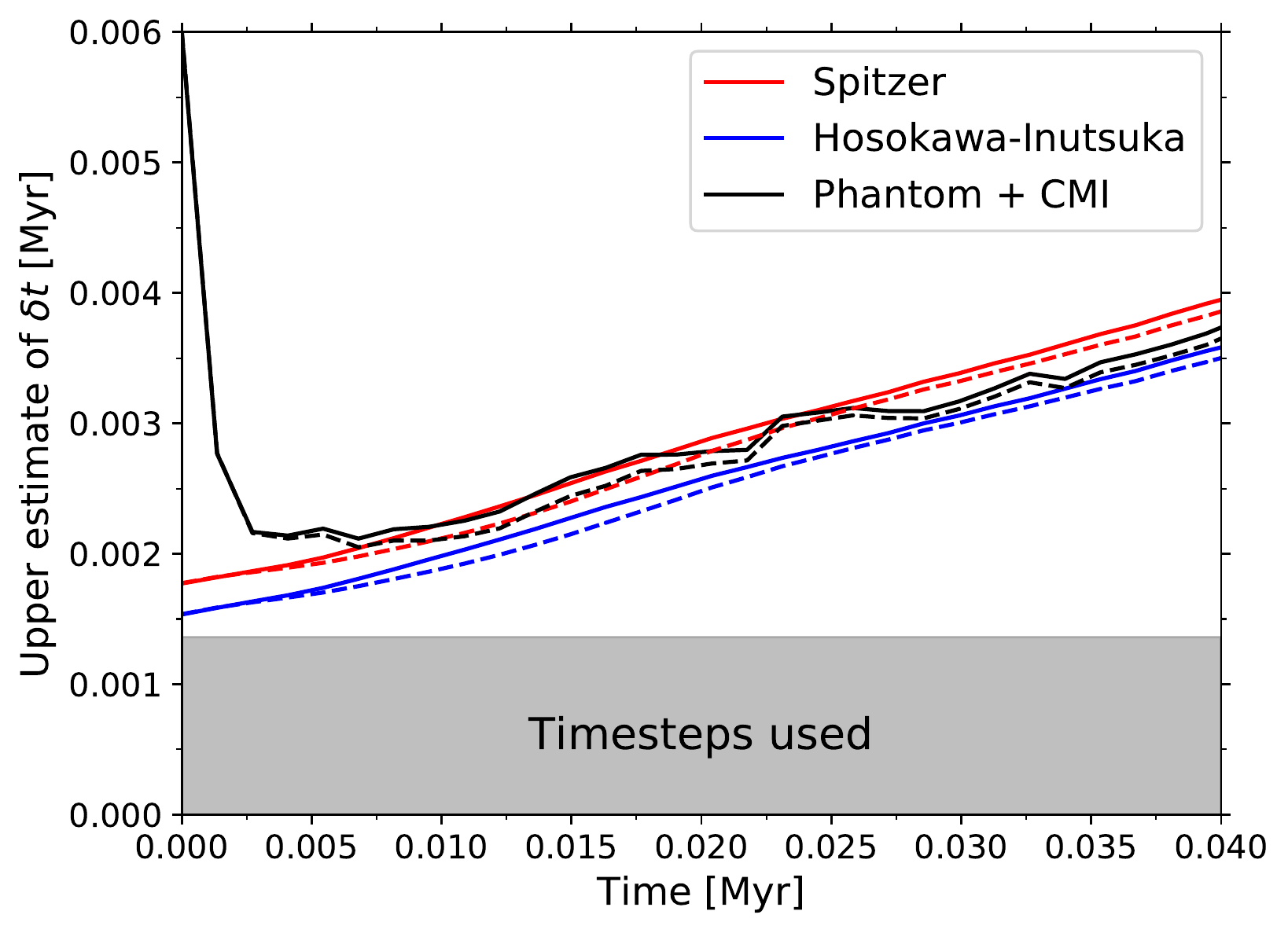}
	\caption{\textit{Top:} velocity of the ionisation front as a function of time. The coloured lines use the analytic form of the ionisation front speed of the Spitzer (red; eq.\ref{eq:Sp-vel}) and the Hosokawa--Inutsuka (blue; eq.~\ref{eq:HI-vel}) solutions, while the black line has been empirically calculated from `SB VI'. \textit{Middle:} cell size immediately outside the ionisation front as a function of time. The solid line represents the average cube root of the cell volumes and the dash line is the average cube root of the particle mass over cell densities. \textit{Bottom:} upper estimate of the allowed ionisation time step, using eq.~\ref{eq:dt-constraint}. The solid and dashed lines have the same meaning as in the above panel. The grey shaded area contains all of the time steps used for the simulations.}
	\label{fig:timesteps-explained}
\end{figure}

Figure~\ref{fig:timesteps} presents the expansion curves of the aforementioned simulations, and it shows good agreement between them. Within the first few time steps ($t < 0.002$ Myr) the curves overlap completely, and after that ($t < 0.007$ Myr) a small offset develops between them. The time period of offset development corresponds to the most rapid expansion of the H~II region within the simulations (see the top panel of Figure~\ref{fig:timesteps-explained}). Once formed, the small offset remains between the curves for the rest of the simulation time. 

We can explain this good convergence by considering a timestepping condition for the ionisation similar to the Courant criterion \citep{Courant1928}. In order to ensure numerical convergence, we want the ionisation front to expand by at most the local particle/Voronoi cell separation between two consecutive ionisation time steps, i.e.:
\begin{equation}
    \delta t < \frac{l_i}{v_{\rm{IF}}}\approx \frac{h_a}{v_{\rm{IF}}}.
    \label{eq:dt-constraint}
\end{equation}
In the above $v_{\rm{IF}}$ is the speed of the ionisation front, $h_a$ is the smoothing length of particle 'a', located at the periphery of the ionisation front, and $l_i$ is the size of the co-spatial Voronoi cell. The difference between eq.~\ref{eq:dt-constraint} and the Courant criterion is subtle. The latter uses the (absolute or relative) velocity of a particle, while the former considers the speed of the ionisation front, which can advance independently from the particles.

In Figure~\ref{fig:timesteps-explained} we have tested that the above condition holds for all of the timestepping test simulations. In the top panel we have plotted $v_{\rm{IF}}/c_{\rm i}$ as a function of time, both calculated as average speed between consecutive snapshots, and estimated theoretically from the Spitzer and Hosokawa-Inutsuka solutions. The estimates have been obtained using equations \ref{Spitzer} and \ref{Hosokawa-Inutsuka}, from which we can write the time derivatives of the H~II region radius as:

\begin{equation}
\dot{R}_{\rm{Sp}}(t)= c_{\rm i} \left (1 + \frac{7}{4}\frac{c_{\rm i}t}{R_{\rm{St}}}\right )^{-3/7},
\label{eq:Sp-vel}
\end{equation}

\begin{equation}
\dot{R}_{\rm{HI}}(t)= \sqrt{\frac{4}{3}} c_{\rm i} \left (1 + \frac{7}{4}\sqrt{\frac{4}{3}}\frac{c_{\rm i}t}{R_{\rm{St}}}\right )^{-3/7}.
\label{eq:HI-vel}
\end{equation}
This results in $\dot{R}_{\rm{Sp}}(t=0)= c_{\rm i}$ and $\dot{R}_{\rm{HI}}(t=0)= \sqrt{4/3} c_{\rm i}$, and decreasing expansion rates for $t>0$. Numerically (and physically), however, the gas is initially at rest. As a result, the initial expansion of the H~II region does not happen as rapidly as predicted, and the calculated value of $v_{\rm{IF}}$ remains lower than $c_{\rm i}$ at all times.

The middle panel of Figure~\ref{fig:timesteps-explained} presents the Voronoi cell size immediately outside the ionisation front as a function of time. These cells are located around the inner edge of the shock and have been selected as having ionic fractions between 0.3 and 0.5. Due to the irregular shapes of Voronoi grid cells, the average cell size has been calculated as the average cube root of the cell volumes or the average cube root of the particle mass over cell densities. We can see in this panel that there is a gentle increase of cell sizes with time, however, their values remain close to the initial conditions.

Finally, in the bottom panel of Figure~\ref{fig:timesteps-explained}, we have calculated upper estimates for the ionisation time step using equation~\ref{eq:dt-constraint}. The shaded box contains all of the $\delta t$ values used for the simulations and this figure demonstrates that the criterion posed in equation~\ref{eq:dt-constraint} is always met. Using the results of Figure~\ref{fig:timesteps-explained} we also propose using the Courant criterion as a conservative timestepping requirement:
\begin{equation}
\delta t < \frac{1}{c_{\rm i}} \left ( \frac{m_{\rm{part}}}{\rho_{\rm o}} \right ) ^{1/3},
\label{eq:dt-criterion}
\end{equation}
which establishes a direct relationship with the initial physical setup of each simulation.

The timestepping criterion derived above ensures the convergence of the StarBench test as the particle mass is altered. In order for this RHD scheme to be reliably applied to more complex astronomical problems, however, we need to be able generalise the choice of time step beyond this simple test. The difficulty here arises from the fact that typically there are particle motions which are independent from the pressure-driven expansion of the H~II region, and these particle motions can rapidly alter the volume of the ionised material. For example, a clump of dense gas can move at the periphery of the H~II region, changing the local 3D geometry and causing a rapid ionisation of previously unavailable low-density gas, or alternatively, restricting radiative access to some parts of the H~II region and thus shrink it.

There is no analytic way of determining the size of the H~II region in non-uniform medium (or else we would not need to perform the radiative transfer altogether), and hence we cannot determine the timestepping criterion precisely. Linking the ionisation time step to the Courant criterion \citep{Courant1928}, however, is probably a reasonable estimate. Ideally this should happen during the simulation runtime and the ionisation time step should be allowed to vary dynamically.

\subsection{Choice of grid and density mapping}
\label{sec:grid&dens}
\begin{figure}
	\centering
	\includegraphics[width=\columnwidth]{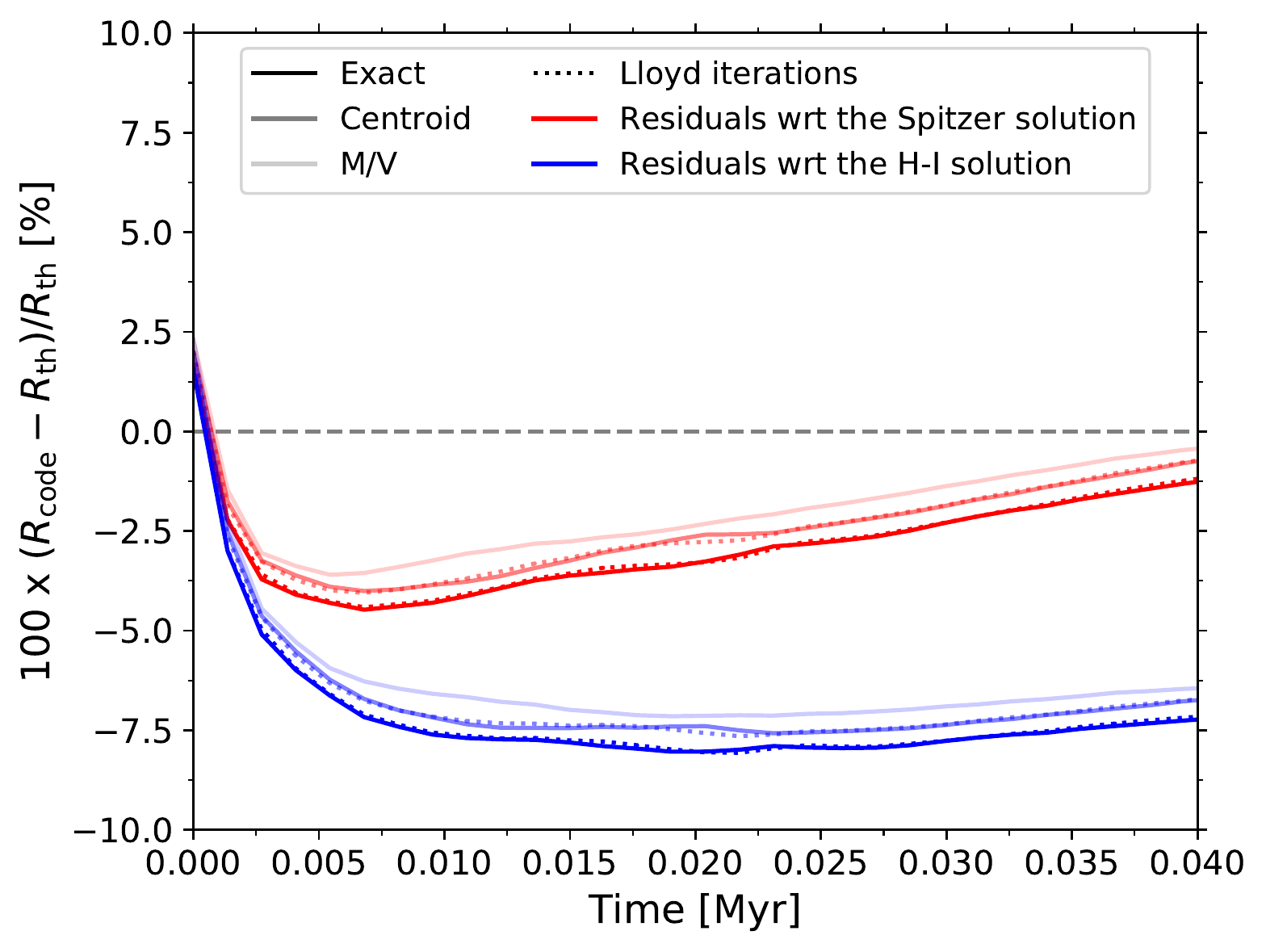}	
	\includegraphics[width=\columnwidth]{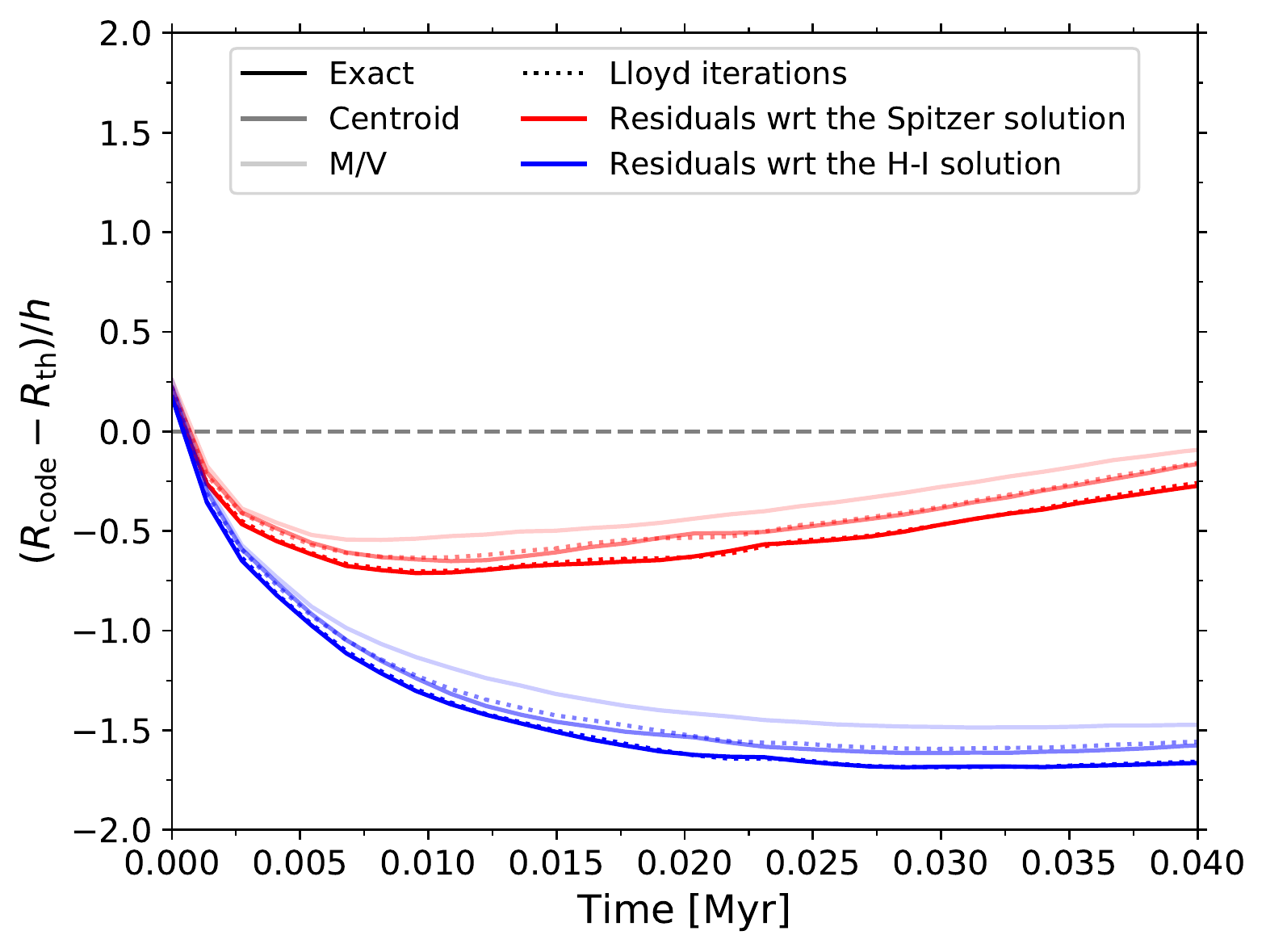}	
	\caption{The early D-type expansion of an H~II region, using different combinations of Voronoi grids and density mapping methods (`SB VI--X'; see Table \ref{table:sim-params}). The plots show the relative error of the \textsc{Phantom} + \textsc{CMacIonize} solutions compared to the Spitzer (red) and Hosokawa-Inutsuka (blue) solutions, expressed as a percentage of the ionisation radius (\textit{top}) and as a multiple of the local smoothing length (\textit{bottom}). The solid lines represent simulations using a basic Voronoi grid, while the dotted lines use a Voronoi grid, which is regularised with Lloyd's iterations. The different shades of each colour correspond to different density mapping methods.}
	\label{fig:early-expansion-grids-and-dens}
\end{figure}

In order to study how the choice of Voronoi grid and type of density mapping affect the H~II region expansion, we have performed a set of simulations (`SB VI' -- `SB X'; see Table \ref{table:sim-params}) with $N = 64^3$, $m_{\rm{part}} = 10^{-3}$ M$_{\rm{\odot}}$ and $\delta t = 1.36 \times 10^{-4}$ Myr (the latter in compliance with the timestepping criterion from equation~\ref{eq:dt-criterion}). Since only two of the three types of density mapping (`Exact' and `Centroid') are compatible with both the basic and the regularised Voronoi grid, combining all grids and density mapping methods result in five simulations.

Figure~\ref{fig:early-expansion-grids-and-dens} shows a comparison between the five simulation runs. In the top panel of the figure we can see that all five expansion curves follow the theoretical solutions of Spitzer and Hosokawa-Inutsuka within 5\% and 8\% respectively. Furthermore, the individual curves of the simulations show only minor variation with respect to each other, and hence we have only plotted the residuals with respect to the theoretical solutions to emphasise these small variations.

In the top panel of Figure~\ref{fig:early-expansion-grids-and-dens} we can see that each type of density mapping produces D--type expansion, happening at a slightly different rate (`M/V' happens most rapidly, followed by `Centroid' and then finally by `Exact'). The choice between basic and regularised Voronoi grid, however, results in no difference in the expansion curve (compare the dotted lines with the solid lines of the same shade). This is not surprising, since these simulations start with a `glass' distribution of particles, which in turn create regularly shaped Voronoi grid cells.

It remains to be determined if the differences between the expansion curves, caused by the different types of density mapping, are significant. For this purpose, we have expressed the difference between the radius of the ionisation front of the simulation and that of the theoretical models in terms of the local smoothing length, $h$ (see the bottom panel of Figure~\ref{fig:early-expansion-grids-and-dens}). The smoothing length used in the figure is the average smoothing length of all particles with ionic fractions between 0.2 and 0.8 (which are the same particles used to determine the position of the ionisation front). In the bottom panel of the figure, we can see that all expansion curves are within $\sim 0.2 h$ of each other. This makes the difference between them sub--resolution, and hence we can deduce that neither the density mapping method, nor the choice of Voronoi grid have a significant impact of the D--type expansion of an H~II region in uniform medium.

\subsection{Resolution}
\label{sec:resolution}
\begin{figure*}
	\centering
	\includegraphics[width=\textwidth]{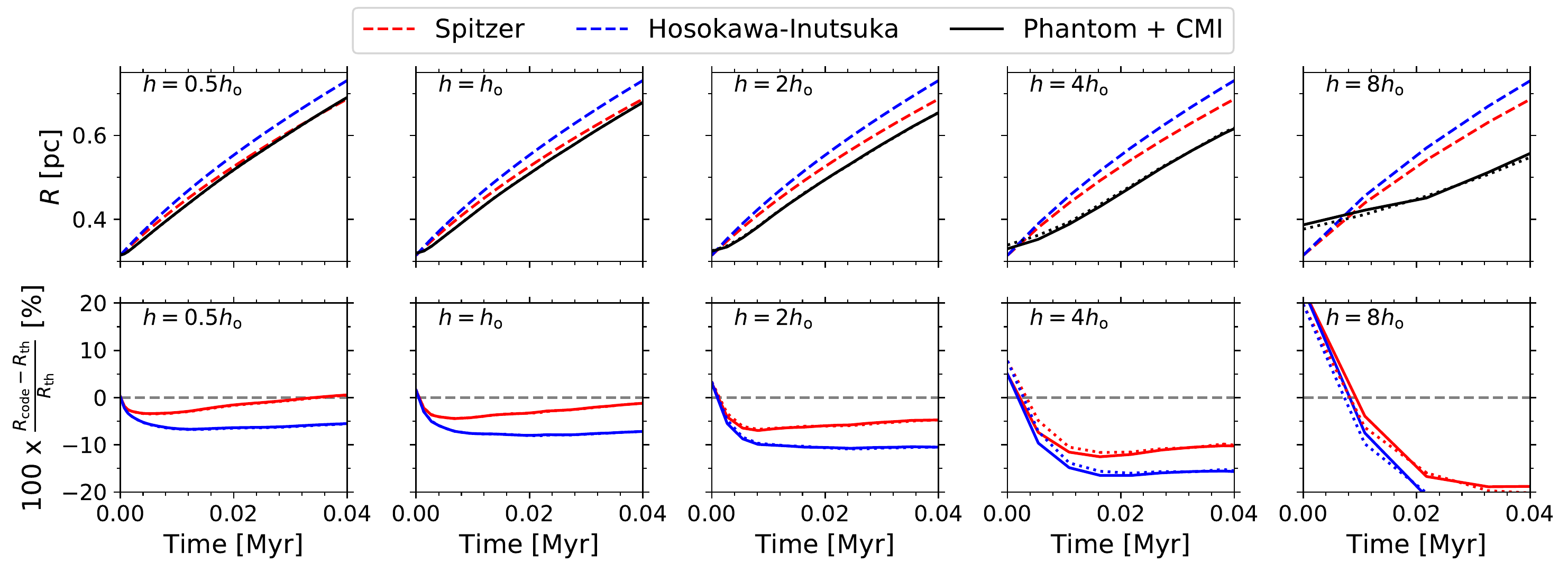}	
	\caption{The early D-type expansion of an H~II region, using initial setup with different spatial resolution (`SB VI--VII',`SB XI--XVIII'; see Table \ref{table:sim-params}). We express the resolution in terms of the smoothing length, where the default resolution is referred to as $h_{\rm o}$. \textit{Top panels:} ionisation front radius as a function of time. The Spitzer and Hosokawa-Inutsuka solutions are shown in dashed red and blue lines. The black lines represent the \textsc{Phantom} + \textsc{CMacIonize} radiation hydrodynamics scheme, with (dotted) and without (solid) using Lloyd's algorithm. \textit{Bottom panels:} the relative error of the \textsc{Phantom} + \textsc{CMacIonize} solutions compared to the Spitzer (red) and Hosokawa-Inutsuka (blue) solutions for simulations with and without Lloyd's algorithm.}
	\label{fig:mass-resolution}
\end{figure*}

\begin{figure*}
	\centering
	\includegraphics[width=\textwidth]{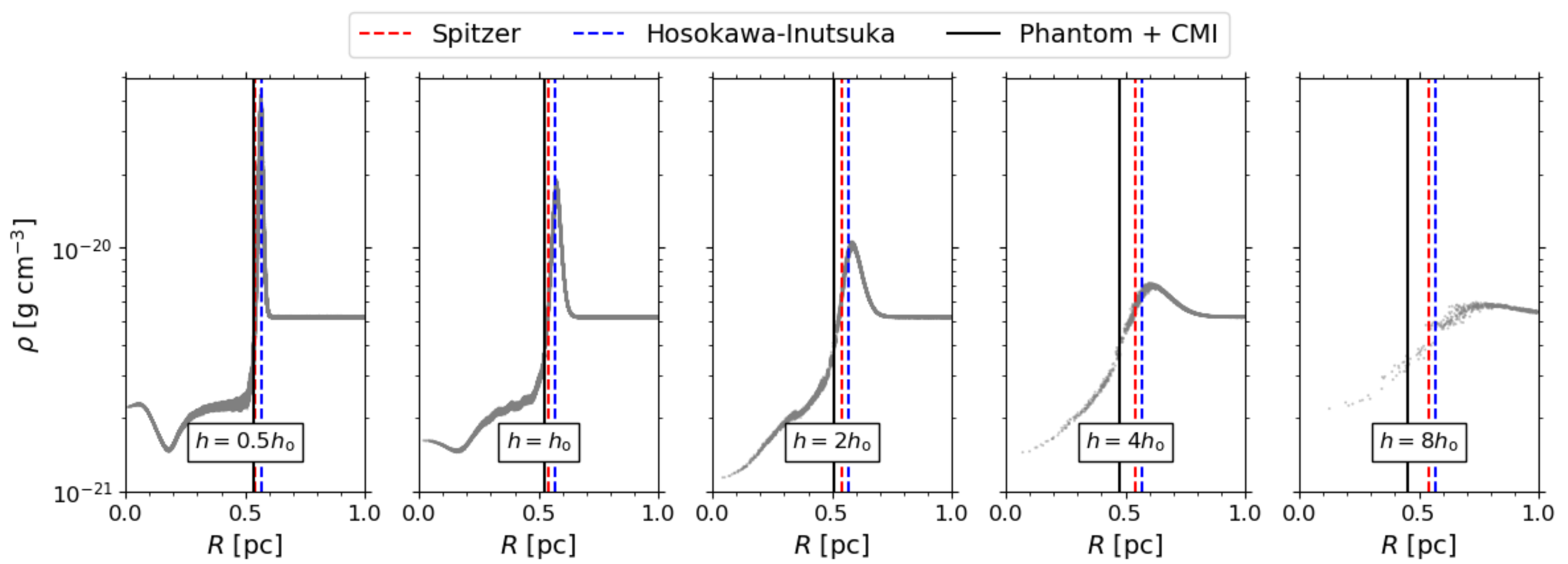}	
	\caption{Radial density profile of the simulations presented in Figure~\ref{fig:mass-resolution} at $t\approx 0.02$ Myr (without using Lloyd's iterations). The grey data points are individual gas particles (from left to right: `SB XVII', `SB VI', `SB XV', `SB XIII', `SB XI'). The Spitzer and Hosokawa-Inutsuka solutions are shown in dashed red and blue lines, and the black lines represent the \textsc{Phantom} + \textsc{CMacIonize} radiation hydrodynamics scheme.}
	\label{fig:density-profile}
\end{figure*}

\begin{figure}
	\centering
    \includegraphics[width=\columnwidth]{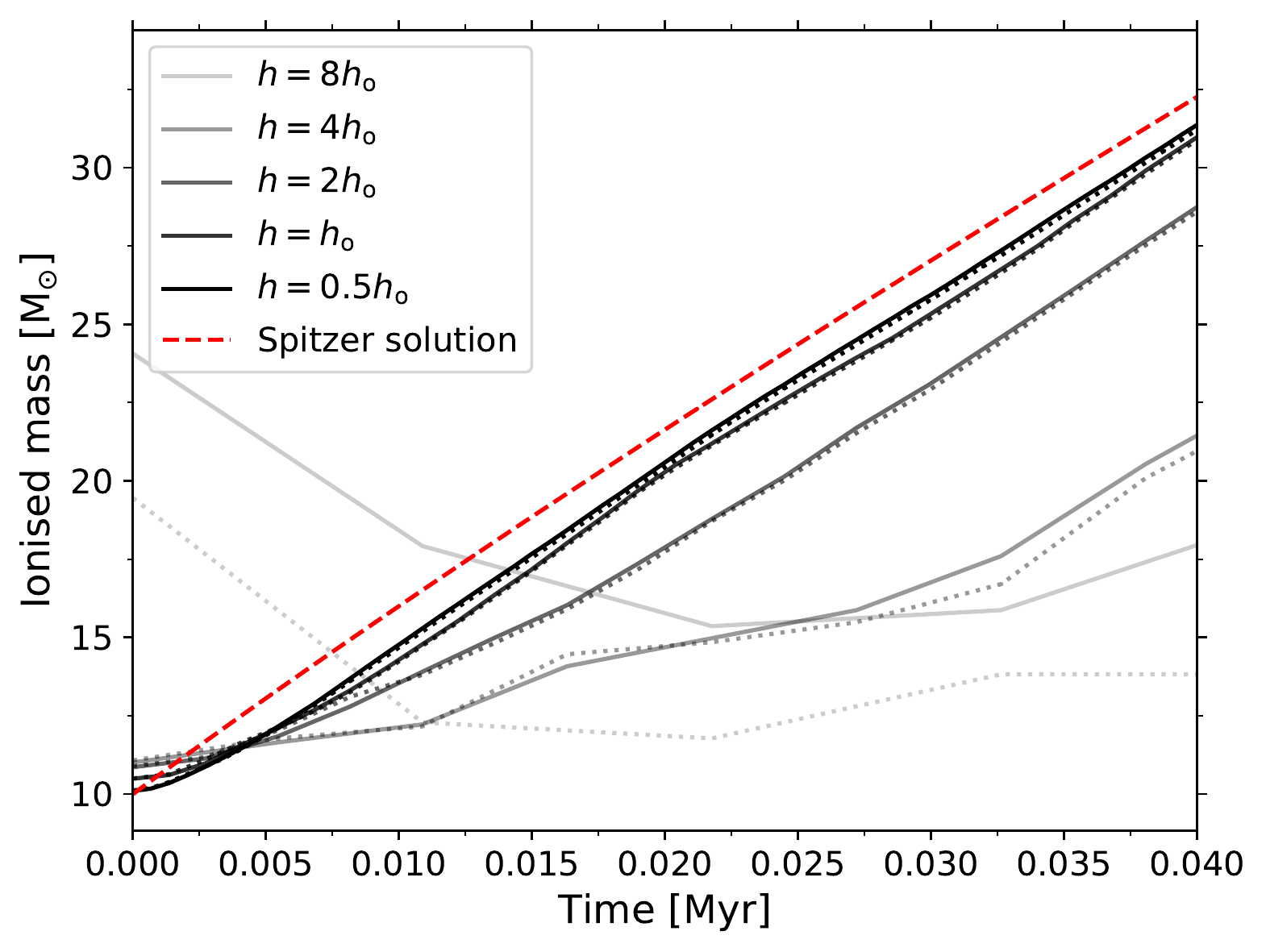}
    \caption{Ionised mass as a function of time for the different resolution tests (`SB VI--VII', `SB XI--XVIII'). The ionised mass is calculated by counting the SPH particles with ionic fractions over 0.5, and assuming that they are fully ionised. The solid lines represent the simulations with basic Voronoi grids, while the dotted lines show the simulations with regularised Voronoi grids. Note that the lines converge to a solution as the spatial resolution is increased. A theoretical prediction of the ionised mass as a function of time, based on the Spitzer solution, is shown in a red dashed line.}
	\label{fig:ionised-mass-sb}
\end{figure}

\begin{table}
  \centering
  \begin{tabular}{cccrl}
    \hline
    Simulation & $m_{\rm{part}}$  & $h$ & \multicolumn{2}{|l|}{Initially ionised particles}\\
    name & $[$M$_{\rm{\odot}}]$ & [pc] & with Lloyd & w/o Lloyd \\
    \hline
    \hline
    SB XI--XII & 0.512 & $2.26 \times 10^{-1}$ & 47 & 38\\
    SB XIII--XIV & 0.064 & $1.13 \times 10^{-1}$ & 172 & 173 \\ 
    SB XV--XVI & 0.008 & $5.65 \times 10^{-2}$ & 1357 & 1361 \\ 
    SB VI--VII & 0.001 & $2.82 \times 10^{-2}$ & 10510 & 10508 \\ 
    SB XVII--XVIII & 0.000125 & $1.41 \times 10^{-2}$ & 80781 & 80898 \\ 
    \hline
  \end{tabular}
\caption{Parameters of the simulations exploring the effect of different mass resolution. The last two columns contain the number of particles that are initially ionised.}
\label{table:mass-res}
\end{table}

Next we study the effects of resolution on the early--time StarBench test. We have achieved that by keeping the same initial gas density, and changing the particle mass, which naturally leads to variations in the smoothing length size (`SB VI', `SB VII', `SB XI' -- `SB XVIII'; see Table \ref{table:sim-params}). We have performed two simulation runs per resolution, one with basic and one with regularised Voronoi grid. The resolution in SPH is expressed in terms of particle mass, however, we report our results in terms of average initial smoothing length, since spatial resolution is more intuitive. The correspondence between mass resolution and spatial resolution is summarised in Table \ref{table:mass-res}. Another implication of the varied resolution is that the initial Str\"omgren sphere is represented by a different number of particles at each simulation setup (see Table \ref{table:mass-res}). We have adopted ionisation time steps according to equation~\ref{eq:dt-criterion}, which are a function of $m_{\rm{part}}$.

Figure~\ref{fig:mass-resolution} summarises the results of the five different spatial resolution tests. {In it, $h_{\rm o}$ denotes the average smoothing length of the SPH particles in the glass distribution of the original StarBench setup. The figure shows} that using $h = 0.5 h_{\rm o}$ and $h = h_{\rm o}$ produces quite similar outcomes. However, as we pick larger smoothing lengths, the solutions start to diverge. In particular, $h = 4 h_{\rm o}$ and $h = 8 h_{\rm o}$ no longer closely follow the Spitzer and Hosokawa-Inutsuka solutions, and their H~II region expansion happens more slowly than expected. Additionally, the choice of basic or regularised Voronoi grid does not significantly alter the simulation outcome at any chosen resolution.

Figure~\ref{fig:density-profile} presents the radial density profile of the simulations at $t\approx 0.02$ Myr. In the figure we can see that the runs with larger smoothing lengths do not reproduce the expected thin shock (that is to say, a shock much thinner than the size of the ionised region). Since all of the theoretical solutions assume a thin shock, the numerical results naturally deviate from these solutions. Furthermore, the simulation with $h = 8 h_{\rm o}$, in particular, has a significantly higher average density of the region internal to the shock, which additionally suppresses the radius of the ionised sphere.

Finally, Figure~\ref{fig:ionised-mass-sb} shows the mass of the ionised gas as a function of time for all of the resolution tests. The masses have been calculated by counting all of the particles with ionic fraction over 0.5 as fully ionised. \textcolor{black}{This definition of the ionised mass gives us the exact amount of hot gas in the hydrodynamics, which has direct effect on the dynamics of the gas, and it is, arguably, the more relevant definition when we look at the time evolution of the H~II region. An alternative approach would be to sum the ionic fractions multiplied by the particle mass. This latter definition produces the amount of ionised gas, as calculated by the radiative transfer (for the density mapping methods that preserve mass).} The figure demonstrates a clear progression between the different simulations as the resolution is increased, and indicates a convergence towards a specific mass curve (see $h=h_{\rm o}$ and $h=0.5 h_{\rm o}$). We can also see that all runs overestimate the initial ionised mass (as they overestimate the Str\"omgren radius). This can be noticed most prominently with the lowest resolution simulations, which overshoot the mass by $\sim 100\%$. This very high initial mass subsequently decreases with time, suppressed by a very thick shock, until it begins to increase after about 0.02 Myr. The initial overestimate of ionised mass happens because the simulations with $h=8h_{\rm o}$ initially ionise only about 40 particles, and hence have a substantial numerical error of the ionised mass.

\citet{Bisbas2015} have expressed the mass of the H~II gas from theoretical considerations, as:

\begin{equation}
M_{\rm{ion}} = \frac{4\pi}{3} \rho_{\rm o} R_{\rm{St}}^{3/2} R_{\rm{IF}}(t)^{3/2}.
\label{eq:mass-ion}
\end{equation}
If we substitute the Spitzer (eq.~\ref{Spitzer}) or the Hosokawa--Inutsuka solution (eq.~\ref{Hosokawa-Inutsuka}) for $R_{\rm{IF}}(t)$, we find that $M_{\rm{ion}} \propto t^{6/7}$. We have plotted the theoretically predicted ionised mass using the Spitzer solution in Figure~\ref{fig:ionised-mass-sb} as a red dashed line. We have used the Spitzer solution and not that of Hosokawa--Inutsuka, because the former provides a better fit to our simulations. We can see from the line that even though the dependence of the mass on the time is sub-linear, for the small values of $t$ which we consider, the curve cannot be visibly distinguished from a straight line. Figure~\ref{fig:ionised-mass-sb} shows that the slope of our simulations (and especially the higher resolution ones) follow the slope of the theoretical solution after a period of initial adjustment. This adjustment period is longer for the lower resolution simulations. In particular, for $h=2h_{\rm o}$ this period is $\approx 0.017$ Myr, for $h=4h_{\rm o}$ it is $\approx 0.033$ Myr, and for $h=8h_{\rm o}$ the slope is not reached at all within the first $\approx 0.040$ Myr of development. The agreement between the simulated and the theoretically predicted slope is expected, since the slope is set by the sound speed, the initial gas density and the Str\"omgren radius, which match between simulations and theory. Note that we do not expect the simulations to converge fully to the red line, as we do not expect them to converge fully (especially at very early times <0.007 Myr) to the Spitzer solution.

Using the above results, we can estimate the resolution necessary for producing a converged initial size of the ionised region. If we assume that reproducing the Str\"omgren sphere radius within a few percent results in satisfactory accuracy, then we need to choose the resolution such that we have at least a few thousand initially ionised particles. Such resolution is sufficient for a uniform density simulation, however we cannot easily test if this estimate holds for simulations of clumpy medium.

By contrast, if we want to accurately model the structure of the shock (provided that one exists) swept up by an H~II region, we need to resolve the thickness of the shock. We can see in Figure~\ref{fig:density-profile} that even the lowest particle smoothing length which has been considered (corresponding to $\sim 8 \times 10^4$ initially ionised particles) does not guarantee a converged shock profile.

\section{Simulations in a clumpy medium}
\label{sec:clumpy}
We expand the parameter space exploration of the previous section, by now considering simulations of a single source in a clumpy medium (see simulations labelled as `B' in Table~\ref{table:sim-params}).
To recreate an astronomically realistic density structure, we have used a simulation of the galactic centre cloud known as the Brick \citep{Dale2019,Kruijssen2019}. We have chosen a simulation snapshot corresponding to the present day location of the Brick along its orbit, and we have selected the particles contained in a $R_{\rm{cl}}=6$ pc cube around the densest region. We have also shifted the origin of the coordinate system of the simulation to the centre of the cube for simplicity. This results in a setup with $N=101128$ particles, $m_{\rm{part}} = 0.4463$ M$_{\rm{\odot}}$ and a total mass of about $M_{\rm{cl}}=45000$ M$_{\rm{\odot}}$. The mean particle velocity has been subtracted from all individual particles' velocities in order to remove the orbital motion of the cloud. All particles have an initial temperature $T_{\rm o} = 65$~K, which was used by \citet{Dale2019} and \citet{Kruijssen2019}, and is also consistent with the observed temperature of the Brick cloud \citep{Ao2013,Ginsburg2016,Krieger2017}. We have used one source of ionising radiation with $\dot Q = 5 \times 10^{50}$ s$^{-1}$, which corresponds to a young stellar population with a mass of $\sim10^4~\rm{M_{\odot}}$ \citep{leitherer14}. The source is located either in a density peak ($x=1.5$ pc, $y=-0.5$ pc, $z=-0.51$ pc; local density $3.6 \times 10^{-19}$~g/cm$^3$), or in lower density environment ($x=0$ pc, $y=0.3$ pc, $z=-0.51$ pc; local density $1.8 \times 10^{-22}$~g/cm$^3$). The positions of the sources are marked in Figure~\ref{fig:clumpy-hii-IC} with stars.

Similarly to the first set of simulations, the source ionises a certain volume of gas which expands. In contrast to the first set of simulations, the ionised volume is not spherically-symmetric. We have explored the expansion of the H~II region using two different positions of the source, as well as different grid types and density mapping methods. Unlike in the previous section, we have not varied the ionisation time step or the resolution. The former choice has been made because we have already derived a timestepping criterion from the uniform density simulations (see eq.~\ref{eq:dt-criterion}). The latter choice is due to the fact that the resolution of a clumpy medium could not be easily altered. Additionally, we vary the ionising luminosity of the source in the range $\dot Q = 10^{50}-10^{51}$ s$^{-1}$, and calculate the corresponding initially ionised mass. Finally, we discuss the physical implications of the presence of an ionising source on the host cloud.

\subsection{Time evolution}
\label{sec:clumpy-timestep}
\begin{figure*}
	\centering
	\includegraphics[width=\columnwidth]{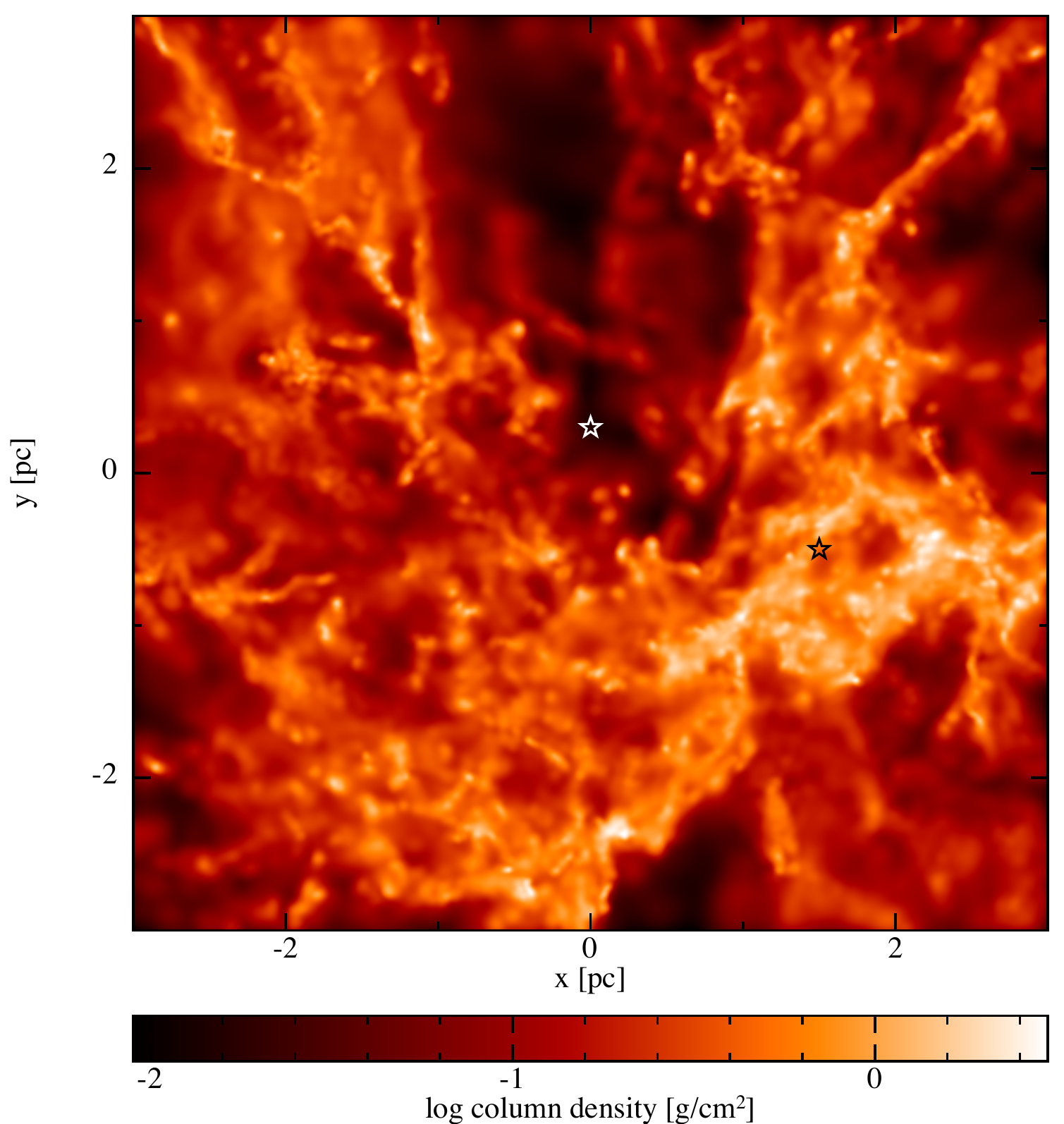}
	\includegraphics[width=\columnwidth]{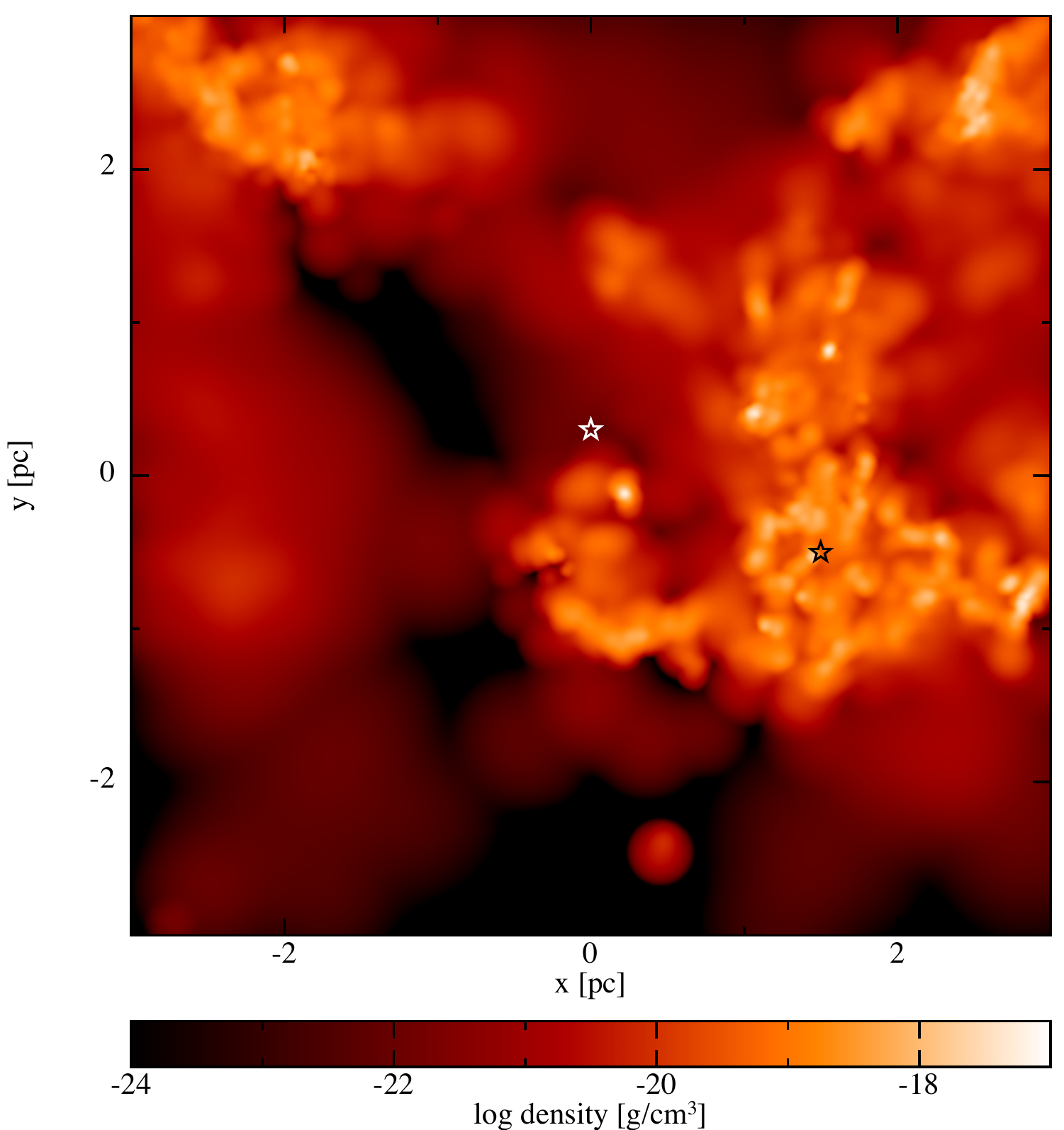}
	\caption{Selected cubic volume from a simulation snapshot of the Brick, used as initial conditions for the RHD scheme \citep{Dale2019,Kruijssen2019}. \textit{Left:} gas column density integrated along the z-axis. \textit{Right:} volume density of the gas in the plane of $z=-0.51$ pc. The two stars mark the positions of the ionisation sources that have been used.}
	\label{fig:clumpy-hii-IC}
\end{figure*}

\begin{figure*}
	\centering
	\includegraphics[width=\columnwidth]{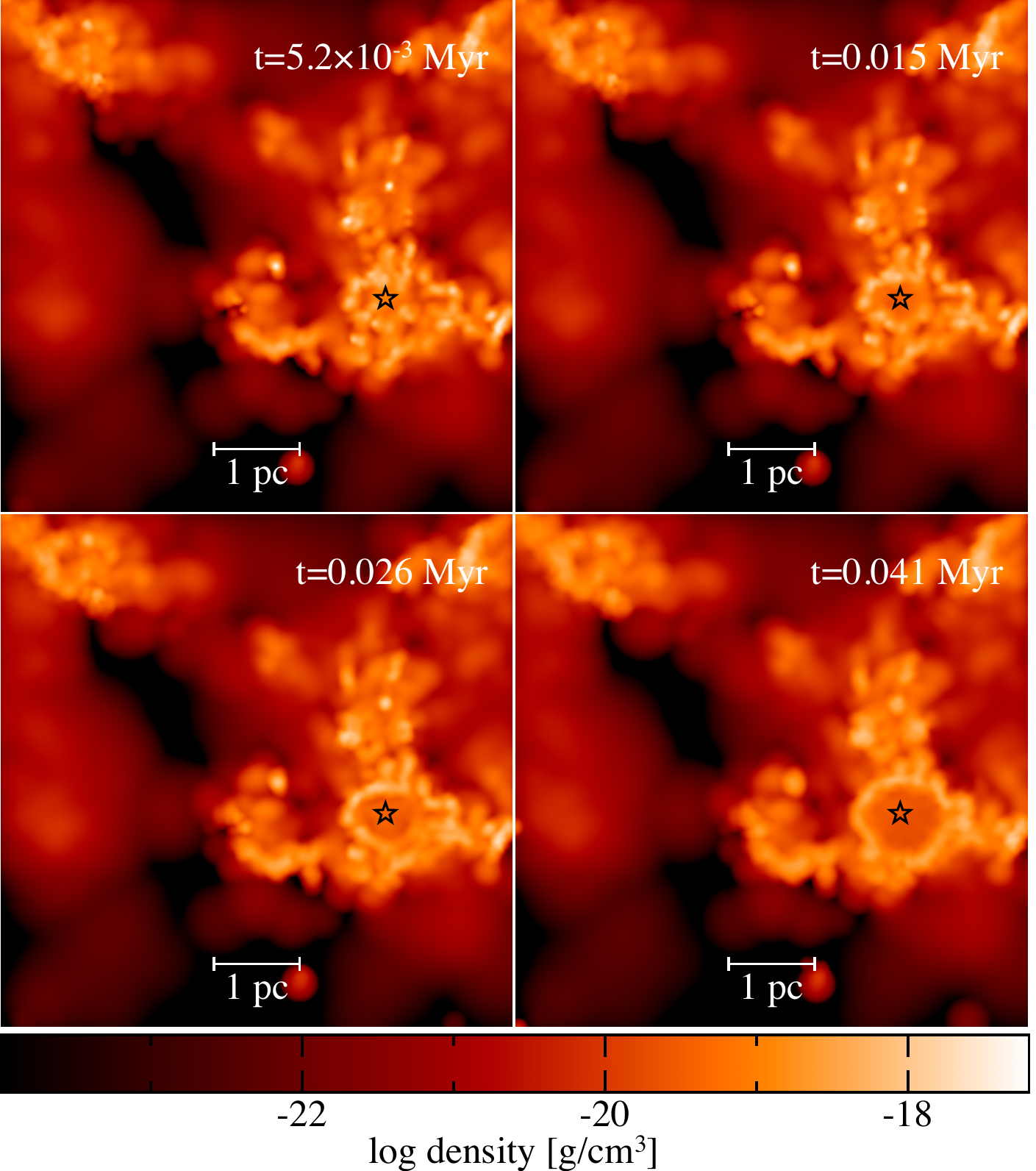}
	\includegraphics[width=\columnwidth]{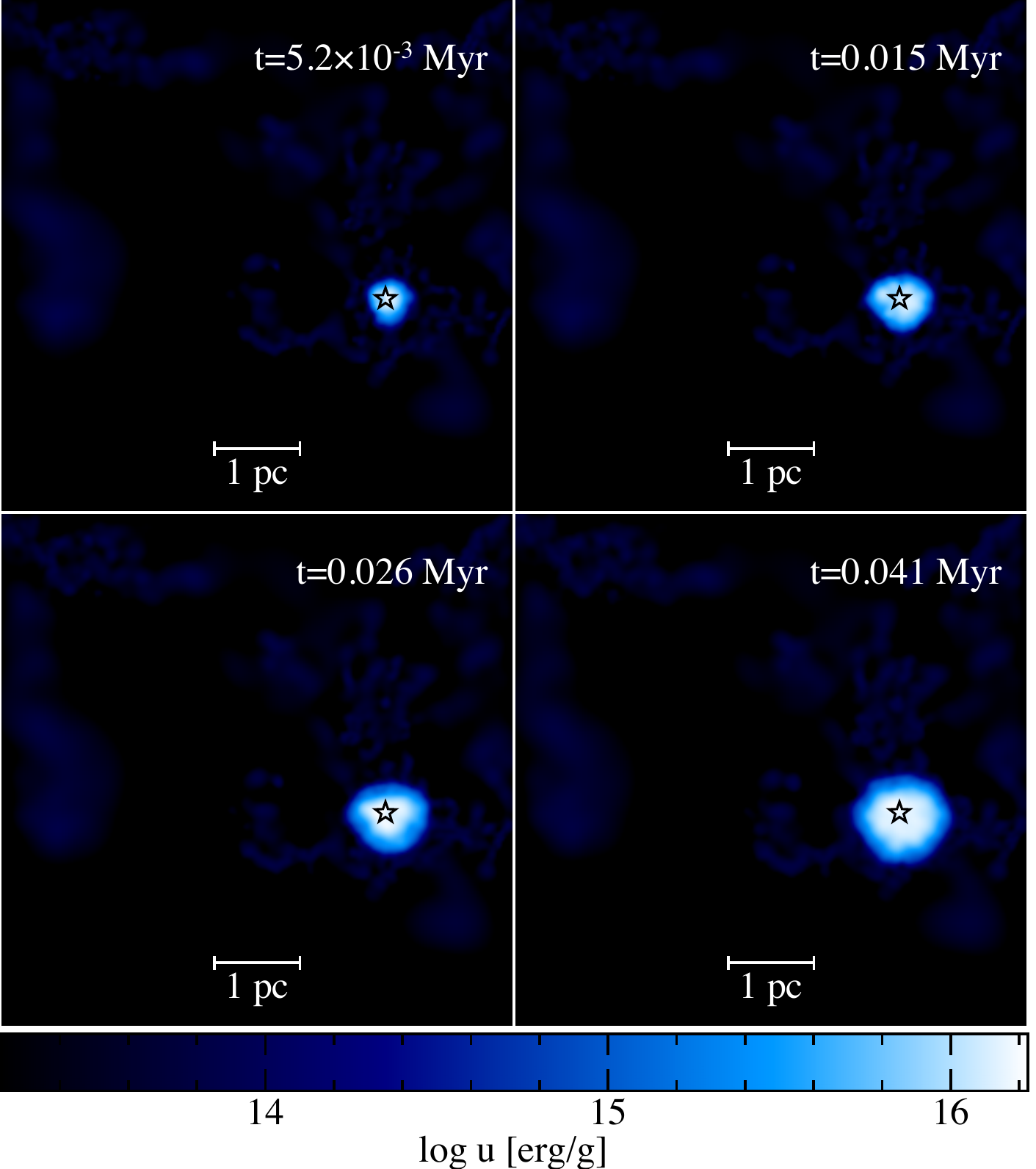}
	\caption{Cross-section images at $z=-0.51$ pc of volume density (\textit{left side}) and specific internal energy (\textit{right side}) of the Brick RHD simulation `B II' (see Table \ref{table:sim-params}) at different times, shown on the panels. The source has been embedded in a dense clump and its location is marked with a star.}
	\label{fig:clumpy-hii-expansion}
\end{figure*}

\begin{figure*}
	\centering
	\includegraphics[width=\columnwidth]{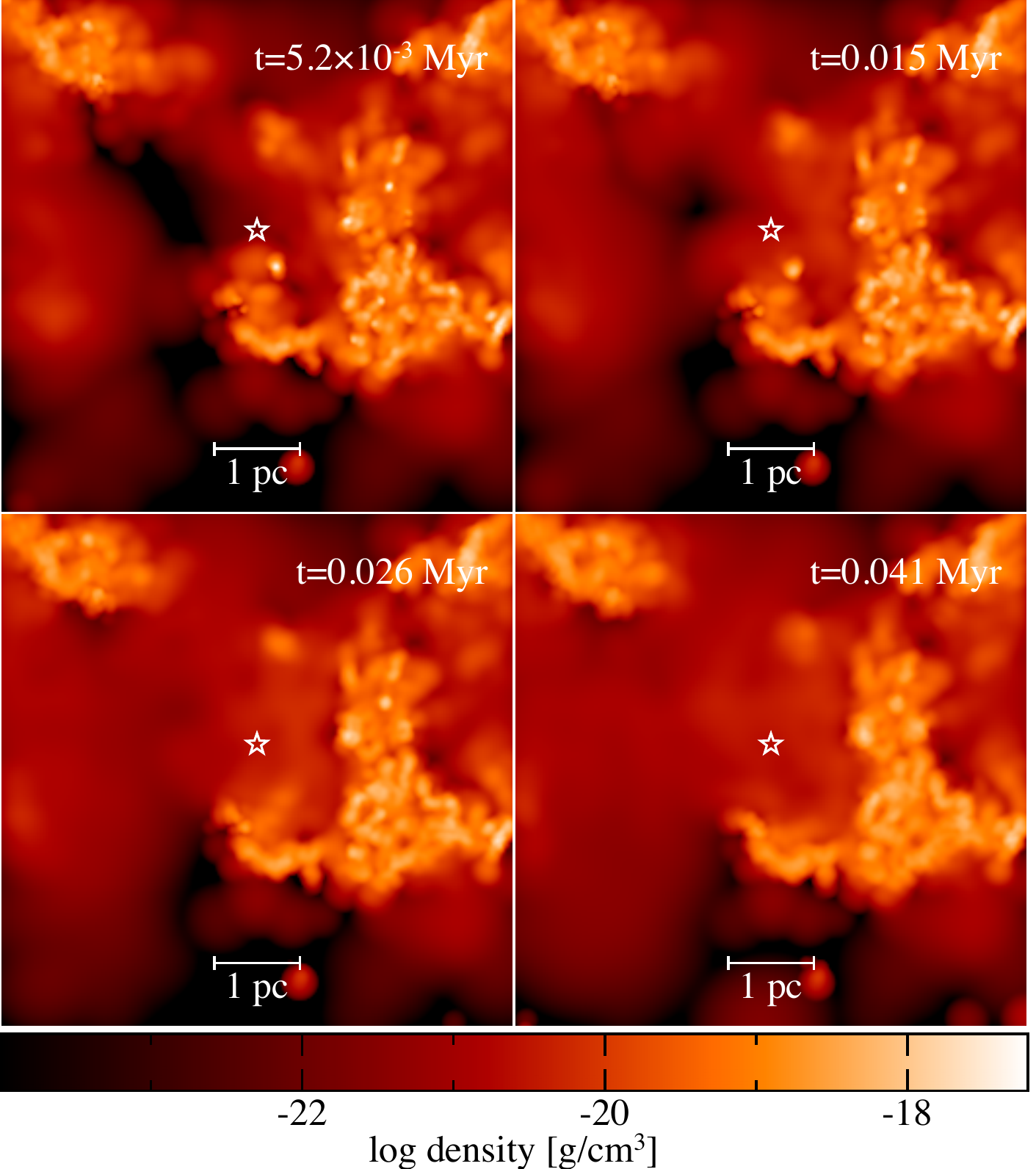}
	\includegraphics[width=\columnwidth]{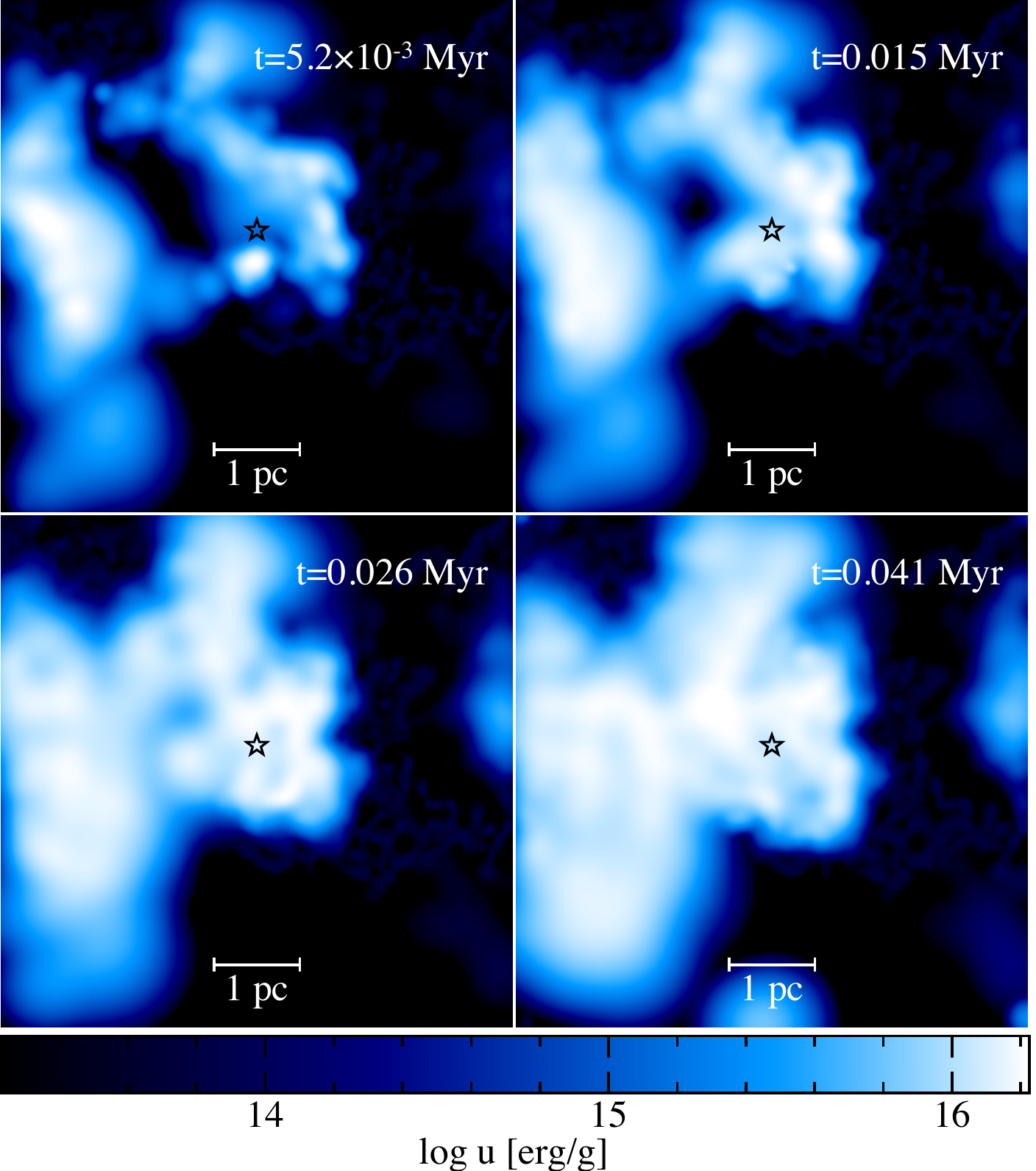}
	\caption{Same as Figure~\ref{fig:clumpy-hii-expansion}, with the exception that the source has been located in low-density environment.}
	\label{fig:clumpy-hii-expansion-ld}
\end{figure*}

\begin{figure}
	\centering
	\includegraphics[width=\columnwidth]{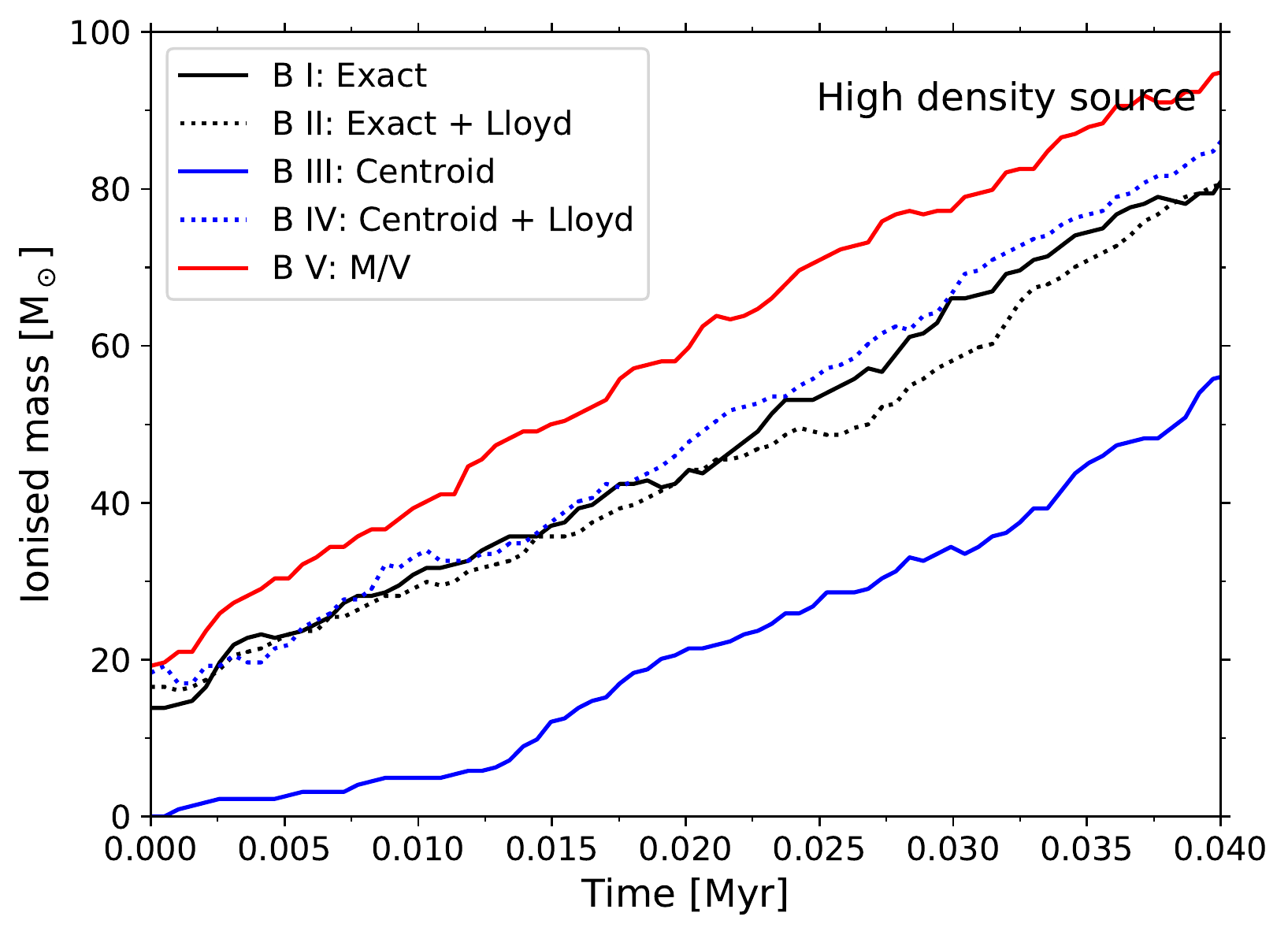}
	\includegraphics[width=\columnwidth]{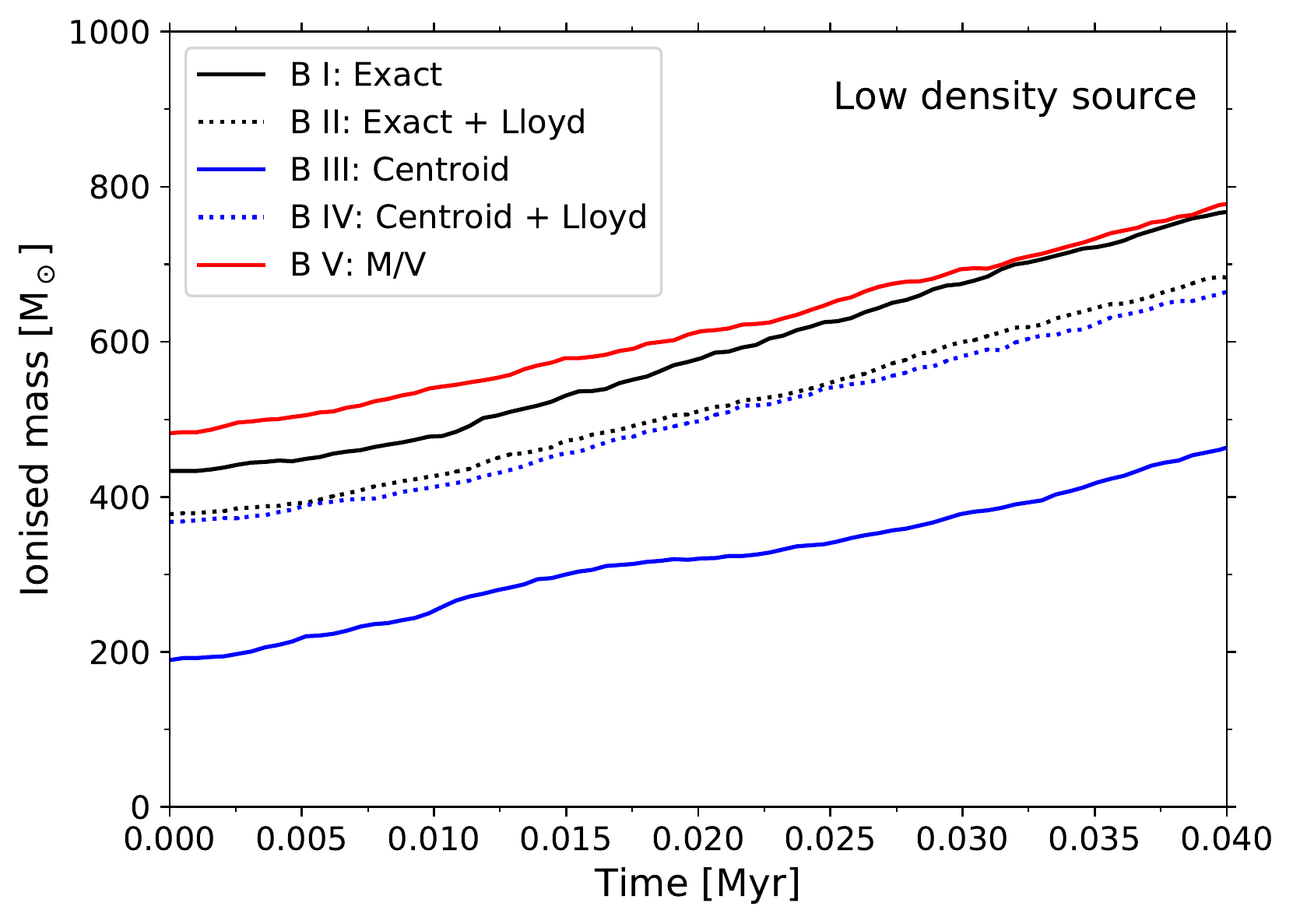}
	\caption{Ionised mass as a function of time for the Brick simulations (`B I--V'; see Table \ref{table:sim-params}) with embedded source (\textit{top}) and the simulations with a source in a low-density environment (\textit{bottom}). The solid lines represent runs using the basic Voronoi grid, while the dotted lines correspond to runs with regularised Voronoi grids. The different colours show the type of density mapping used, with $\rho_{\rm p}$ (eq.~\ref{eq:exact}) being black, $\rho_{\rm c}$ (eq.~\ref{eq:centroid}) being blue, and $\rho_{\rm m}$ (eq.~\ref{eq:MonV}) being red.}
	\label{fig:clumpy-ionised-mass}
\end{figure}

We have used the Brick setup to perform ten different simulations, which repeats the simulations labelled as `B` in Table \ref{table:sim-params} twice. For half of them the source is embedded in a clump, and for the other half the source is in a lower--density environment (see Figure~\ref{fig:clumpy-hii-IC}). The five simulations per source probe the two different Voronoi grids combined with the three different density mapping methods. We have not altered the resolution of the simulations, as this is difficult to achieve with non--uniform initial conditions. The non--uniform setup has also created difficulties in estimating the ionisation time step in advance. We have used the value of $\delta t = 5 \times 10^{-4}$ Myr. In order to ensure the suitability of this choice, we have computed the right-hand-side of equation~\ref{eq:dt-criterion} for each fully or partially ionised particle during runtime (using the particle's density) and have confirmed that these values were always larger than our time step. 

Figures \ref{fig:clumpy-hii-expansion} and \ref{fig:clumpy-hii-expansion-ld} illustrate the expansion of the simulated ionised region with time, when we use the exact density mapping and a regularised Voronoi grid (`B II'). We can see that when the radiation source is located in a dense clump, it initially ionises a small amount of 
high density gas. Through its expansion the ionised gas sweeps up a shock, similarly to the StarBench test. In contrast, the source located in lower density environment immediately fills up the inter--clump space with ionised material and its H~II region undergoes gentle expansion.
We can see the dynamic signatures of these two scenarios by looking at the gas density (left hand panels of Figure~\ref{fig:clumpy-hii-expansion} and~\ref{fig:clumpy-hii-expansion-ld}). For clarity, we have also included the specific internal energy of the particles in the right hand panels of Figure~\ref{fig:clumpy-hii-expansion} and~\ref{fig:clumpy-hii-expansion-ld}. Since the ionised areas have significantly higher specific internal energies than the neutral ones, the former appear in light blue and white.

Additionally, if we look at the density peaks in Figures~\ref{fig:clumpy-hii-expansion} and~\ref{fig:clumpy-hii-expansion-ld}, we notice that they diffuse out with time. This behaviour is not caused by the ionisation, but it is a result of the physical assumptions in our simulations, and the way in which the initial conditions have been constructed. The Brick simulation from \citet{Dale2019} and \citet{Kruijssen2019} uses an external gravitational potential and self--gravity. In our setup we have turned off the gravitational forces and we have additionally set all particle velocities to zero. As a result, the density peaks have higher pressure than the surrounding low density regions, and no force to hold them together. This leads to the observed diffusion, which affects (and likely serves to enable) the expansion of the H~II region at later times. This does not hamper the numerical experiment performed here, in which we are exclusively interested in the impact of sub-structure, rather than the full, self-consistent evolution of the region.

To quantify the expansion of the H~II regions, we have computed the ionised mass as a function of time by adding up the masses of the particles with ionic fractions over 0.5 (see Figure~\ref{fig:clumpy-ionised-mass}).
Despite representing two different modes of H~II region expansion (pressure-confined and depressurised), the two panels of Figure~\ref{fig:clumpy-ionised-mass} look very similar to each other. In both panels we see that the ionised mass increases approximately linearly for all simulation setups. This behaviour is also in agreement with the StarBench results (see Figure~\ref{fig:ionised-mass-sb}). The difference between the top panel (high density source) and the bottom panel (low density source) of Figure~\ref{fig:clumpy-ionised-mass} is mostly in the noise level of the curves and the magnitude of their slopes. The H~II region embedded in a clump has noisier curves than the H~II region in low density environment, which can be explained by the clumpy structure of the high density environment (see Figure~\ref{fig:clumpy-hii-expansion}). Additionally, the expansion curves of the embedded H~II region have shallower slopes than those of the low density H~II region. This is expected as the embedded ionised gas needs to overcome the higher pressure of its surrounding material.

\subsection{Voronoi grids and cell densities}
\label{sec:discussion-grids}
\begin{figure}
	\centering
	\includegraphics[width=\columnwidth]{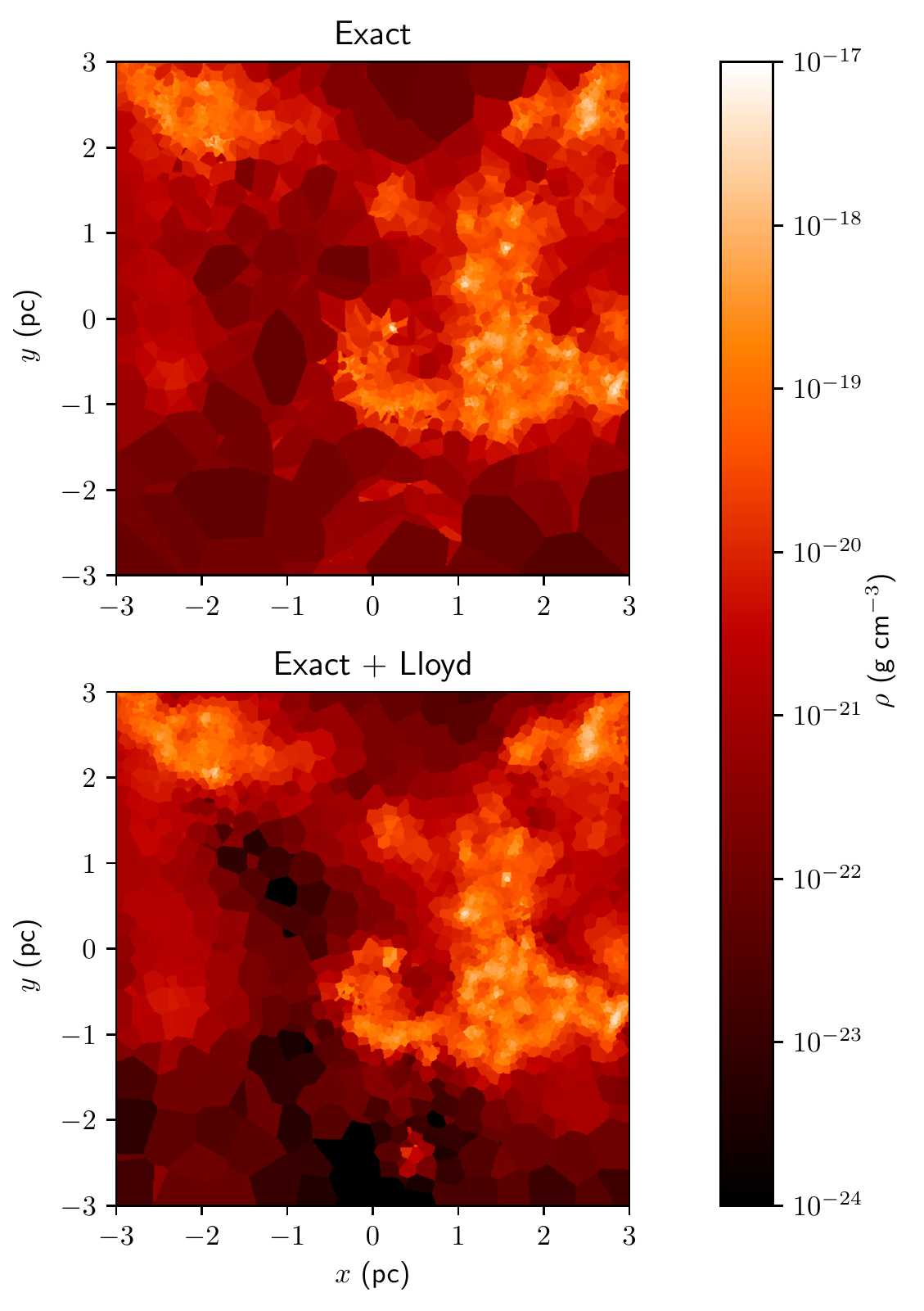}
	\caption{Cross-section images of the starting time Voronoi grids at $z=-0.51$ pc. The colour bar corresponds to the exact volume density, $\rho_{\rm p}$ (see equation~\ref{eq:exact}), mapped onto a Voronoi grid built around the particles' positions (\textit{top}) and a regularised Voronoi grid (\textit{bottom}).}
	\label{fig:clumpy-voro-grid}
\end{figure}

\begin{figure}
	\centering
	\includegraphics[width=\columnwidth]{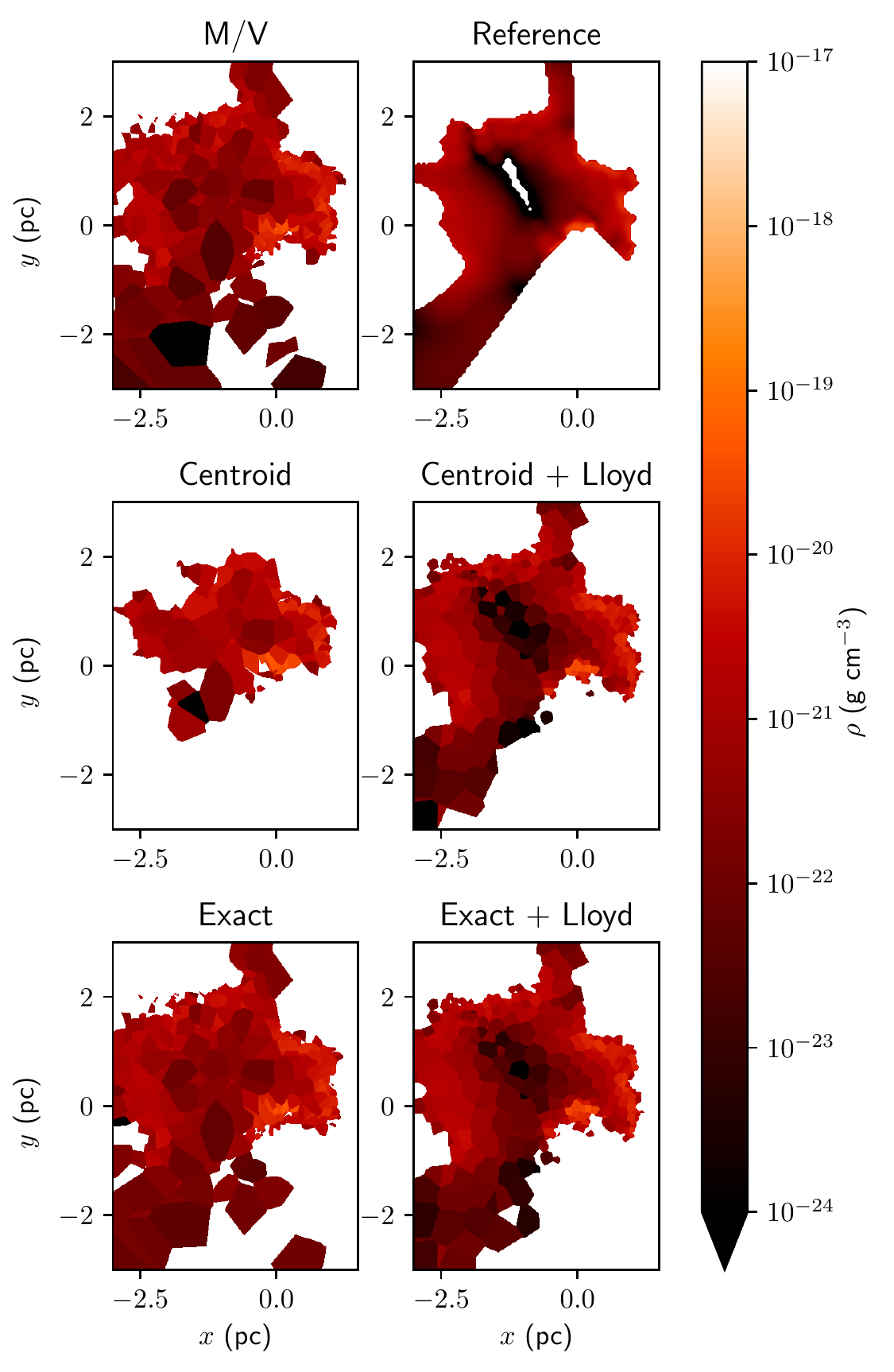}
	\caption{Cross-section images of the initially ionised grid cells at $z=-0.51$ pc with $\dot Q = 5\times 10^{50}$ s$^{-1}$ and a source positioned at low density environment within the Brick. The colours correspond to the density of the ionised gas only. The different panels correspond to the five different simulations described in Section~\ref{sec:clumpy}, with an additional reference run. The reference simulation has been performed on a high-resolution ($1024^3$) Cartesian grid. For comparison, each Voronoi grid consists of 101128 cells.}
	\label{fig:clumpy-ionised-density}
\end{figure}

While the choice of grid structure and density mapping prescription does not significantly affect the StarBench test, in the Brick simulations we see that the amount of ionised gas can vary greatly between the different configurations. We examine these variations, both in terms of the time evolution of the H~II region (see Figure \ref{fig:clumpy-ionised-mass}), and in terms of the initial ionised mass.

As already discussed in Section \ref{sec:clumpy-timestep}, the different simulation setups produce 
qualitatively
the same expansion of their respective H~II regions (with 
some exception for
`B III' with the high density source). Furthermore, if we compare the five simulations with the low density source, we see that the ionised mass curves appear to be systematically offset from one another. The simulations with initially lower or higher ionised mass, continue to have lower or higher ionised mass at later times, relative to the other simulations. This may indicate that the differences between the time evolution of the simulations are primarily caused by differences in the initial ionised mass. Alternatively, the conditions which create the initial mass discrepancies (grid and density variations) may persist in time, and continuously maintain the discrepancies. 
Similar systematic behaviour is present in the simulations with an embedded source. However, since the expansion curves are noisier, there is no clear distinction between the results of `B~I', `B~II' and `B~IV'. Overall the relative order of the expansion curves of the simulations with an embedded source is similar to that of the curves of the low density source,
which favours the interpretation that the different grid and density configurations maintain differences in the expansion continuously.

We will now focus on comparing the individual simulations with one another. Let us first consider the basic Voronoi grid and the regularised one with the same type of density mapping. Visually, the regularised grid resolves the low- and intermediate--density gas better (see Figure~\ref{fig:clumpy-voro-grid}). Due to the regularisation, however, the grid cells become more evenly distributed and hence the density peaks are represented by fewer cells, causing them to be under--resolved. This means that an embedded ionising source
may experience
a smoothed density profile in its surroundings, with lower peak densities. As a result this source initially ionises a larger region compared to if it was coupled with a basic Voronoi grid structure. 

In addition to the structure of the Voronoi grid, the type of density mapping also plays an important role in the outcome of the Brick simulations. The greatest outlier in Figure~\ref{fig:clumpy-ionised-mass} is the curve, corresponding to the Centroid method, combined with the basic Voronoi grid (`B III'). This simulation setup significantly over-predicts the cell mass (by a factor of $\sim 3$; see Appendix~\ref{sec:dens-mapping}), and it results in highly suppressed ionised mass, especially when the source is embedded in a clump. The inaccuracy of the Centroid method is less pronounced when paired with a regularised grid (`B IV'; only $\sim 9$\% higher than the total particle mass), due to the cell shapes. As a result, the behaviour of the simulation is similar to that with the exact density mapping (see the two dashed lines in Figure~\ref{fig:clumpy-ionised-mass}). 
Finally, the density mapping in which a particle mass is divided by the cell volume (`M/V'; `B V') agrees reasonably well (within a few tens of percent) with the ionised mass produced by the exact density mapping on a basic Voronoi grid (`B I'). This can be attributed to the fact that while the local cell density may have inaccuracies, on larger scale the average density is correct, since the method conserves mass.

The disagreement between the five different simulations raises the question of which one gives the most accurate solution to the problem. Since we do not have an analytic solution in the general case of non-uniform density medium, we have attempted to answer this question numerically. We have applied \textsc{CMacIonize} to a high-resolution Cartesian grid without performing any subsequent hydrodynamics calculations due to high computational cost. The density mapping has been performed using the Centroid method (eq.~\ref{eq:centroid}). Since the method does not necessarily conserve mass, as previously discussed, the resolution of the Cartesian grid has been gradually increased until both the total mass of the medium and the ionised mass had converged. The convergence is achieved for a grid with $1024^3$ cells. Note that the cell size of this Cartesian grid is $0.006$ pc, while the smallest smoothing length in the Brick simulation is $0.014$ pc, which allows the grid to capture features that might not be resolved in the hydrodynamics.

A comparison of the initial ionised gas density between the five simulations together with the Cartesian grid is shown in Figure~\ref{fig:clumpy-ionised-density}. The figure contains a density slice at $z=-0.51$ pc, and the source has been positioned in the lower density location ($x=0$ pc, $y=0.3$ pc, $z=-0.51$ pc). We can see that the simulations with regularised Voronoi grids are in better agreement with the Cartesian grid (labelled as `Reference'\footnote{Note that the `Reference' image contains a white region in the middle of the ionised volume. This does not indicate a clump of neutral gas, but it is instead a region, which is completely empty of gas (ionised or neutral).}). Furthermore, if we compare the simulations with basic Voronoi grid, we can see that the `Exact' and `M/V' runs agree quite well with each other, and are still relatively similar to the Cartesian grid result, while the `Centroid' run differs the most from all the others.

\begin{table}
\caption{Initial ionised mass in the Brick simulations. The ionised mass is calculated by multiplying the ionic fraction of a particle by the particle's mass and summing over all particles. The reference value is obtained by performing the radiative transfer on a high-resolution ($1024^3$) Cartesian grid.}
  \centering
  \begin{tabular}{ccc}
    \hline
    Simulation & $M_{\rm ion}$~[M$_{\odot}$] & $M_{\rm ion}$~[M$_{\odot}$]\\
    name & (high density) & (low density)\\
    \hline
    \hline
    B~I & 22.1 & 573 \\
    B~II & 30.5 & 499 \\
    B~III & 2.0 & 240 \\
    B~IV & 33.0 & 469 \\
    B~V & 25.6 & 549 \\
    Reference & 18.3 & 340 \\
    \hline
  \end{tabular}
\label{table:brick-initial}
\end{table}

Additionally, we have calculated the total ionised mass (including contributions from the partially ionised cells) obtained with the Cartesian grid. In the case of the embedded source, the total ionised mass is 18.3~M$_{\rm{\odot}}$, while in the case of the low--density source it is 340~M$_{\rm{\odot}}$. Both of these values are lower than the numbers obtained by four of the five RHD simulation setups (see Table~\ref{table:brick-initial}). In particular, the simulations using the exact density mapping and the two types of Voronoi grids, which are expected to produce the most accurate results, overestimate the ionised mass relative to the grid. The only setup which produces lower ionised mass is the `Centroid' run, which overestimates the total mass due to lack of mass conservation, as discussed above. The disagreement between ionised mass computed on the Cartesian grid and that on the Voronoi grids likely originates from the limited resolution of the latter, and it will be discussed further in Section~\ref{sec:clumpy-res}. Note that here we compare the total ionised mass as "seen" by the raditative transfer (i.e. the sum of ionic fraction times particle mass), as opposed to the initial ionised mass in Figure~\ref{fig:clumpy-ionised-mass}, where we have the total ionised mass as "seen" by the hydrodynamics (i.e. the total mass of all particles with ionic fraction over 0.5). This choice has been made for consistency with the Cartesian test.

\begin{figure}
	\centering
	\includegraphics[width=\columnwidth]{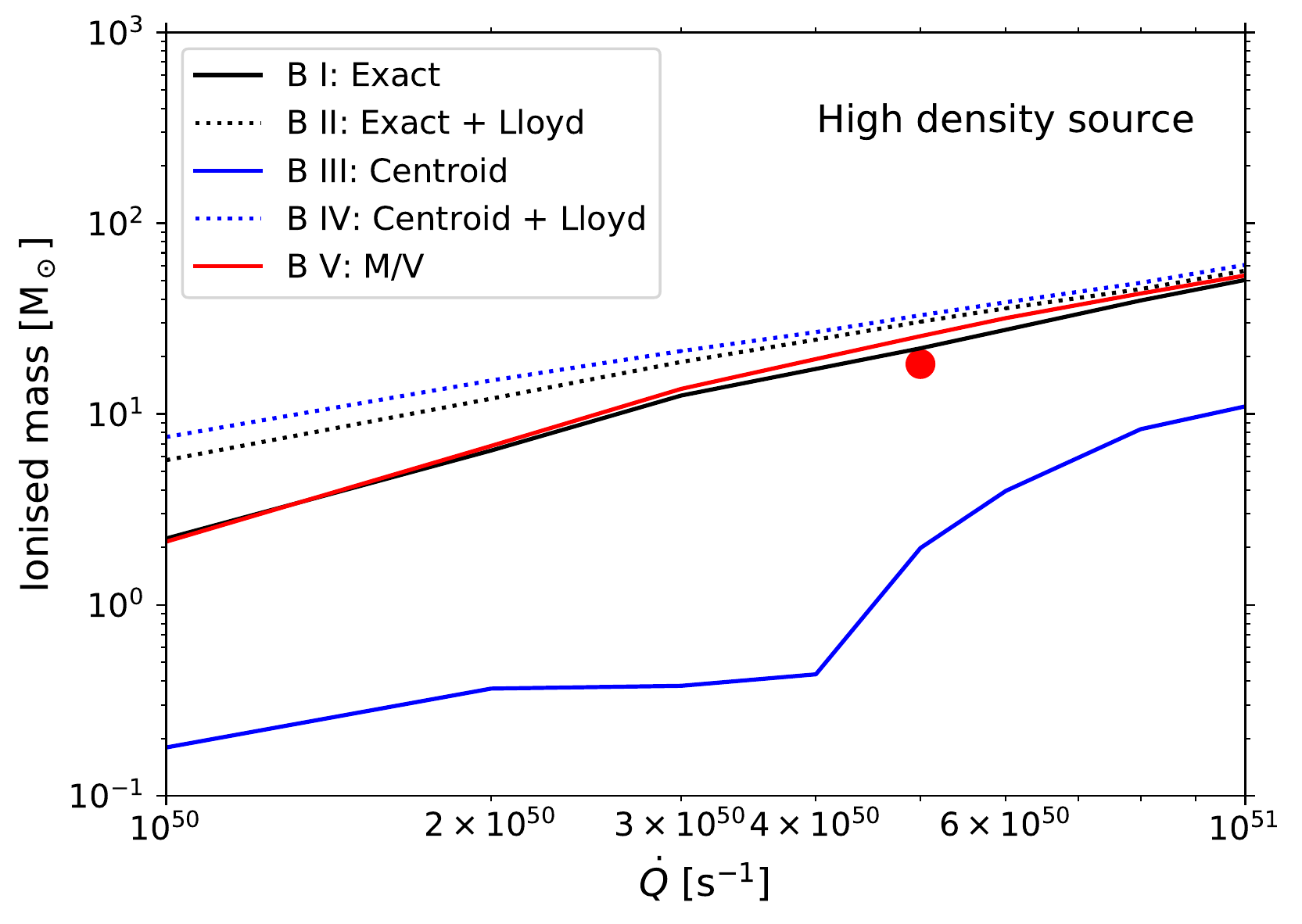}
	\includegraphics[width=\columnwidth]{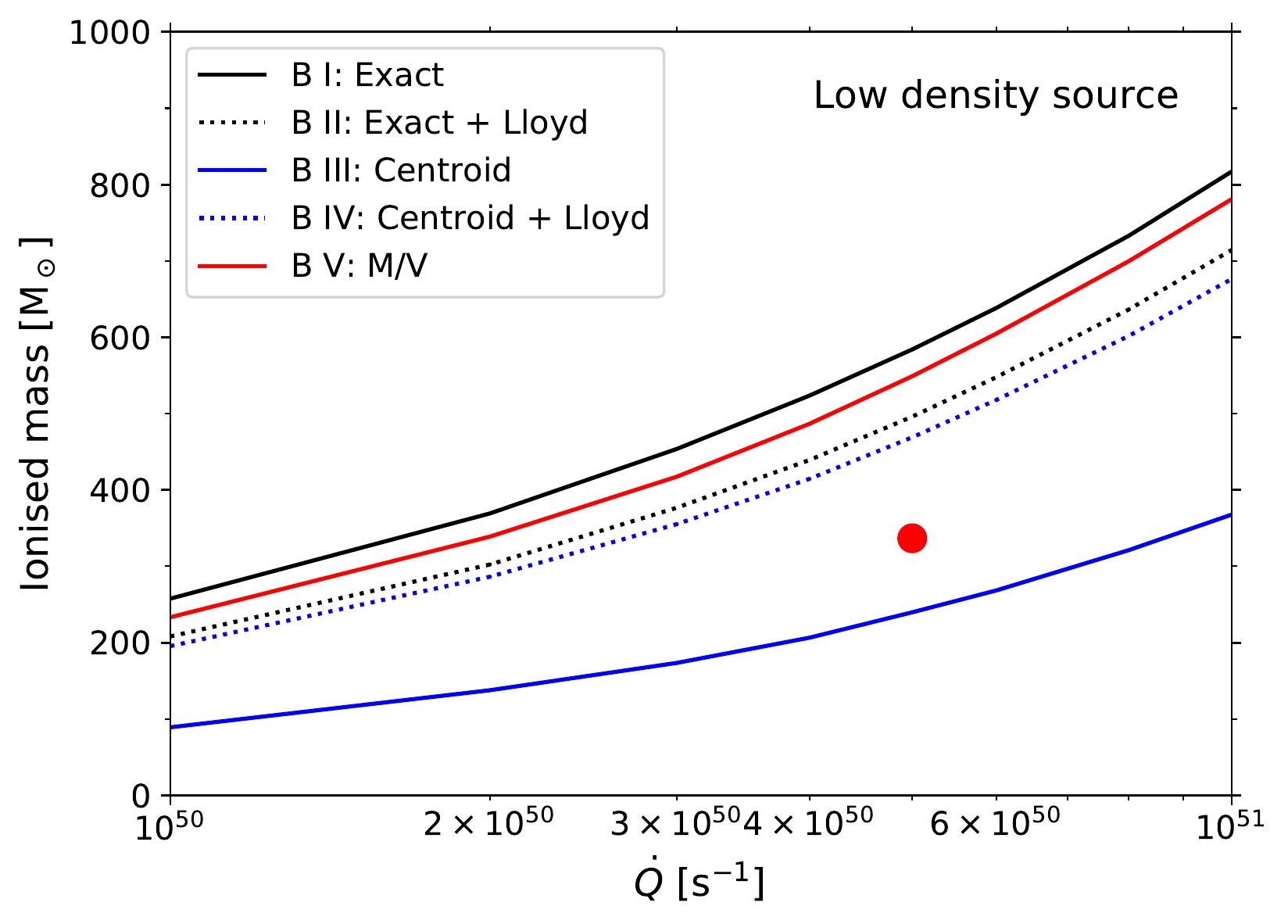}
	\caption{Initial ionised mass as a function of ionising luminosity ($\dot Q$) for the simulations with embedded source (\textit{top}) and the simulations with a source in a low-density environment (\textit{bottom}) within the Brick. The different lines represent runs with different Voronoi grids and types of density mapping, and have the same meaning as in Figure~\ref{fig:clumpy-ionised-mass}. The red symbols have been obtained using high resolution ($1024^3$) Cartesian grid.} 
	\label{fig:clumpy-lum-function}
\end{figure}

 Finally, we also study the initial ionised mass (from the radiative transfer calculation) for different values of the ionising luminosity, $\dot Q$ (see Figure~\ref{fig:clumpy-lum-function}). This is achieved by performing individual MCRT runs without evolving the cloud in time. Figure~\ref{fig:clumpy-lum-function} demonstrates that as we increase $\dot Q$, the initial ionised mass also increases, as expected. Note that the red data points represent the results of the Cartesian grid tests, mentioned above (and shown in Table~\ref{table:brick-initial}).
 We also see that the different simulation setups produce curves that do not cross each other as $\dot Q$ is varied. This demonstrates that there are systematic differences, resulting from the choice of grid and density mapping. The order of the curves (from higher to lower mass) in the two panels of Figure~\ref{fig:clumpy-lum-function} differs between the embedded and the low density source, which emphasises the importance of the gas density distribution around the source.

\subsection{Resolution}
\label{sec:clumpy-res}
Despite the fact that we have not altered the resolution of the Brick setup, it is crucial to consider the resolution when discussing our radiation-hydrodynamics results. Since the H~II region has an irregular shape, the most reliable measure of resolution is the number of initially ionised particles. For the simulations with an embedded source this number is under 50, while for the low density source ones it is under 1000. Compared to the uniform density tests, these simulations fall within the lower end of the resolutions that have been considered (see Table~\ref{table:mass-res}). Therefore, from a numerical perspective, we expect that the 
expansion of the H~II regions within the Brick is likely to be underpredicted. Note that in our lower resolution StarBench tests the initial ionisation radius and mass are typically overpredicted (see Figure~\ref{fig:mass-resolution} and~\ref{fig:ionised-mass-sb}), but despite that the expansion is slower than in the higher resolution runs. This may be caused by the time scale required for the formation of the shock, which is more prolonged at lower resolution.

The overprediction of the initial ionised mass present in the low resolution StarBench test is consistent with overpredicted initial ionised mass in the Brick, when compared to the Cartesian reference test (see Table~\ref{table:brick-initial} and Figure~\ref{fig:clumpy-lum-function}). Indeed, the curves of the simulations with the exact density mapping (`B~I' and `B~II'; black lines), which we believe to be the most accurate estimate of the ionised mass, lie above the red data points in Figure~\ref{fig:clumpy-lum-function}.

These results indicate a limitation of the RHD scheme, related to the fact that SPH is optimised for resolving fluid elements of equal mass, whereas solving for the ionisation state of a diffuse medium requires high resolution at the ionisation front (this is illustrated by the outlines of the ionised regions in Figure~\ref{fig:clumpy-ionised-density}). This limitation can be addressed either by improving the resolution of the grid structure, or by sufficiently increasing the mass resolution in SPH. The first approach can be achieved by inserting additional generating sites in the Voronoi grid via methods such as the one proposed by \cite{Koepferl2017}, or by choosing a different high-resolution grid structure altogether, such as a Cartesian grid or a grid constructed with an active mesh refinement scheme. The increase of resolution can, however, significantly increase the computational cost of the RHD scheme to the point where it becomes prohibitive. Additionally, increasing the resolution of the grid while keeping the number of SPH particles the same is unlikely to improve the time evolution of the ionised gas. Therefore, it is crucial that we select appropriately small SPH particle mass for the specific problem that we consider.

\subsection{Physical implications for the Brick}
In addition to using the Brick setup for numerical tests, our limited radiation-hydrodynamics simulation can already be used towards answering scientific questions. 
As mentioned in the introduction, we want to find out (i) the ionised mass of a newly formed H~II region and (ii) its rate of increase.
The simulations of \citet{Dale2019} and \citet{Kruijssen2019} provide a natural starting point for answering these questions numerically, since they are currently the only existing models of the Brick capturing the effects of the galactic gravitational potential on the structure of the cloud, and reproducing key observable properties, such as the aspect ratio, the column density, the line of sight velocity dispersion and the gradient of the line of sight velocity \citep{Kruijssen2019}. Our RHD runs of the Brick use a snapshot of one of these simulations as initial conditions. However, we have made crucial physical simplifications (such as using a sub-region of the cloud and omitting gravity
) for the purpose of our tests. Therefore, we will discuss the impact of these simplifications while addressing the above questions.

The first of the above questions can be directly answered from our results, and is not affected by our choice of physical assumptions. The size/mass of a newly formed H~II region depends solely on the luminosity and position of the source, as well as the density distribution of the gas. However, as discussed in Section~\ref{sec:discussion-grids}, the density distribution of the gas is sensitive to the choice of grid type and density mapping method. Using Figure~\ref{fig:clumpy-lum-function}, we can conclude that the low density source initially ionises 340--600~M$_{\rm{\odot}}$, while the embedded source ionises only 18--31~M$_{\rm{\odot}}$, for an ionising luminosity of $\dot Q = 5 \times 10^{50}$ s$^{-1}$. The ranges for these values have been obtained by considering the results of the Cartesian grid (the lower value of each range), and those of the simulations with exact density mapping (`SB I' and `SB II'). Even though the results of the Cartesian grid are more accurate due to its higher resolution at the ionisation front, it is useful to consider the full ranges, as they give estimates of the ionised mass uncertainty due to resolution limitations in the RHD simulations. Keeping these uncertainties in mind, we will be able to better estimate the ionised mass for other luminosities, where we do not have Cartesian results.

The ionising luminosities that we have considered in Figure~\ref{fig:clumpy-lum-function} are quite high for single sources (under the assumption of a well-sampled stellar initial mass function, they span stellar population masses of $2{-}20\times10^3~\rm{M_{\odot}}$, \citealt{leitherer14}). This choice has been made because of the embedded source, of which the H~II region otherwise remains sub--grid. Even with these high values of $\dot Q$, the embedded source typically ionises only a few SPH particles (with $m_{\rm{part}}=0.4463$ M$_{\rm{\odot}}$). This is an indicator that a young, high-mass star in this simulated cloud would likely have a very compact and trapped H~II region. Therefore, it appears that only a star which has accreted its natal clump or has been expelled from its birth environment to a low gas density location can substantially ionise the cloud. This result is consistent with the findings of \citep{Sartorio2021}, who studied single-source ionisation in a turbulent box, and found that placing the source next to a filament produced a small H~II region and only had a minor impact on the gas dynamics. 

A possible limitation of our simulations is the fact that, in practice, embedded stars accrete material through a disk, and the lower density environment above and below the disk can facilitate a larger H~II region. Furthermore, earlier feedback in the form of stellar wind can carve out bipolar low density cavities, which subsequently become ionised \citep{Kuiper2018}. All of these considerations appear on size (and mass) scales much smaller than our particle resolution, and, hence, are not captured by our RHD simulations. However, while these sub-resolution considerations can affect the initial size and shape of the H~II region of the embedded source, our results provide at least an order of magnitude estimate of the initial ionised mass. Note that for the low density source, we do not expect any significant H~II region variation due to sub-grid geometry. 

The second question that we have posed (on how rapidly the  mass of the H~II region increases) is also answered by our RHD simulations. From Figure~\ref{fig:clumpy-ionised-mass}, we can measure that the expansion rate of the H~II region in low density environment is $\sim 7-9 \times 10^{-3}$~M$_{\rm \odot}$~yr$^{-1}$,
while that of the embedded H~II region is $\sim 1.6-1.9 \times 10^{-3}$~M$_{\rm \odot}$~yr$^{-1}$. In this case, however, our assumed physical simplifications do affect the results. The fact that we have neglected the effects of gravity means that the denser structures cannot be held together and diffuse out over time (see Figures \ref{fig:clumpy-hii-expansion} and \ref{fig:clumpy-hii-expansion-ld}). As a result, more of the gas enters the outskirts of high-density regions and becomes ionised, implying that the time-evolution results in Figure~\ref{fig:clumpy-ionised-mass} should represent upper limits on the H~II region expansion. However, we can see in Figures~\ref{fig:clumpy-hii-expansion} and \ref{fig:clumpy-hii-expansion-ld} that the majority of the high density structures survive the H~II region expansion and remain neutral, which suggests that the diffusion is likely to have a minor impact on the ionised mass, as the H~II region is expanding in a low-density environment. Some uncertainty remains about whether or not the embedded source would be able to disperse its natal clump on the time-scale that we observe in our simulations when gravity is present and able to disrupt the H~II region expansion. By contrast, from a numerical perspective, the measured H~II region expansion is suppressed due to insufficient spatial resolution (as discussed in Section~\ref{sec:clumpy-res}), and thus represents a lower limit on the expansion. However, according to Figure~\ref{fig:ionised-mass-sb}, this has an effect on the size and mass of the H~II region, but not on its rate of expansion, and hence it does not interfere with our measurement. 

Despite the uncertainty in the size of the H~II regions, we see a physically plausible spatial distribution of the ionised gas. The H~II region of the low density source evolves to fill up the low density environment, with only minor destruction done on some of the clumpy structures. Similar behaviour has been observed by \cite{Dale2011} and \citep{Sartorio2021}, who find that ionising feedback can sometimes be inefficient at carving out voids and the H~II region would spread into pre-existing low density regions instead. In contrast, the embedded H~II region manages to expand within its clump, but our limited setup does not allow us to determine if it would be able to destroy the cloud.

An interesting additional question is whether or not the H~II region expansion can trigger star formation. This question is more relevant for the embedded source, where the expansion of the ionised gas sweeps up a shock. Unfortunately, due to the lack of self-gravity, the pre-existing dense cores diffuse out, as previously discussed. As a result, the shock is not able to compress them beyond the tipping point of star (or sink) formation. Furthermore, the relatively low resolution of the embedded H~II region means that the shock is quite thick, and hence its density never reaches high enough values to trigger fragmentation via the collect-and-collapse mechanism. Therefore, the question of triggered star formation will be addressed in a further study.

\section{Conclusions and recommendations}
\label{sec:conclusion}
We have introduced a novel radiation-hydrodynamics scheme combining SPH and MCRT, and linking them via a Voronoi grid. We have demonstrated that our scheme can correctly model the hydrodynamics of an expanding H~II region in uniform density environment. Furthermore, we have applied the RHD scheme to a clumpy cloud model of the Brick \citep{Dale2019,Kruijssen2019}, which is a dense and massive cloud in the CMZ of the Milky Way, thought to be the progenitor of a young massive cluster \citep{Longmore2012}.

As part of our work, we have explored varying the ionisation time step, the SPH resolution, and we have used two different kinds of Voronoi grid structure and three types of density mapping from the particles onto the grid cells. The results of these experiments allow us to formulate recommendations for future simulations, involving SPH, MCRT and Voronoi grids, which we include below.

We recommend the use of a timestepping criterion, which we have derived in eq.~\ref{eq:dt-criterion}. The criterion is similar to that of \citet{Courant1928} and other timestepping conditions that are otherwise employed in SPH codes \citep[e.g.][]{Price2018}. The criterion can ensure convergence in non-dynamical initial conditions. However, additional care should be taken when running simulations with a significant velocity dispersion.
 
We have studied the effects of varying the spatial resolution of the uniform-density simulations by increasing or decreasing the particle mass. We have found that lowering the resolution causes suppressed expansion of the H~II region due to thicker shocks. Based on our results, we can deduce that we need at least a few thousand initially ionised particles in order to have a good model of D-type expansion in uniform medium. Based on our Brick simulations, we recommend that a similar criterion is also used in simulations in a clumpy medium.

We have performed our simulations with two different types of Voronoi grids and three different types of density mapping (see Section~\ref{sec:numerics}). When comparing the basic to the regularised Voronoi grid, some preference should be given to the latter, as it represents the overall structure of the diffuse medium better. However, when we consider the simulation outcome, the better choice of Voronoi grid is situation dependent. Choosing one grid over the other makes no significant difference to the expansion of the uniform density simulations. In the case of the Brick simulations, regularising the Voronoi grid improves the outcome when the source is in low-density environment and worsens it when the source is embedded, as the regularised grid underresolves density peaks. In terms of the density mapping method, we always recommend using the exact density mapping (see eq.~\ref{eq:exact}), when the computing time is not an issue. If it is not possible to use the exact mapping, the choice between dividing the particle mass by the cell volume (eq.~\ref{eq:MonV}), or using the centroid method (eq.~\ref{eq:centroid}) depends on the choice of Voronoi grid. In the Brick simulations, we find that using the former approximate density mapping method gives similar results to the exact mapping on the basic Voronoi grid. By contrast, the centroid method produces a similar outcome to the exact mapping on the regularised Voronoi grid. Note that all types of density mapping result in very similar simulation results in uniform medium.

In our simulations of the Brick with a source in a low-density environment we can see that the ionised material is preferentially located in pre-existing low-density regions. This does not have a strong effect on the cloud morphology, and it is a phenomenon which has been seen in previous simulations.
In contrast, our simulations of the Brick with a source embedded in a dense clump result in H~II region expansion that sweeps up a shock.
Our results also demonstrate that we need the ionising luminosity of a stellar population of several $10^3~\rm{M_{\odot}}$ in order to ionise material outside the clump of an embedded source, at the extreme gas densities observed in the CMZ. However, the necessary ionising luminosity may be overestimated due to insufficient resolution. In particular, resolving the accretion disc of a young star, and/or including stellar winds, may accelerate the ionisation by allowing ionising photons to escape via low density bipolar cavities \citep{Kuiper2018}.

The RHD scheme proposed in this paper contains a careful and accurate treatment of ionising radiation in combination with the large dynamical range advantages of SPH. As a result, it can be a useful tool for future studies of star formation and ionising feedback. Furthermore, the results presented here will serve as a guide for choosing time steps, resolution, type of Voronoi grid and density mapping for future simulations.

\section*{Acknowledgements}
MAP would like to thank Daniel Price and James Wurster for their technical assistance with \textsc{Phantom}, as well as Jim Dale for useful discussions.

MAP and IAB acknowledge funding from the European Research Council for the FP7 ERC advanced grant project ECOGAL. MAP and JMDK gratefully acknowledge funding from the European Research Council (ERC) under the European Union's Horizon 2020 research and innovation programme via the ERC Starting Grant MUSTANG (grant agreement number 714907). JMDK gratefully acknowledges funding from the German Research Foundation (DFG) in the form of an Emmy Noether Research Group (grant number KR4801/1-1). BV acknowledges partial funding from STFC grant ST/M001296/1 and currently receives financial support from the Belgian Science Policy Office (BELSPO) through the PRODEX project "SPICA-SKIRT: A far-infrared photometry and polarimetry simulation toolbox in preparation of the SPICA mission" (C4000128500).

Part of this work was performed using the DiRAC Data Intensive service at Leicester, operated by the University of Leicester IT Services, which forms part of the STFC DiRAC HPC Facility (www.dirac.ac.uk). The equipment was funded by BEIS capital funding via STFC capital grants ST/K000373/1 and ST/R002363/1 and STFC DiRAC Operations grant ST/R001014/1. DiRAC is part of the National e-Infrastructure.

The SPH figures in this paper were created using SPLASH (\citealt{Price2007}).

\section*{Data Availability}
The data underlying this article will be shared on reasonable request to the corresponding author.




\bibliographystyle{mnras}
\bibliography{rad-hydro} 



\appendix

\section{Computing time}
\label{sec:computing-time}
\begin{figure}
	\centering
	\includegraphics[width=\columnwidth]{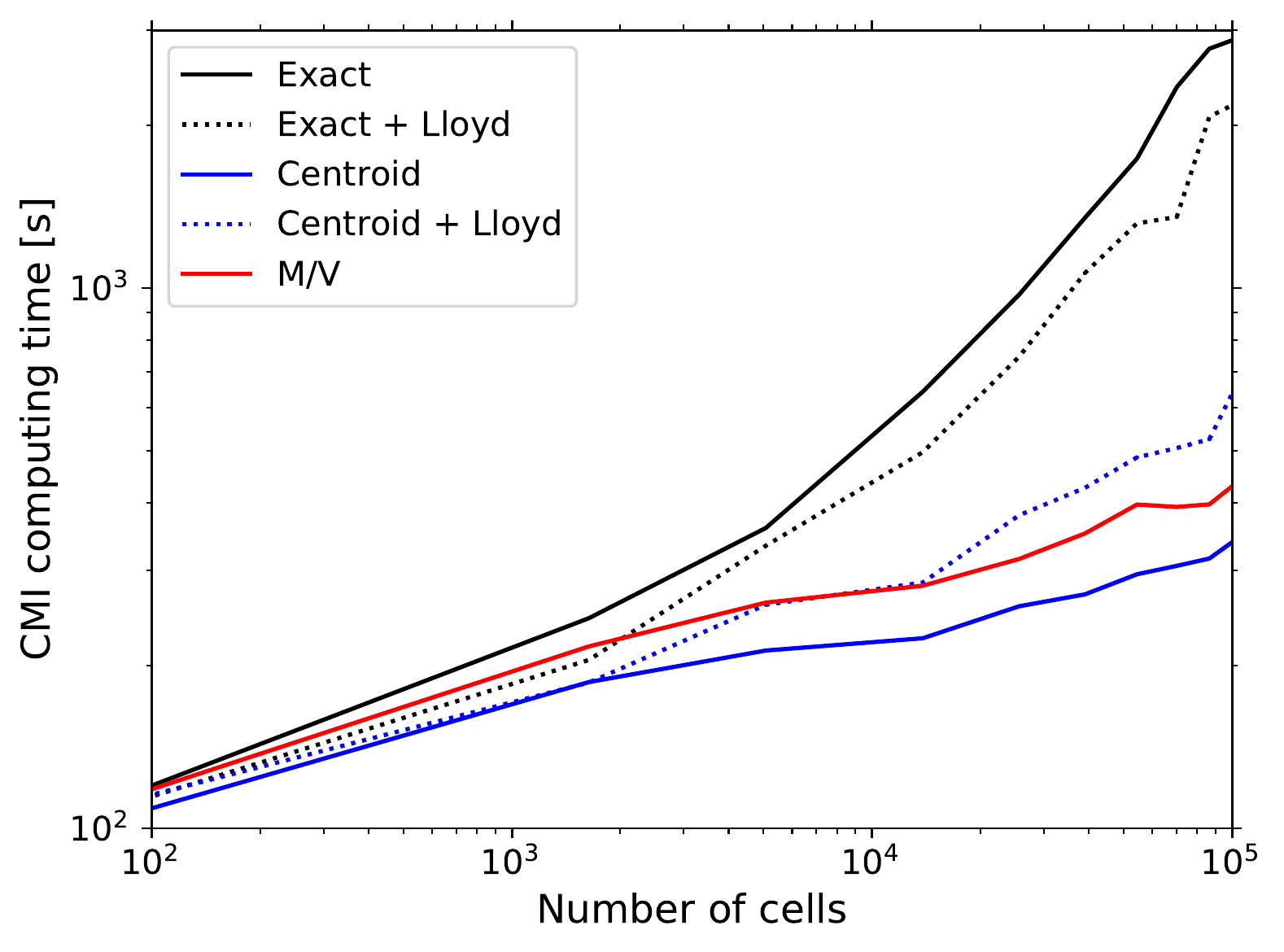}
	\includegraphics[width=\columnwidth]{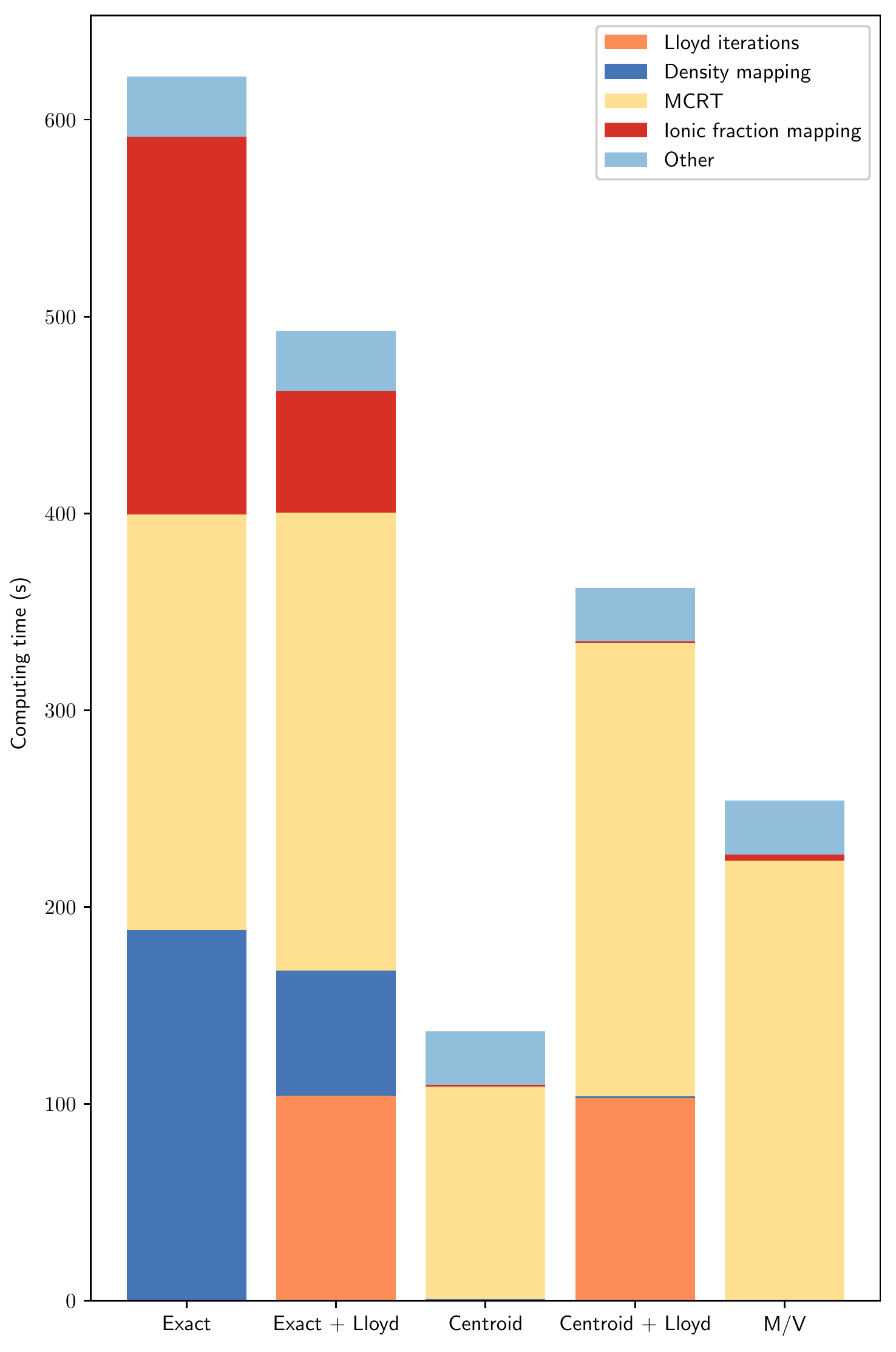}
	\caption{Computing time of one \textsc{CMacIonize} call for the different combinations of density mappings and Voronoi grids. \textit{Top:} the computing time as a function of number of particles/cells that \textsc{CMacIonize} has been applied to. \textit{Bottom:} breakdown of the computing time for different parts of the CMI library.}
	\label{fig:timing}
\end{figure}

The computing time of the simulations presented in this paper is largely dominated by the radiative transfer. This is due to the fact that we have used very simplified treatment of the hydrodynamics by assuming pure atomic hydrogen with no cooling and by neglecting gravity. As a result here we will focus on the computing time of the CMI library.

Figure~\ref{fig:timing} shows the difference in computing time (normalised to a single CPU) between the different density mapping methods and Voronoi grids used within \textsc{CMacIonize}. These are obtained for simulations using the clumpy initial conditions of the Brick. In order to plot the computing time as a function of cell number, we have repeated the CMI runs for a series of progressively larger cubical volumes, enclosing different fractions of the simulated region.

In the top panel of Figure~\ref{fig:timing} we can see that while all methods have similar performance when there are fewer particles and grid cells, the use of the exact density mapping (given by eq.~\ref{eq:exact}) results in considerably longer computing time for the case of large number of particles/cells. 
The difference in slope between the curves can be explained by considering the main computational cost of each method. As the number of cells is increased, the simulations using the exact method become more significantly dominated by the mapping algorithms, which depend roughly linearly on the number of cells. This can be seen in the upwards turn of the black curves towards a slope of one. The simulations using the other two methods, however, are dominated by the cost of the radiative transfer. Since the number of ionised particles increases sub-linearly with the number of cells (and eventually saturates), the slope of the blue and red curves remains shallower.

This behaviour is emphasised again in the bottom panel of Figure~\ref{fig:timing}, where we have included a breakdown of the different components of the CMI runs for the full simulation volume\footnote{Note that there is a small discrepancy between the computing times in the top and the bottom panel. This is due to the fact that the simulations for each panel were performed on different machines. Despite that, the trends with the grid and density mapping choices persist.}. For the exact method the computing time of the CMI library is dominated by the mapping of density from the particles to the grid cells (red bar), together with the mapping of ionic fraction from the grid cells to the particles (dark blue bar). Meanwhile, when one of the other two methods is used, these mappings add a negligible cost to the CMI computing time.

Furthermore, the use of Lloyd's iterations to regularise the Voronoi grid, alters the total CMI run time. In simulations with the exact mapping, the total CMI time is decreased due to the reduced cost of the mapping. This is due to the fact that the mapping method scales with the number of cell faces, and more regular grid cells have, on average, fewer faces. On the other hand, when the Lloyd's iterations are performed together with the centroid method (see eq.~\ref{eq:centroid}), the cost of the mapping remains the same and the time spent on the regularisation (orange bar) becomes an overhead. 

For all five simulations the computing time of the radiative transfer itself (yellow bar) remains roughly the same. The slight fluctuations come from the fact that the different density distribution and cell arrangement result in different path length (in terms of number of visited cells) of the photon packets. In particular, when we use Lloyd's iterations more cells are positioned in regions of low and intermediate density, which is where most of the material becomes ionised. This is why the use of Lloyd's iterations can slightly increase the cost of the radiative transfer. Note that when we apply the centroid method without Lloyd's iterations the total cell mass is much greater than the mass contained in the particles (see Appendix~\ref{sec:dens-mapping}). This results in overall fewer ionised particles, and therefore the radiative transfer takes less time compared to the other simulations.

\section{Volume density distribution of the Brick}
\label{sec:dens-mapping}
\begin{figure}
	\centering
	\includegraphics[width=\columnwidth]{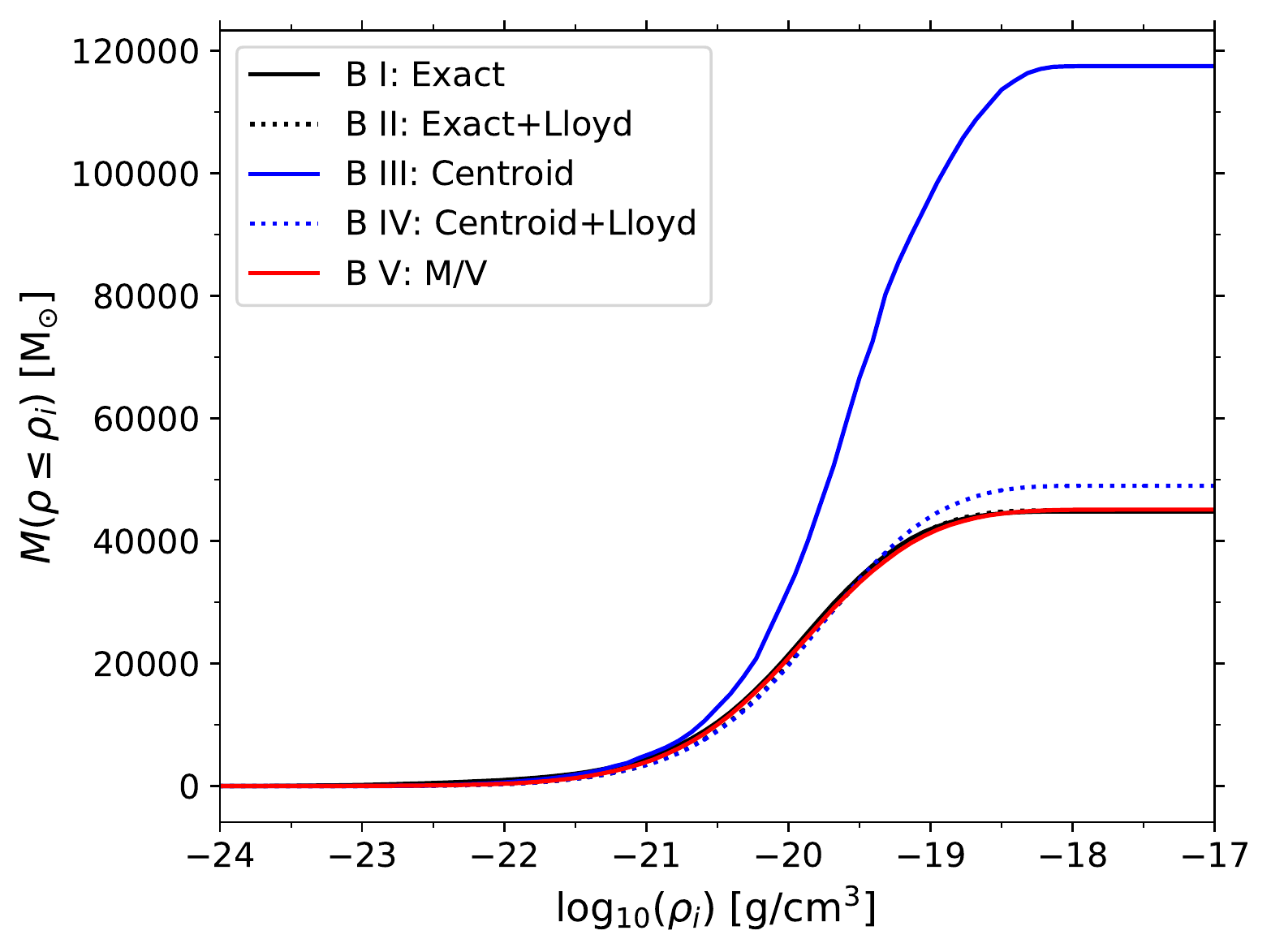}
	\caption{Cumulative distribution of mass contained within different density bins of the initial snapshot of the Brick simulation volume. The different lines correspond to the different combinations of Voronoi grid types and density mapping methods. Note that we have used 100 density bins, which have a constant width in log space.}
	\label{fig:mass-dens}
\end{figure}

To further illustrate how the different combinations of Voronoi grid type and density mapping method affect the structure of the Brick simulation when mapped onto a grid, we present the cumulative cell mass as a function of density in Figure~\ref{fig:mass-dens}. The mass has been calculated by assigning grid cells to density bins, and summing up the products of cell densities and cell volumes. Note that we have used 100 density bins, spanning the range $10^{-25}-10^{-16}$~g~cm$^{-3}$, where the width of each density bin is constant in log space. Figure~\ref{fig:mass-dens} demonstrates that the centroid method of density mapping does not conserve mass between the SPH dataset and the Voronoi grid. This effect is most pronounced in the `B III' setup (solid blue curve), where the total mass contained in the grid is $\sim 3$ times higher than that in SPH. Moreover, within the density range of a few $10^{-21}$~g~cm$^{-3}$ and a few $10^{-19}$~g~cm$^{-3}$, the blue solid curve is steeper than all of the others, which indicates that the mass is continuously overestimated for these densities. A similar, but much milder behaviour is seen by the blue dotted curve (`B IV'), which corresponds to the centroid method being applied to a regularised Voronoi grid. In this setup, the total mass is overestimated by $\sim 9\%$, and most of this excess mass is located at the mid- to high-density range ($\rho_i > few \times 10^{-20}$~g~cm$^{-3}$). The rest of the curves correspond to density mapping methods which conserve the total SPH mass, and Figure~\ref{fig:mass-dens} shows that they have very similar mass distributions.


\bsp	
\label{lastpage}
\end{document}